\documentclass[11pt, a4paper]{article}
\usepackage[pdfauthor={P. P. Cook, M. Fleming},
pdftitle={Gravitational Coset Models}]{hyperref}
\usepackage{amsmath}
\usepackage{graphicx}
\usepackage{youngtab}
\usepackage{ytableau}
\usepackage{amssymb}
\usepackage{amsthm}
\usepackage{chngcntr}
\usepackage{xy}
\usepackage{longtable}
\usepackage{bm}
\usepackage{mathtools}
\counterwithin{figure}{section}
\counterwithin{table}{section}
\newcommand{\comment}[1]{}
\renewcommand{\vec}[1]{{\bm{#1}}}

\renewcommand{\d}{{\mathfrak d}}

\newfont{\gothic}{ygoth at 12pt} 

\advance\voffset by -1.6cm
\advance\hoffset by -1.7cm
\begin{document}
\title{Gravitational Coset Models}
\author{Paul P. Cook\footnote{email: paul.cook@kcl.ac.uk},Michael Fleming\footnote{email: michael.fleming@kcl.ac.uk} \\ 
{\itshape Department of Mathematics, King's College London \\ 
The Strand, London WC2R 2LS, UK}
}
\begin{titlepage}
\begin{flushright}
KCL-MTH-13-09 \\
\end{flushright}
\vspace{70pt}
\centering{\LARGE Gravitational Coset Models}\\
\vspace{30pt}
\def\thefootnote{\fnsymbol{footnote}}
Paul P. Cook\footnote{\href{mailto:paul.cook@kcl.ac.uk}{email: paul.cook@kcl.ac.uk}} and 
Michael Fleming\footnote{\href{mailto:michael.fleming@kcl.ac.uk}{email: michael.fleming@kcl.ac.uk}}\\
\setcounter{footnote}{0}
\vspace{10pt}
{\itshape Department of Mathematics, King's College London \\ 
The Strand, London WC2R 2LS, UK}\\
\vspace{30pt}
\begin{abstract}
The algebra $A_{D-3}^{+++}$ dimensionally reduces to the $E_{D-1}$ symmetry algebra of $(12-D)$-dimensional supergravity. An infinite set of five-dimensional gravitational objects trivially embedded in D-dimensions is constructed by identifying the null geodesic motion on cosets embedded in the generalised Kac-Moody algebra $A_{D-3}^{+++}$. By analogy with supergravity these are bound states of dual gravitons. The metric interpolates continuously between exotic gravitational solutions generated by the action of the Geroch group but is not a continuously transforming solution of the Einstein-Hilbert action. We investigate mixed-symmetry fields in the brane sigma model, identify actions for the full interpolating bound state and understand the obstruction to the bound state being a solution of the Einstein-Hilbert action.  
\end{abstract}
\end{titlepage}
\clearpage
\newpage

\section{Introduction}

The knowledge that a physical theory possesses a symmetry is a powerful thing. It is frequently so restrictive that solutions to the theory may be constructed using group theory alone. A large symmetry can however be a double-edged sword, its complexity can prove an obstruction to its usefulness. It was conjectured over a decade ago that the non-linear realisation of the generalised Kac-Moody algebra $\mathfrak{e}_{11}$ is an extension of maximal supergravity relevant to M-theory \cite{West:2001as}. The non-linear realisation of $\mathfrak{e}_{11}$ is formulated using a coset group element whose parameters are the vielbein ${e_\mu}^m$, the three-form $A_{\mu_1\mu_2\mu_3}$, the six-form $A_{\mu_1\ldots \mu_6}$, the dual graviton $A_{\mu_1 \ldots \mu_8,\nu}$ as well as infinitely many more fields. To simplify calculation the infinite dimensional sub-algebra may be consistently truncated to a finite dimensional algebra. The non-linear realisation of the truncated symmetry possesses a group element parameterised by only a subset of the original fields. For example the consistent truncation to the three fields ${e_\mu}^m$, $A_{\mu_1\mu_2\mu_3}$ and $A_{\mu_1\ldots \mu_6}$ leads to a non-linear realisation that gives the bosonic sector of eleven dimensional supergravity \cite{West:2000ga}. 

Even very large truncations of the Kac-Moody algebra retain information related to the full Kac-Moody algebra. One of the largest truncations is to restrict $\mathfrak{e}_{11}$ to $\mathfrak{sl}(2,\mathbb{R})$. It was realised in \cite{West:2004st} that the group element appearing in the non-linear realisation of $\mathfrak{sl}(2,\mathbb{R})\in \mathfrak{e}_{11}$, which is parameterised by ${e_\mu}^m$ and just one other tensor field, could be written in a way which encoded the $\frac{1}{2}$-BPS solutions of eleven-dimensional supergravity as well as those of the type IIA and type IIB string theories. This was put into a formal framework when a Lagrangian for the non-linear realisation of $\mathfrak{e}_{11}$ was constructed and it was shown that the fields for the $\frac{1}{2}$-BPS solutions were exact solutions of the equations of motion for the constructed Lagrangian \cite{Englert:2003py}. Previously the Lagrangian for the non-linear realisation of $\mathfrak{e}_{10}$ had been found in \cite{Damour:2002cu} and used to show the appearance of an $\mathfrak{e}_{10}$ symmetry in the vicinity of a cosmological singularity. The solution to the equations of motion for the $\mathfrak{sl}(2.\mathbb{R})\in\mathfrak{e}_{11}$ Lagrangian described a null geodesic on the coset space $\frac{SL(2,\mathbb{R})}{SO(1,1)}$. One might wonder why such large truncations do not trivialise the $E_{11}$ symmetry; they do not as the metric for the supergravity solution is determined by the embedding of the particular $\mathfrak{sl}(2.\mathbb{R})\in\mathfrak{e}_{11}$. Despite the truncation to a finite dimensional sub-algebra information from the full algebra concerning the embedding of the sub-algebra is retained and used to construct the metric.

By truncating the algebra one loses some of the power of the symmetry to identify complex solutions. It is therefore interesting to carry out the non-linear realisation and identify the null geodesic solutions on larger coset groups. Truncations of $\mathfrak{e}_{11}$ to sub-algebras larger than $\mathfrak{sl}(2,\mathbb{R})$ including $\mathfrak{sl}(3,\mathbb{R})$, $\mathfrak{sl}(4,\mathbb{R})$, $\mathfrak{sl}(5,\mathbb{R})$ and $\mathfrak{so}(4,4)$ lead to coset groups whose null geodesics encode bound states of branes \cite{Cook:2009ri,Houart:2009ya,Houart:2011sk,Cook:2011ir}. The dyonic membrane  \cite{Izquierdo:1995ms} and other bound state solutions were encoded as a group element in \cite{Cook:2009ri}. It was subsequently shown in \cite{Houart:2009ya} that these solution-encoding group elements could be systematically derived from a Lagrangian formulated on cosets of groups embedded in $E_{11}$. The dyonic membrane in eleven space-time dimensions, for example, is encoded by a null geodesic on the coset $\frac{SL(3,\mathbb{R})}{SO(1,2)}$  \cite{Houart:2009ya}. Many further examples of bound states were constructed in this fashion in type IIA and type IIB string theory in \cite{Cook:2011ir}.

Ultimately one might aspire to work with the full non-linear realisation of $\mathfrak{e}_{11}$. A stepping stone in this direction would be the complete understanding of the solutions described by null geodesics on cosets of affine groups embedded in $\mathfrak{e}_{11}$. The associated solution would be described by infinitely many parameters and would approach the complexity of the full non-linear realisation of $\mathfrak{e}_{11}$. Early work on affine cosets in this setting was carried out in \cite{Kleinschmidt:2005bq} where the cosets on $A_{D-2}^{++}$, the over-extension of $SL(D,\mathbb{R})$, were investigated by restricting the algebra to an interesting infinite subset of generators which were argued to correspond to polarised Gowdy cosmologies. The role of the affine group $E_9$ which is a sub-group of both $E_{10}$ and $E_{11}$ was elegantly investigated in \cite{Englert:2007qb}, where it was shown that the Weyl reflections of  affine $SL(2,\mathbb{R})$ contained within $E_9$ discretely mapped supergravity solutions to exotic supergravity solutions. This infinite dimensional solution generating group was identified as the Geroch group \cite{Geroch:1970nt,Geroch:1972yt,Breitenlohner:1986um} which was originally discovered as a solution-generating group in four-dimensional gravity. An affine $\mathfrak{sl}(2,\mathbb{R})$ sub-group within $E_{9}$ was shown to act similarly on the M2 and M5 branes of M-theory as well as the gravitational sector \cite{Englert:2007qb}. The infinite towers of solutions are constructed using the Weyl reflections of affine $\mathfrak{sl}(2,\mathbb{R})$ which form a discrete sub-algebra of $A_{1}^{+}$. In light of the recent successes in identifying continuous symmetry groups with bound state solutions of supergravity and string theory, it is timely to investigate whether the discrete solution generation associated with the Weyl reflections of the Geroch group might be extended to a continuous group. 

\begin{figure}[h]
\centering
\includegraphics[scale=0.6,angle=0]{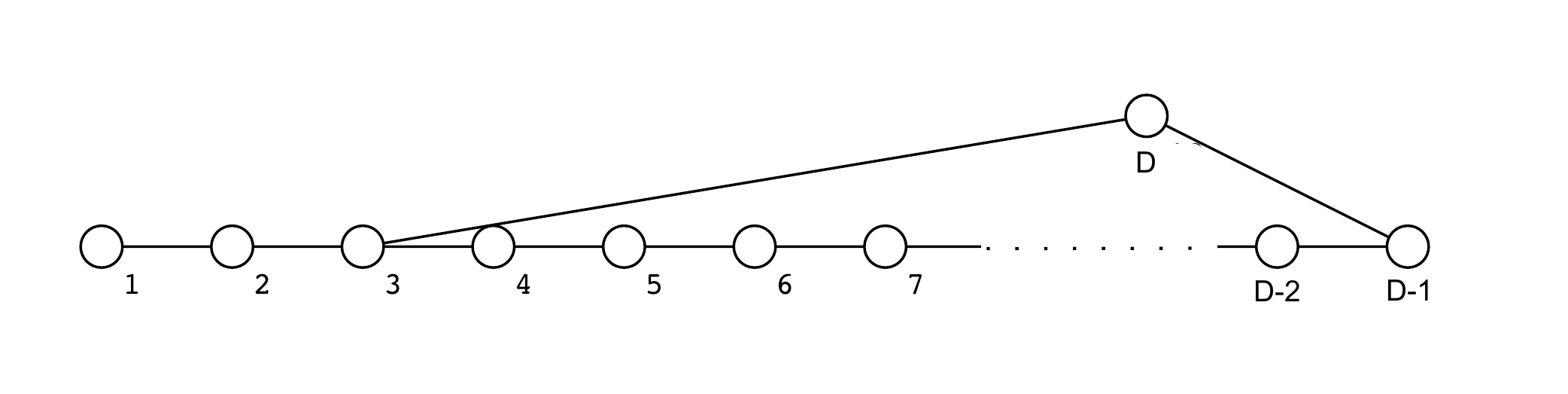}
\caption{The Dynkin diagram for $A_{D-3}^{+++}$ where $D>3$.} \label{Dynkindiagram}
\end{figure}

In the present paper we investigate the class of generalised Kac-Moody algebras denoted $A_{D-3}^{+++}$, the very-extended algebra whose non-linear realisation was proposed to describe gravity \cite{Lambert:2001he} and further investigated in \cite{West:2002jj}. An infinite tower of roots was identified within $A_{8}^{+++}\subset E_{11}$ in \cite{Riccioni:2006az} associated with dual gravitons and dual actions for eahc of these dual fields were constructed in \cite{Boulanger:2012df,Boulanger:2012mq}. Before we describe our investigation it will be useful to motivate the study of these algebras and explain their connection to $E_{11}$. The Dynkin diagram for this class of algebras is shown in figure \ref{Dynkindiagram}. For the case where $D=4$ the Geroch group $A_1^+$ is manifestly embedded within $A_{1}^{+++}$. The non-linear realisation of $A_{D-3}^{+++}$ is a theory containing only gravitational degrees of freedom. 
Deletion of node $D-1$ in figure \ref{Dynkindiagram} leaves the Dynkin diagram of $E_{D-4}^{+++}\equiv E_{D-1}$. The $E_n$ series of symmetries that appear upon compactification of eleven dimensional bosonic supergravity all appear in this manner by deleting node $D-1$ from the $A_{D-3}^{+++}$ Dynkin diagram to give the hidden $E_{n}$ symmetry that appears upon dimensional reduction to $12-D$ dimensions as summarised in table \ref{table:Decomposition_of_A+++}. The dimensional reduction of $A_{D-3}^{+++}$ is akin to Kaluza-Klein dimensional reduction: a purely gravitational theory in $D+1$ dimensions whose dimensional reduction gives a gravitational theory and a gauge theory in $D$-dimensions as well as a tower of KK states. 
\begin{table}[] 
\centering 
\scalebox {0.9} {
\begin{tabular}{| c | c | c |  } 
\hline $D$ &  Dynkin diagram of $A_{D-3}^{+++}$ & Following the deletion of node $(D-1)$ \\ [0.5ex]	
\hline  
$12$ & \includegraphics[scale=0.35,angle=0]{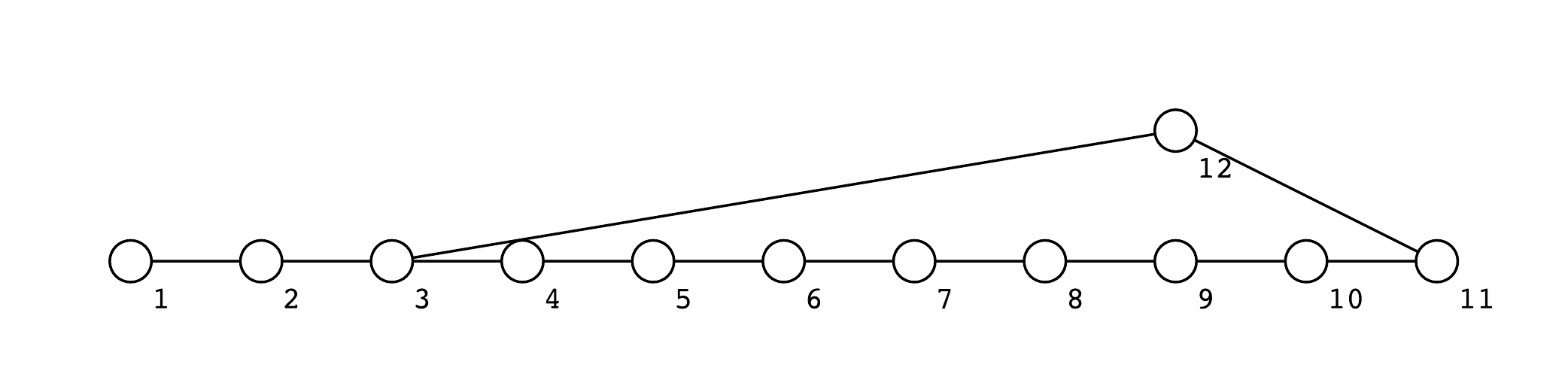} & \includegraphics[scale=0.35,angle=0]{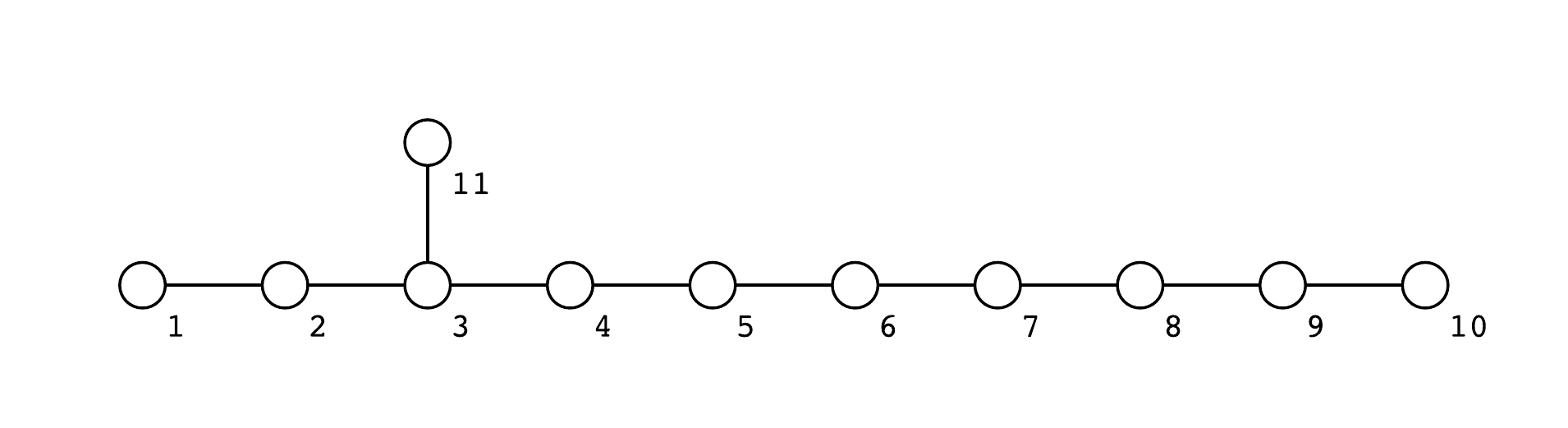}  \\ 
&&$E_{11}$\\
\hline  
$11$ & \includegraphics[scale=0.35,angle=0]{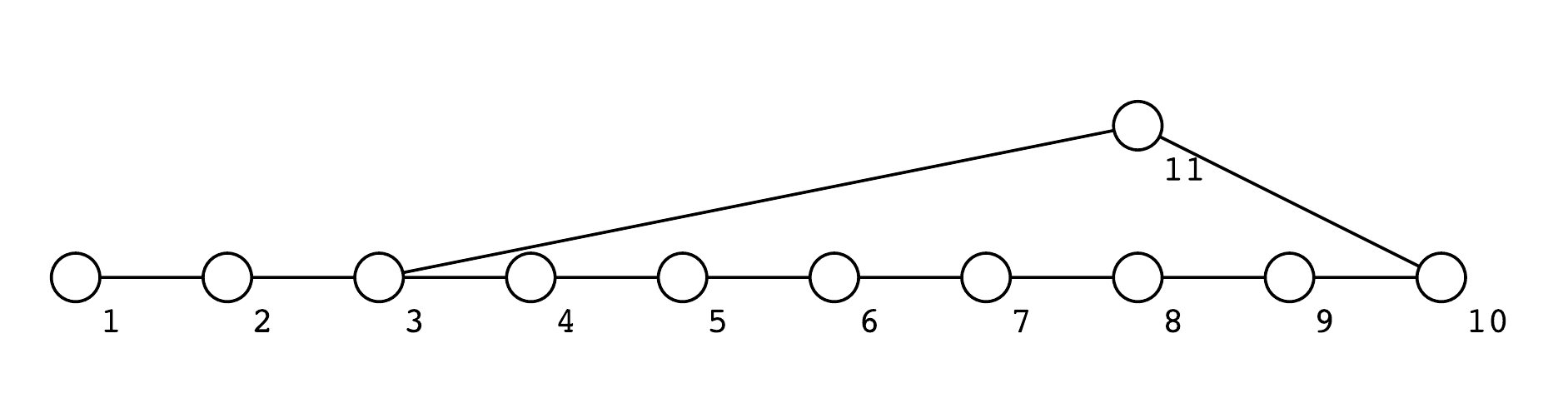} & \includegraphics[scale=0.35,angle=0]{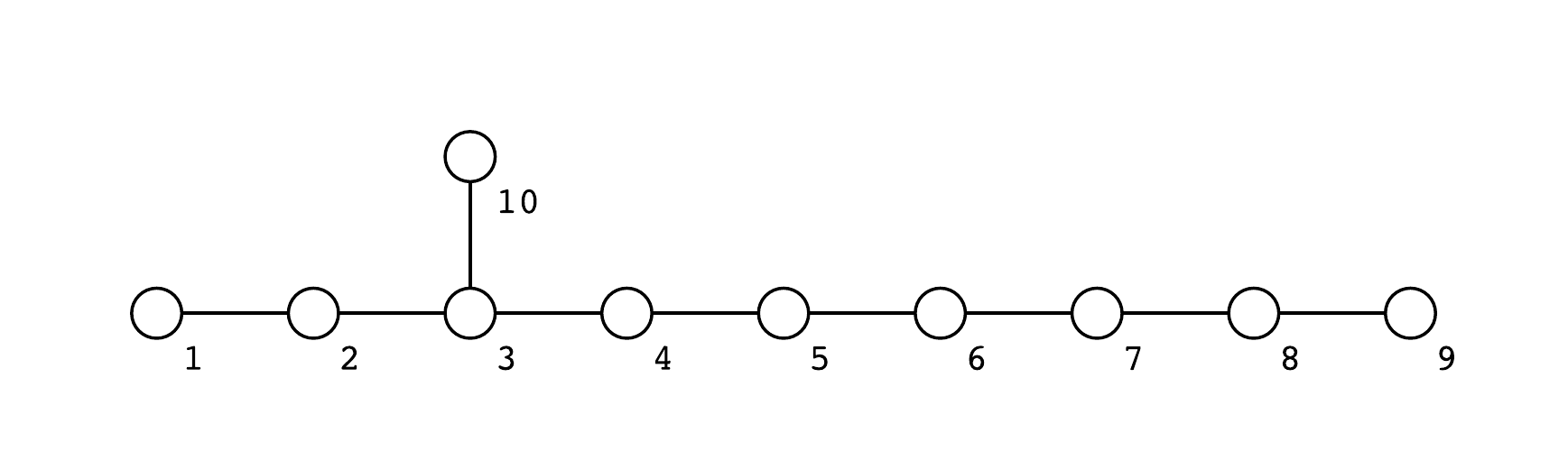}  \\ 
&&$E_{10}$\\
 \hline 
 $10$ & \includegraphics[scale=0.35,angle=0]{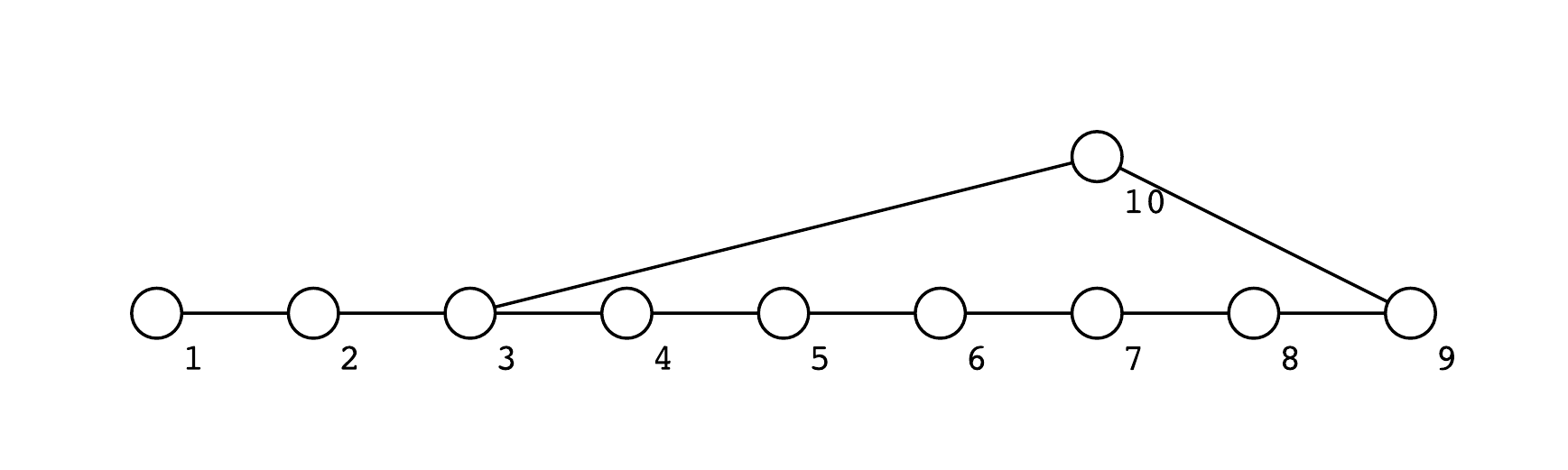} & \includegraphics[scale=0.35,angle=0]{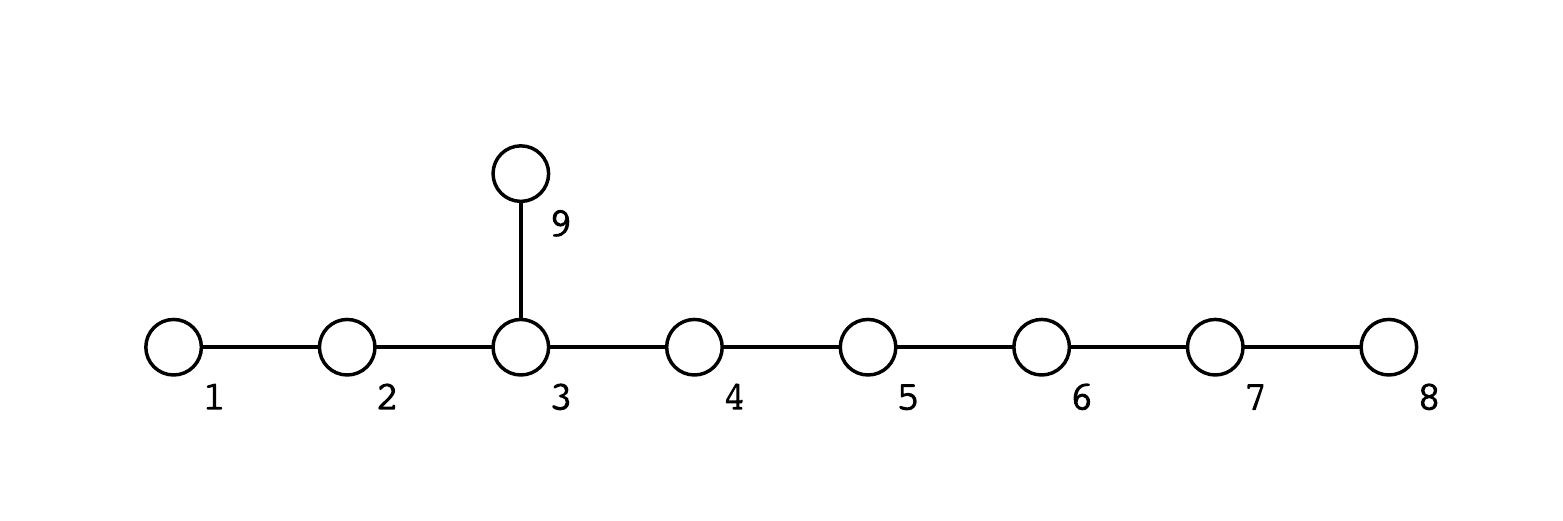}  \\ 
&&$E_{9}$\\
\hline
 $9$ & \includegraphics[scale=0.35,angle=0]{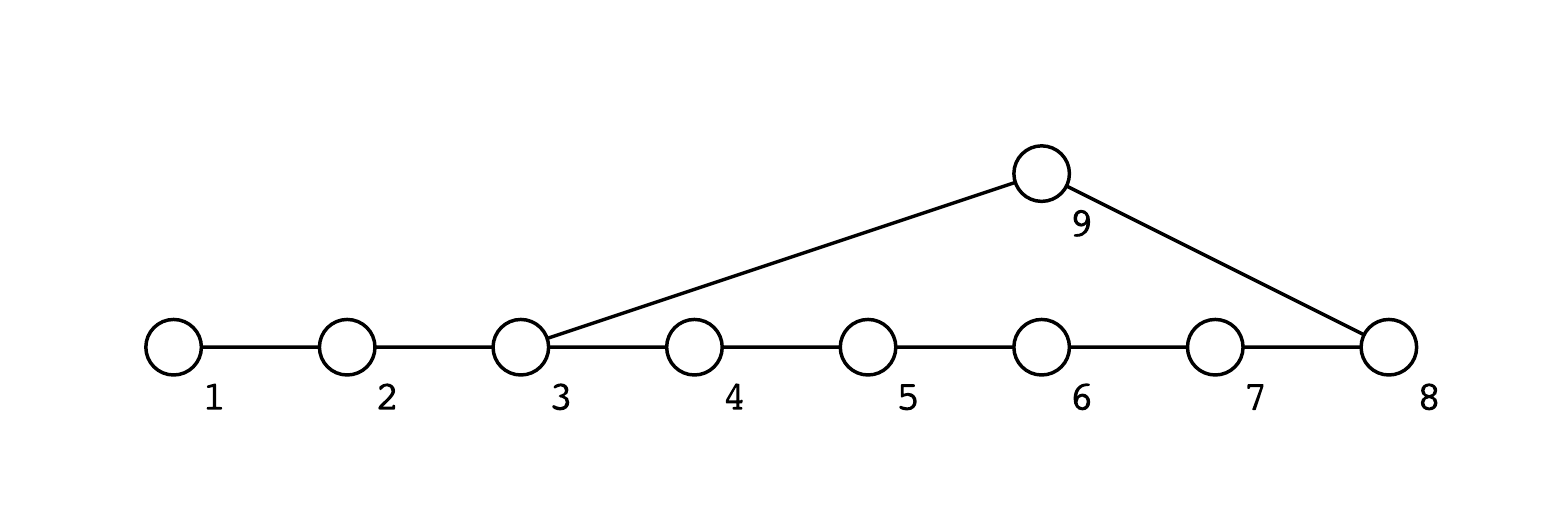} & \includegraphics[scale=0.35,angle=0]{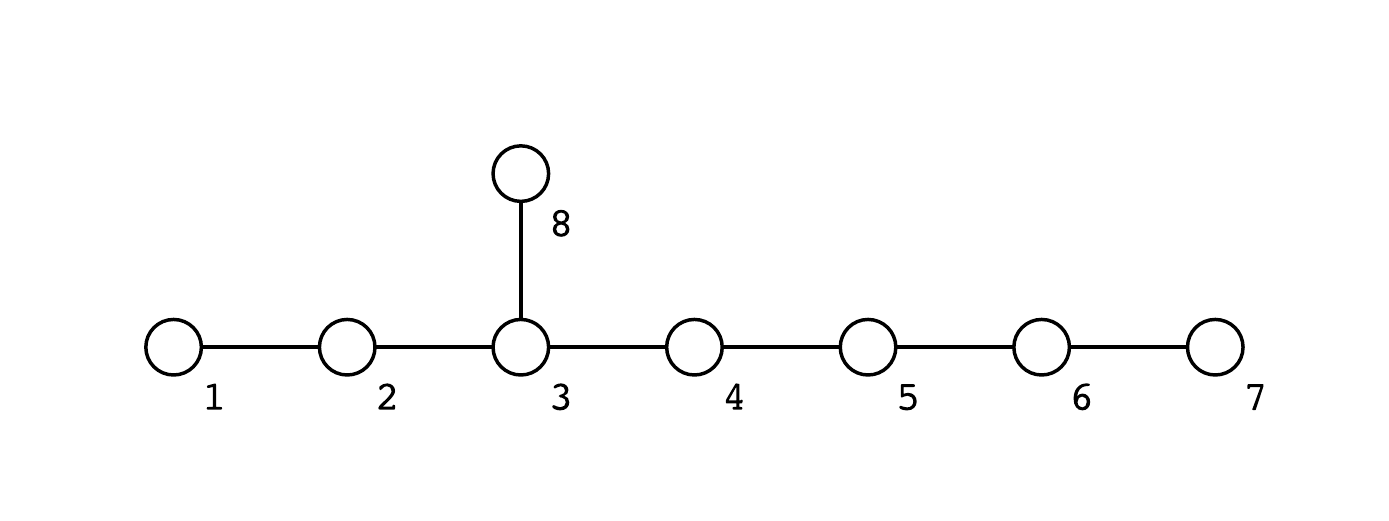}  \\ 
&&$E_{8}$\\
\hline
$8$ & \includegraphics[scale=0.35,angle=0]{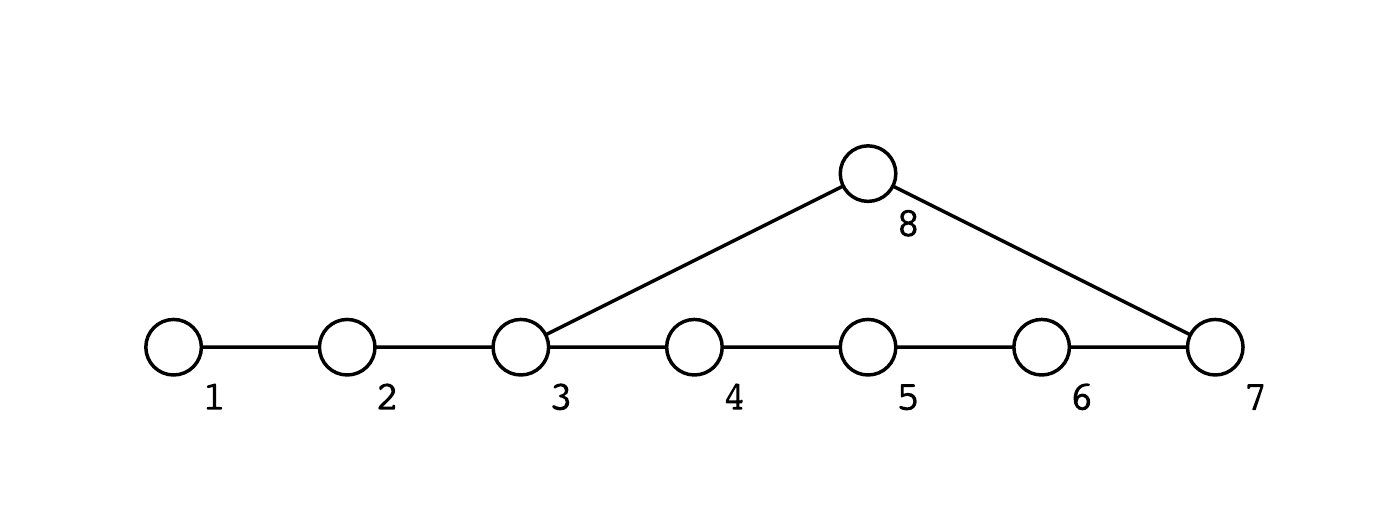} & \includegraphics[scale=0.35,angle=0]{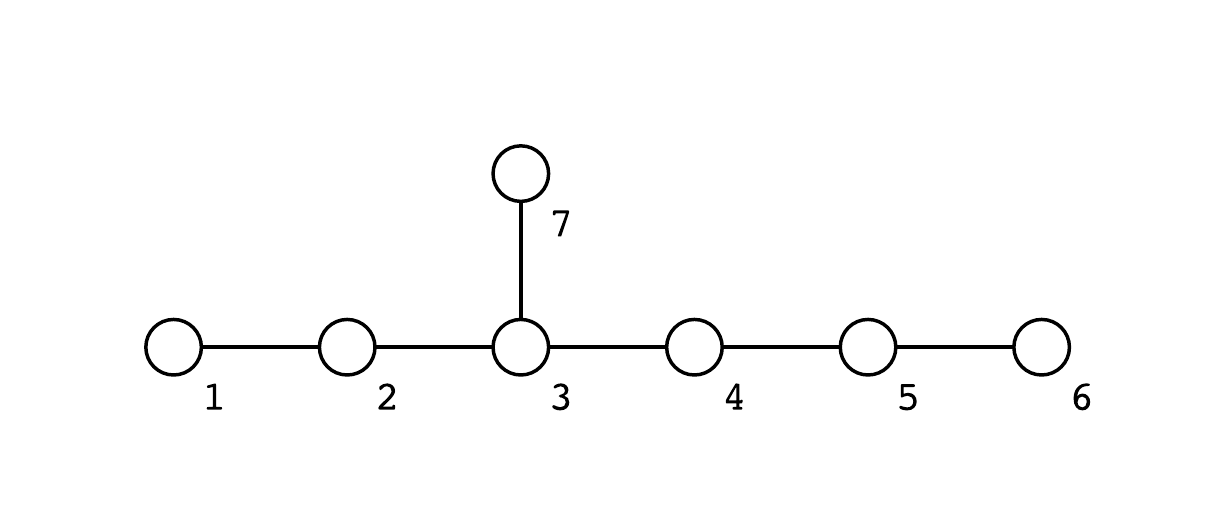}  \\ 
&&$E_{7}$\\
\hline
$7$ & \includegraphics[scale=0.35,angle=0]{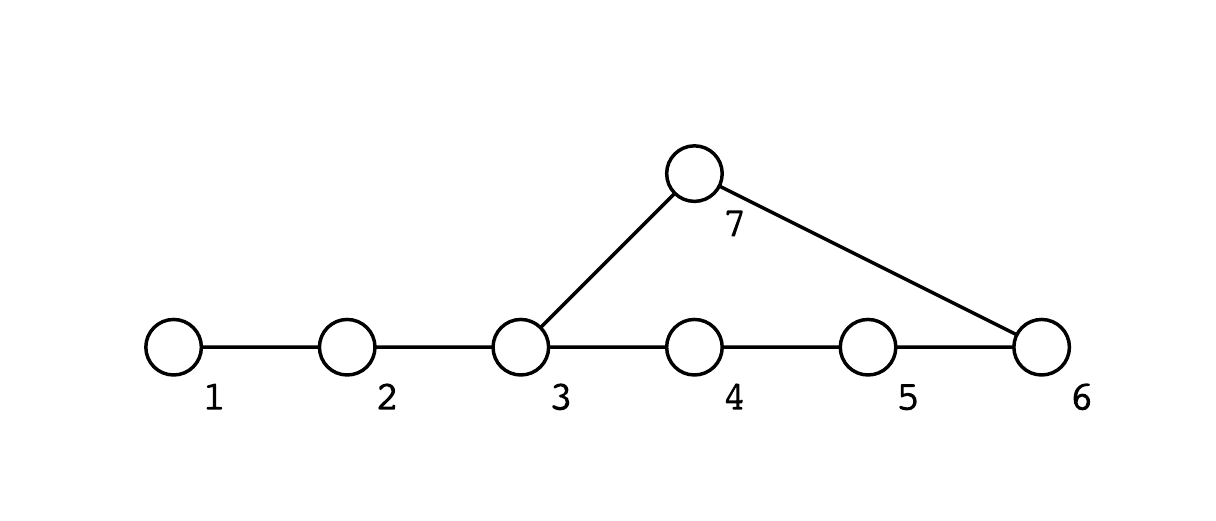} & \includegraphics[scale=0.35,angle=0]{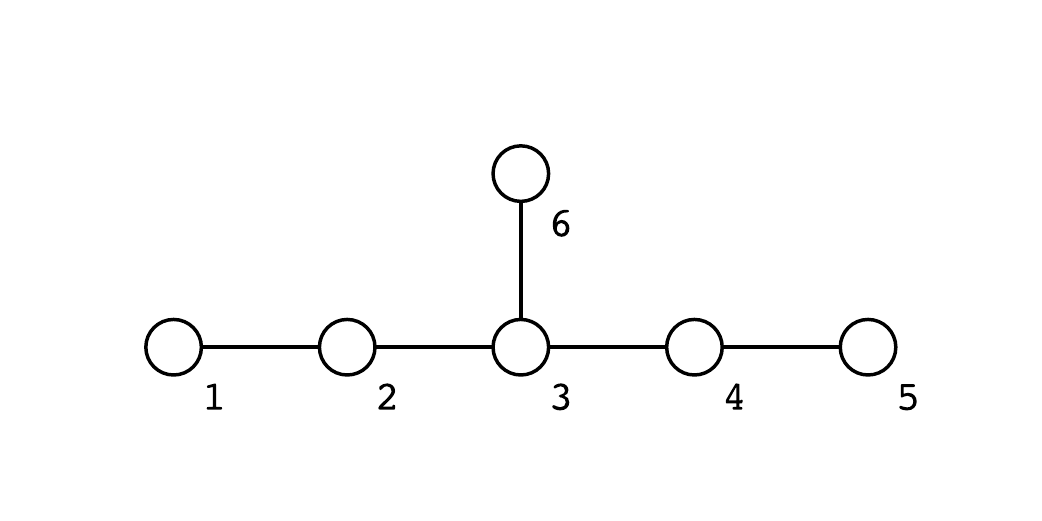}  \\ 
&&$E_{6}$\\
\hline
$6$ & \includegraphics[scale=0.35,angle=0]{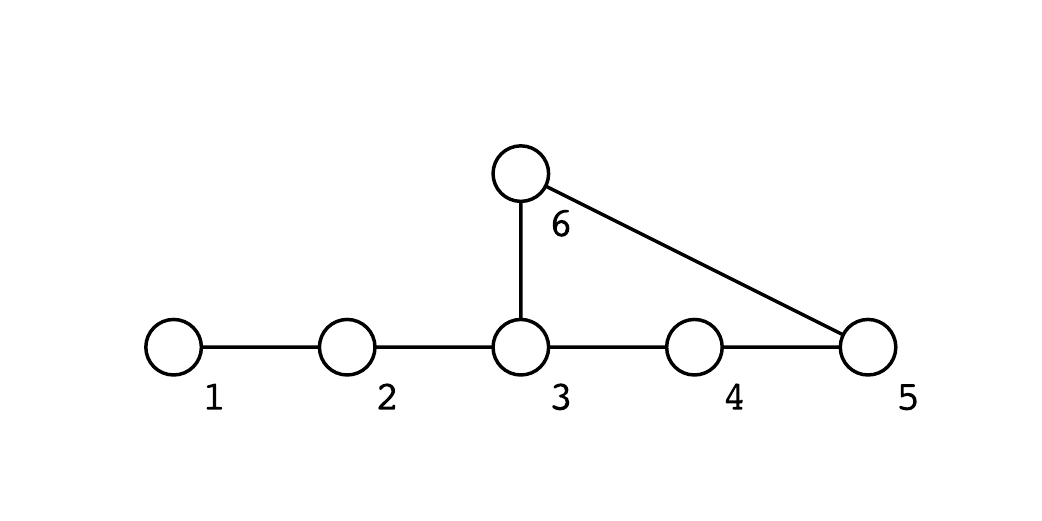} & \includegraphics[scale=0.35,angle=0]{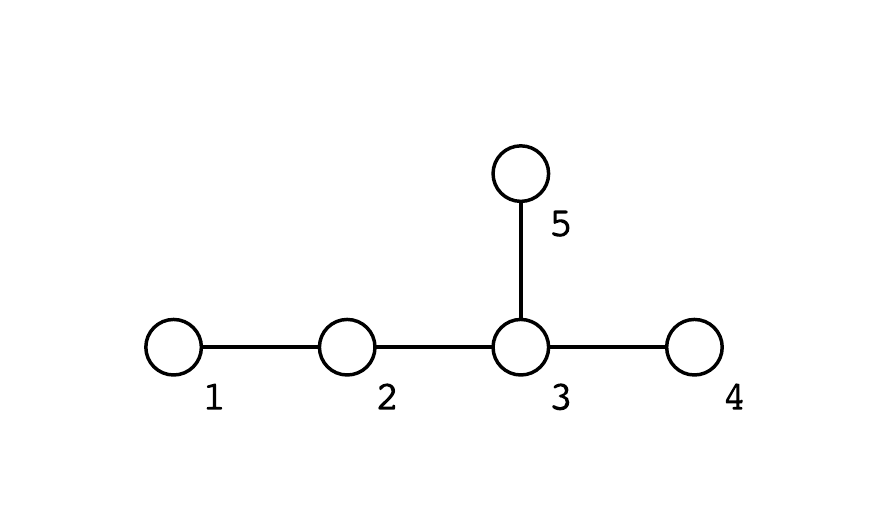}  \\ 
&&$D_{5}$\\
\hline
$5$ & \includegraphics[scale=0.35,angle=0]{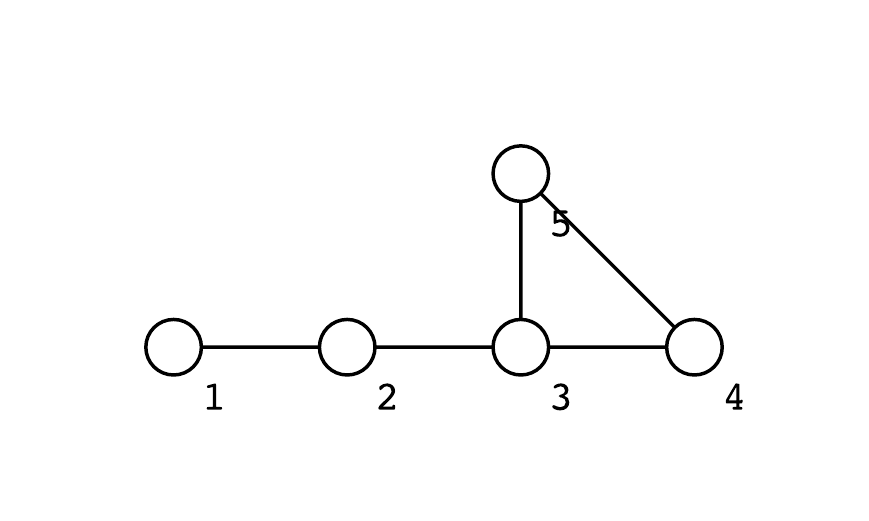} & \includegraphics[scale=0.35,angle=0]{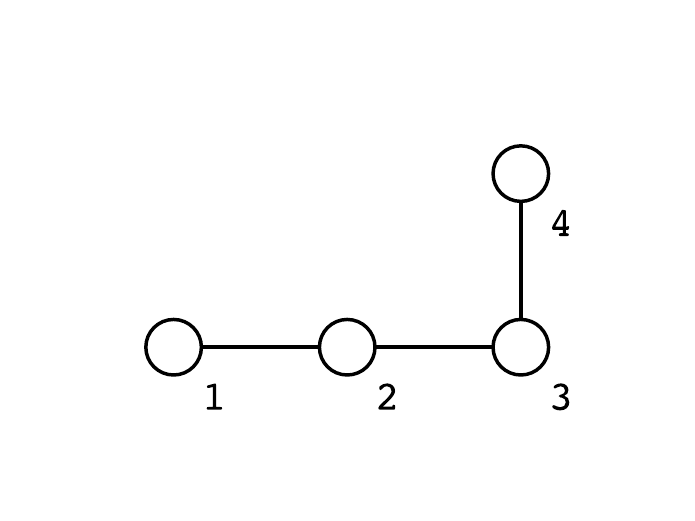}  \\ 
&&$A_{4}$\\
\hline
$4$ & \includegraphics[scale=0.35,angle=0]{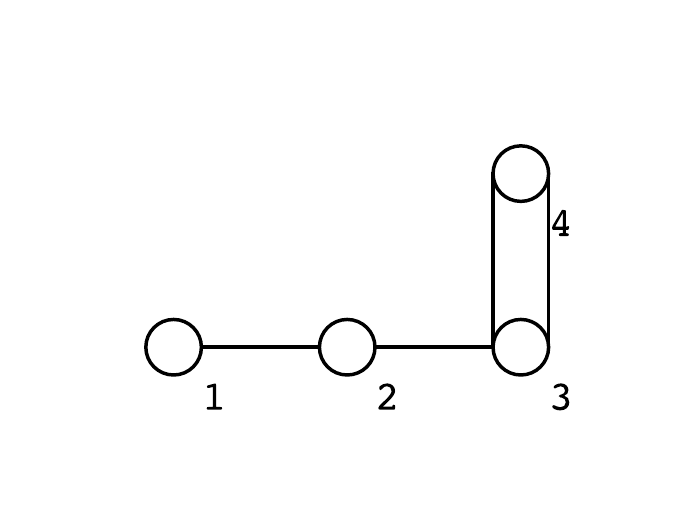} & \includegraphics[scale=0.35,angle=0]{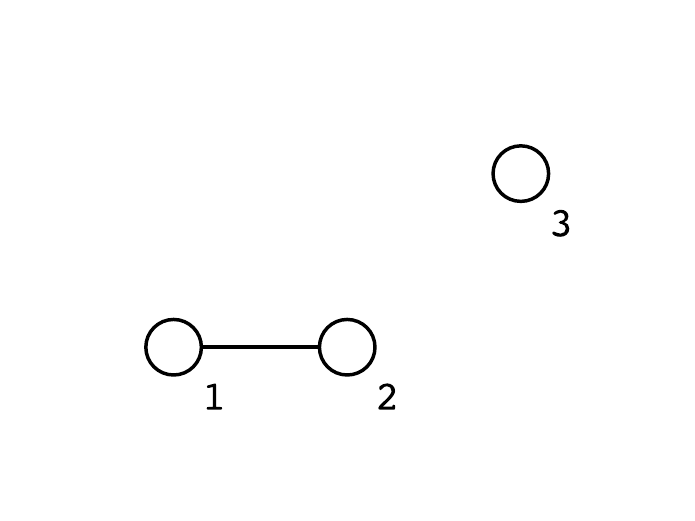}  \\ 
&&$A_{2}\otimes A_1$\\
\hline
 \end{tabular} }
 \caption{The decomposition of $A_{D-3}^{+++}$ containing the $E_{D-1}$ series of algebras.} \label{table:Decomposition_of_A+++} 
\end{table}
When $D=12$ the relevant algebra is $A_{9}^{+++}$ which upon dimensional reduction leaves a theory with a manifest $E_{11}$ symmetry which is conjectured to be M-theory. Moreover $E_{8}^{+++}\equiv E_{11} \supset A_{8}^{+++}$, while $E_{8}^{++}\equiv E_{10} \supset A_{8}^{++}$ and $E_{8}^{+}\equiv E_9 \supset A_{8}^{+}\supset E_8$ which identifies a sequence of inclusions of infinite algebras terminating with the finite algebra $E_8$:
\begin{equation}
A_9^{+++} \supset E_{11} \supset A_{8}^{+++}\supset E_{10} \supset A_8^{++} \supset E_9 \supset  A_8^{+}\supset E_8.
\end{equation} 
The fact that $E_{11}$ lies between $A_9^{+++}$ and $A_8^{+++}$ in this sequence coupled with the expectation that the $A_{D-3}^{+++}$, being gravitational algebras rather than p-form algebras, should provide new ways to investigate $E_{11}$ motivates the present focus on $A_{D-3}^{+++}$ algebras.

Our main aim in this work is to use the non-linear realisation to investigate continuous solution generating sub-algebras within $A_{D-3}^{+++}$. Although we are ultimately motivated to understand the non-linear realisation of the affine sub-groups we will in this paper identify finite sub-algebras which are embedded within $A_{D-3}^{+++}$ and which may be combined to give a set of continuous transformations which cover a set of solutions generated by an affine group. We will identify truncations of the algebra in the first instance to $\mathfrak{sl}(2,\mathbb{R})$ sub-algebras encoding gravity solutions and in the second instance to $\mathfrak{sl}(3,\mathbb{R})$ sub-algebras which we will interpret as bound states of pp-waves, KK-branes and other exotic gravitational objects. An infinite set of sequential $\mathfrak{sl}(3,\mathbb{R})$ sub-algebras will be presented which interpolate between the solutions discretely mapped to by the Geroch group. For the case when $D=11$ we find that the null geodesic motion on the cosets of $SL(3,\mathbb{R})$ reproduce the gravitational tower of solutions identified in \cite{Englert:2007qb}, but now the continuous interpolation between these solutions is identified as a solution to an extension of the Einstein-Hilbert action. A corollary of the work presented here is that the M2-M5 infinite tower of solutions found in \cite{Englert:2007qb} may be understood to originate from the (gravitational) Geroch group acting within $A_{9}^{+++}$.

The paper is organised as follows in section 2 we define the algebra $A_{D-3}^{+++}$ and interpret each of the roots as Young tableaux carrying the symmetries of their associated generator. In section 3 we set up our notation by briefly reviewing the construction of the sigma model for bound states found in \cite{Houart:2009ya}. In section 4 the main body of our work is presented. In section 4.1 the null geodesic motion on cosets of $SL(2,\mathbb{R})$ and $SL(3,\mathbb{R})$ embedded within $A_{D-3}^{+++}$ are studied. We give particular emphasis to understanding the appearance of the fundamental gravitational solutions, the pp-wave and the KK$(D-5)$ brane, before we investigate more exotic solutions including the interpolating $SL(3,\mathbb{R})$ gravitational solutions which continuously map between the Geroch solutions. These one dimensional coset-model solutions do not lift to solutions of the Einstein-Hilbert action, so in section 5 we investigate the supergravity dictionary used for mixed-symmetry fields and identify gravity and exotic matter actions which do admit the full interpolation as a solution to their $D$-dimensional equations of motion. In section 6 we focus on the problem of dualising these exotic actions to the Einstein-Hilbert action and understanding why the full interpolating solution is lost. Section 7 is devoted to a discussion of the results. 

\section{The $A_{D-3}^{+++}$ algebra.}\label{decompositionsection}

The collection of very-extended Kac-Moody algebras $A^{+++}_{D-3}$ where $D\geq 4$ has a root space spanned by the simple roots $\left\{\bm\alpha_{1},\bm\alpha_{2},\ldots,\bm\alpha_{D}\right\}$ which may be conveniently embedded in $\mathbb{R}^D$. Let $\left\{\vec e_{1},\vec e_{2},\ldots,\vec e_{D}\right\}$ be an orthonormal basis for $\mathbb{R}^D$ and an embedding of the basis of the root space is 
\begin{eqnarray}
\nonumber \bm\alpha_{i} &=& \vec e_{i}-\vec e_{i+1} \quad \mbox{where \,\,$i<D$}\\
\bm\alpha_{D} &=& \vec e_{D} + \sum^{D}_{i=3}\vec e_{i} \label{rootembedding}
\end{eqnarray}
\noindent For root vectors $\vec{a}\equiv \sum^{D}_{i=1}a_{i}\vec e_{i}$ and $\vec{b}\equiv \sum^{D}_{i=1}b_{i}\vec e_{i}$ the inner product on the root space is given by
\begin{equation}
\left\langle \vec{a},\vec{b}\right\rangle = \sum^{D}_{i=1}a_{i}b_{i} -\frac{1}{D-2}\sum^{D}_{j=1}a_{j}\sum^{D}_{k=1}b_{k} \label{innerproduct}
\end{equation}
\noindent One can confirm that this inner product acting on the positive simple roots embedded in $\mathbb{R}^D$ reproduces the Cartan matrix of $A_{D-3}^{+++}$ when $D\geq 4$. 

An indefinite Kac-Moody algebra may be decomposed into an infinite set of highest weight representations of a classical Lie algebra each labelled by the level at which they occur in the decomposition. $A_{D-3}^{+++}$ may be decomposed into a set of highest weight representations of $SL(D,\mathbb{R})$ corresponding to nodes $1$ to $D-1$ in figure 1.1 and the level specified by a single integer. If a generic root of the algebra is given by $\vec \beta \equiv \sum_{i\leq D}m_i\bm\alpha_i$ in the simple root basis then the decomposition amounts to partitioning the Dynkin labels $(m_1,m_2,\ldots,m_D)$ into two sets of labels $(m_1,m_2,\ldots,m_{D-1})$ and $(m_D)$. The first set labels a highest weight representation of $SL(D,\mathbb{R})$; the $D-1$ numbers are the Dynkin labels of the root $\hat{\vec{\beta}}=\sum_{i\leq D-1} m_i\bm\alpha_i$ associated to the highest weight in the representation. The remaining integer $m_D$ is called the level in the decomposition and labels where each representation of $SL(D,\mathbb{R})$ occurs in the decomposition of $A_{D-3}^{+++}$. 

The example of $A^{+++}_{8}$, occurring when we choose $D=11$, is relevant to M-theory and we will emphasise this particular example. Its field content at low levels was first found in \cite{Kleinschmidt:2003mf}. At levels 0 and 1 of the decomposition of $A^{+++}_{8}$ into representations of $SL(11,\mathbb{R})$ ${K^{a}}_{b}$ and $R^{a_{1}\cdots a_{8},b}$ are associated with the KK-wave and KK6 brane solutions of M-theory. Indeed when $D=11$ the inner product (\ref{innerproduct}) coincides with the inner product of $E_{11}$ embedded in $\mathbb{R}^{11}$ and the algebra $A^{+++}_{8}$ is a sub-algebra of $E_{11}$. The results in this paper will be readily adapted to $A^{+++}_{8}$. For the case of $A^{+++}_{8}$ the low level roots are shown in table \ref{rootsofA8+++} up to level $m_{11}=3$ which was produced using \cite{Nutma:2012p4929}. These same representations at levels $0$, $1$, $2$ and $3$ occur within the decomposition of $E_{11}$ but at levels $0$, $3$, $6$ and $9$ and the pattern continues for all higher levels i.e. an $SL(11,\mathbb{R})$ representation at level $m_{11}$ in the decomposition of $A_8^{+++}$ is also always found at level $3m_{11}$ in the decomposition of $E_{11}$.

Reproducing all the information of table \ref{rootsofA8+++} is computationally challenging. A very efficient way to reproduce the highest weight representations appearing in the decomposition relies on the embedding of the root space within $\mathbb{R}^D$ advocated in equation (\ref{rootembedding}). The choice of the embedding is such that the covariant and contravariant index structure of the $SL(D,\mathbb{R})$ highest weight representations associated to each of the generators is encoded in the coefficients of the $\vec{e}_i$ in a manner that we will now describe. Each highest weight $SL(D,\mathbb{R})$ representation may be represesented by a Young tableau whose columns are antisymmetrised and whose rows have widths $w_i$ where $1\leq i\leq D$, where $w_1\leq w_2\leq w_3 \leq \ldots \leq w_D$. For a root of $A_{D-3}^{+++}$ associated with a highest weight representation the Young tableau, carrying the symmetries of the associated generator, has rows of width $w_i$ which may be read from the root by virtue of the embedding in $\mathbb{R}^D$
\begin{equation}
\vec{\beta}\equiv \sum_{i\leq D} m_i\bm\alpha_i = \sum_{i\leq D} w_i \vec e_i. \label{Youngtableau}
\end{equation}
The roots at level zero which make up the algebra of $SL(D,\mathbb{R})$ provide an immediate, but isolated, puzzle for the preceding definition. Namely these roots have the form $e_i-e_j$ which we interpret as a Young tableau having an i'th row of width one but a j'th row of width negative one. The  interpretation is best stated in terms of the generator index structure and a negative coefficient indicates covariant indices on the tensor component while a positive index indicates contravariant indices. The root $\vec e_i - \vec e_j$ is associated to the generator ${K^i}_j$ whose commutator relations are those of the positive generators of the $\frak{sl}(D,\mathbb{R})$ algebra. The simple root at level one $\alpha_{D}$ as given in equation (\ref{rootembedding}) has a mixed-symmetry $\frak{sl}(D,\mathbb{R})$ generator $R^{a_4a_5\ldots a_D,b_D}$. The gauge field associated to this generator has the index structure of a dual graviton, which one may see by dualising the first set of $(D-3)$ indices to find a tensor field related to the vielbein. The highest weight Young table has two columns of $D-3$ and $D$ boxes:
\begin{equation}
\ytableausetup
{mathmode, boxsize=1.2em}
\begin{ytableau}
\mbox{{\scriptsize D}} & \mbox{{\scriptsize D}}  \\
\mbox{{\scriptsize D-1}}  \\
\none[\vdots]  \\
\scriptstyle 5  \\
\scriptstyle 4 \\
\end{ytableau}. \label{dualgravity}
\end{equation}
The Young table above is associated to the particular generator $R^{456\ldots D,D}$, in what follows we will frequently drop the labels in the Young table where we refer to the unique highest weight. The lower weights arise following commutation with the ${K^i}_j$ generators of $\frak{sl}(D,\mathbb{R})$. For $D=4$ the level one Young tableau has the same structure as the metric representing a symmetric two tensor, while For $D>4$ the level one Young tableau are hooked. The associated gauge field component $A_{456\ldots D,D}$ is dualised on its antisymmetric $D-3$ indices to $A_{i,D}$ where $i\in \{1,2,3\}$ which, as we will see, may be associated to the vielbein. The field appearing at level one is therefore referred to as the dual graviton.

The embedding of equation (\ref{rootembedding}) into $\mathbb{R}^D$ furnishes a simple method to uncover the non-trivial commutators of the low-level generators which we will now outline. The commutators of the Kac-Moody algebra are controlled by the Serre relations, these relations may be recast as an algebraic condition on the root length squared \cite{Cook:2009ri}. The commutator of two generators associated to the distinct roots $\vec{\alpha}$ and $\vec{\beta}$ is either trivial or gives rise to a new generator in the algebra depending upon the length squared of the sum of the roots $(\vec{\alpha}+\vec{\beta})^2$:
\begin{equation}
[E_\alpha,E_\beta]=\begin{cases} 0 & \text{if }(\vec{\alpha}+\vec{\beta})^2>2 \quad \mbox{and} \\ E_{\alpha+\beta} & \text{if }(\vec{\alpha}+\vec{\beta})^2\leq 2. \end{cases}
\end{equation}
The generator structure for $E_{\alpha+\beta}$ may be read directly from the root $\bm\alpha+\bm\beta$ using equation (\ref{Youngtableau}). For example we asserted earlier that the generators ${K^i}_j$ associated to the roots $\vec e_i - \vec e_j$ gave rise to the algebra of $\frak{sl}(D,\mathbb{R})$ - we can now confirm this assertion using the inner product (\ref{innerproduct}). To find the general commutator let $\vec{\alpha}\equiv\vec e_i-\vec e_j$ and  $\vec{\beta}\equiv\vec e_k - \vec e_l$. As $(\bm\alpha+\bm\beta)^2=(\vec e_i-\vec e_j + \vec e_k - \vec e_l)^2=4-2\delta_{jk}-2\delta_{il}$ the commutator is only non-trivial if $i=l$ or $j=k$. Note that if both $i=l$ and $j=k$ then the roots are non-distinct as $\bm\alpha =\bm\beta$ and the commutator is trivially zero. Consequently we may write the commutator relation for the level zero roots as:
\begin{equation}
[{K^i}_j,{K^k}_l]=\delta^i_l{K^k}_j-\delta^k_j{K^i}_l.
\end{equation}
The minus sign difference between the terms follows from the antisymmetry of the Lie bracket. The commutators above are recognisable as the commutators of the generators associated to the positive roots of $\frak{sl}(D,\mathbb{R})$.

The criterion for existence of roots (up to vanishing outer multiplicity) may be summarised as: if $\vec \beta^2 = 2, 0, -2, -4, \ldots$\footnote{In a simply-laced algebra all the simple roots have the same normalisation, which we have chosen to be $\sqrt{2}$ here, consequently all roots have an even length-squared which is less than or equal to two.} then $\bm\beta$ is a root in the root lattice of $A_{D-3}^{+++}$. One may find the $SL(D,\mathbb{R})$ Young tableaux at level $L$ by noting that the nested commutator of $L$ copies of the level one generator whose Young tableau has $D-2$ boxes will consist of Young tableaux with $L(D-2)$ boxes. By drawing all the tableaux formed of $L(D-2)$ boxes and projecting out all those whose associated root length squared is greater than two one arrives at a close approximation of the algebraic content of $A_{D-3}^{+++}$. It is, perhaps, simpler to find one Young tableau at each level whose length squared is two and then construct the other Young tableaux at level $L$ by moving the Young tableaux boxes between columns - a transformation which has a simple impact on the root length squared. The movement of a Young tableau box by one column to the left has the effect of lowering the associated root length squared by two, as can be quickly verified, using (\ref{innerproduct}). Consequently the reverse manoeuvre of transferring a box one column to the right raises the root length squared by two. Using these rules on a Young tableau, whose associated root length squared is known, one can quickly construct all Young tableaux in the decomposition at a particular level together with their associated root length squared but without computation. Additionally there always exists a root at any level $L$ whose generator has the symmetries of the highest weight Young table
\begin{equation}
\ytableausetup
{mathmode, boxsize=1.5em}
\begin{ytableau}
\mbox{{\scriptsize D}} & \mbox{{\scriptsize D}} & \none[\dots] & \mbox{{\scriptsize D}}
& \scriptstyle D & \scriptstyle D \\
\mbox{{\scriptsize D-1}} & \mbox{{\scriptsize D-1}}  & \none[\dots]& \mbox{{\scriptsize D-1}}  
& \mbox{{\scriptsize D-1}} \\
\none[\vdots] & \none[\vdots]
& \none[\ldots] & \none[\vdots] & \none[\vdots] \\
\scriptstyle 4 & \scriptstyle 4 & \none[\dots]& \scriptstyle 4 
& \scriptstyle 4 \\
\scriptstyle 3 & \scriptstyle 3 & \none[\dots]
& \scriptstyle 3 \\
\end{ytableau} \label{multidualgravity}
\end{equation}
where there are $(L-1)$ columns of height $(D-2)$, one column of height $(D-3)$ and a single column of height one, whose length squared for any dimension $D\geq 4$ is always two. The construction, in this way, of the Young tableaux in the decomposition of $A_8^{+++}$ may be confirmed at low levels by comparison with table (\ref{rootsofA8+++}). As one can see this construction is almost sufficient to reproduce the decomposed algebra. However the reader will notice that the information in the column headed $mu$ which gives the outer multiplicity of the generators (the number of copies of a particular generator) has not been reproduced. The outer multiplicity is particularly crucial when it is zero, as then, contrary to our expectations, the generator does not appear within the algebra, however the calculation of outer multiplicity is time-consuming. It would be very useful to find a quick computation to determine whether the outer multiplicity of a generator is zero. 

The highest weight generator corresponding to the deleted node defining the level has a Young tableau containing $(D-2)$ boxes. Consequently a generator appearing at level $L$ has a Young tableau formed of $L(D-2)$ boxes, as the generator is defined by $L$ nested commutators involving level one generators.
Note that the inner product of two roots at levels $L_1$ and $L_2$ in the decomposition is given by:
\begin{equation}
\left\langle \bm\alpha^{L_1},\bm\beta^{L_2}\right\rangle = \sum_{i}w(\alpha)_{i}w(\beta)_{i} - L_1L_2\left(D-2\right)
\end{equation}
\noindent where $\vec \alpha^{L_1} = \sum_i w(\alpha)_i \vec e_i$ and $\vec \beta^{L_2} = \sum_i w(\beta)_i \vec e_i$.

\section{The coset model for gravitational solutions.}

In this paper we will be using the coset model approach developed in \cite{Houart:2009ya} to construct gravitational solutions from the $A^{+++}_{D-3}$ algebras. We will restrict our interest to cosets of $A_1$ and $A_2$ sub-groups embedded in $A^{+++}_{D-3}$. The work of \cite{Houart:2009ya} describes a method that associates a one-dimensional solution of M-theory and string theory to a null geodesic motion on cosets $SL(n,\mathbb{R})/H$ where the algebra of $H$ is the fixed point set under some generalised involution. 

\subsection{The generalised involution $\Omega$.}
The Chevalley-Cartan involution $\Omega_C$ is defined on a semisimple Lie algebra $\mathfrak{g}$ of the group $\cal G$ by
\begin{equation}
\Omega_C(H_i)=-H_i, \quad \Omega_C(E_i)=-F_i \quad \mbox{and} \quad \Omega_C(F_i)=-E_i
\end{equation}
where we have used the Chevalley basis to present the algebra so that $H_i$ indicates the Cartan sub-algebra, $E_i$ are the generators associated to the positive simple roots, $F_i$ the generators associated to the negative simple roots and the index $i\in\{1,2,3,\ldots R\}$ where $R$ is the rank of the Lie algebra. The Chevalley-Cartan involution action on the remainder of the algebra is derived from its action on the generators associated to the simple roots as 
\begin{align}
\Omega_C([E_i,E_j])&=[\Omega_C(E_i),\Omega_C(E_j)]\equiv \Omega_C(E_{i+j})\\
\nonumber
\Omega_C([F_i,F_j])&=[\Omega_C(F_i),\Omega_C(F_j)]\equiv \Omega_C(-F_{i+j})
\end{align}
where we have defined $E_{i+j}\equiv[E_i,E_j]$ and $-F_{i+j}\equiv[F_i,F_j]$. The algebra $\mathfrak{g}=\mathfrak{k}\oplus\mathfrak{p}$ is split into a part which is fixed under the involution $\mathfrak{k}$ and the complement $\mathfrak{p}$. The basis generators of $\mathfrak{k}$ are (with normalisation) $k_i\equiv \frac{1}{2}(E_i-F_i)$ while the algebra $\mathfrak{p}$ has basis elements ${\mathcal P}_i\equiv \frac{1}{2}(E_i+F_i)$ and $H_i$. The group $\cal{K(G)}$, found by exponentiating the $\Omega_{C}$-invariant $\mathfrak{k}$, is the maximal compact sub-group of $\cal{G}$. 

A more general involution $\Omega$ can be defined by its action on the Cartan and simple root generators
\begin{equation}
\Omega(H_i)=-H_i, \quad \Omega(E_i)=-\epsilon_i F_i \quad \mbox{and} \quad \Omega(F_i)=-\epsilon_i E_i
\end{equation}
where $\epsilon_i$ is either $-1$ or $+1$. Collecting the $R$ values $\epsilon_{i}$ as a vector $\vec\epsilon$ we can write the Chevalley-Cartan involution as $\vec\epsilon = (+,+,\ldots ,+)$. In the case of $\mathfrak{g}=\mathfrak{sl}(3,\mathbb{R})$ the Chevalley-Cartan involution generates the coset $SL(3,\mathbb{R})/SO(3)$, while the involutions with $\vec\epsilon = (-,+)$ and $(-,-)$ both generate $SL(3,\mathbb{R})/SO(1,2)$. For the general normal real form $A_{n(n)}\cong \mathfrak{sl}(n+1,\mathbb{R})$ each of the possible $\mathcal{K(G)}$ constructed in this way, which are $SO(p,q)$ with $p+q=n+1$, can be obtained by taking the involution with $\vec\epsilon = (+,\ldots +,-_p ,+,\ldots ,+)$. 

\subsection{Solutions as null-geodesics on cosets.}
We define a map from the real line parameterised by $\xi$ into the coset $\frac{\cal{G}}{\cal K(G)}$ in the Borel gauge 
\begin{equation}
g=\mbox{exp}\bigg(\sum^{n}_{i=1}\phi_{i}H_{i}\bigg)\mbox{exp}\bigg(\sum_{E_{\vec{\alpha}}\in \Delta^+}C_{\vec{\alpha}}E_{\vec{\alpha}}\bigg)
\end{equation}
where $\phi_i\equiv\phi_i(\xi)$, $C_{\vec{\alpha}}=C_{\vec{\alpha}}(\xi)$, $H_{i}$ are the Cartan sub-algebra generators and the second summation runs over the generators of the numerator algebra $\mathfrak{g}$ associated with the set of postive roots $\Delta^+$. We can decompose the Maurer-Cartan form 
\begin{equation}
\partial_{\xi}g g^{-1}={\mathcal Q}_{\xi}+{\mathcal P}_{\xi}
\end{equation}
into components of the fixed point algebra ${\mathcal Q}_\xi$ and the complement in $\mathfrak{g}$ denoted ${\mathcal P}_\xi$. The Lagrangian for this model
\begin{equation}
L = \eta^{-1}\left({\mathcal P}_{\xi}|{\mathcal P}_{\xi}\right) \label{CosetModelLagrangian}
\end{equation} 
where $(M|N)=Tr(MN)$ is the Killing form for $\cal G$ and $\eta$ is the lapse function encoding reparameterisation invariance. The Lagrangian is invariant under global $\cal G$ transformations and local $\cal K(G)$ transformations. Its equations of motion are
\begin{eqnarray}
\partial_{\xi}{\mathcal P}_{\xi}-\left[Q_{\xi},{\mathcal P}_{\xi}\right] &=& 0 \\\left({\mathcal P}_{\xi}|{\mathcal P}_{\xi}\right) &=& 0
\end{eqnarray}
where the first set comes from variation of the coset representative $g$ and the second is due to variation of the lapse function $\eta$. The solution to these equations is a null geodesic on a coset. The null geodesics encode all the $\frac{1}{2}$-BPS solutions \cite{West:2004st,Englert:2003py,Cook:2004er} as well as the bound state solutions in supergravity, string theory and M-theory \cite{Cook:2009ri,Houart:2009ya,Houart:2011sk,Cook:2011ir}.The bound state solutions possess a manifest $\cal{G}$ symmetry which permutes the branes in the bound state and an $\cal{K(G)}$ whose compact symmetries interpolate continuously between brane solutions \cite{Houart:2009ya}, the group element associated to the remaining symmetries in $\cal{K(G)}$ are responsible for shifts in the gauge field, Ehlers transformations on the gauge fields and the Cartan sub-algebra acts as a conformal transformation on the metric, for a review of these ideas see \cite{Clement:2008qx} and the references therein.
\section{Null geodesics on cosets of $SL(2,\mathbb{R})$ and $SL(3,\mathbb{R})$.}
Our aim in this work is to understand the continuous symmetries within $A_{D-3}^{+++}$ Kac-Moody algebras when truncated to $\mathfrak{sl}(2,\mathbb{R})$ and $\mathfrak{sl}(3,\mathbb{R})$ sub-algebras. The relevant cosets are $\frac{SL(2,\mathbb{R})}{SO(1,1)}$ and $\frac{SL(3,\mathbb{R})}{SO(1,2)}$ and solutions encoding the null geodesic motion on the coset are presented in \cite{Houart:2009ya}. The major technical challenge is the extension of the dictionary that relates null geodesics on cosets to gravitational solutions. We commence this section with the mapping to the well-known pp-wave and KK$(D-5)$ brane.

\subsection{$SL(2,\mathbb{R})/SO(1,1)$: the pp-wave and KK(D-5) brane.} \label{singlesolutions1}
The algebra of $\mathfrak{sl}(2,\mathbb{R})$ consists of a one-dimensional Cartan sub-algebra spanned by $H$ and a single generator associated to a positive root $E$ and its negitive root counterpart $F$. The coset representative in the Borel gauge of $\frac{SL(2,\mathbb{R})}{\cal K}$, where $\cal K$ is either $SO(1,1)$ or $SO(2)$, is
\begin{equation}
g=\mbox{exp}(\phi H)\mbox{exp}(C E). \label{sl2cosetrepresentative}
\end{equation}
 The null geodesic solution \cite{Englert:2003py} is given by  
\begin{equation}
\phi=\frac{1}{2}\ln N \quad \mbox{and} \quad C=N^{-1}+K \label{sl2nullgeodesic1}
\end{equation}
where $K$ is a constant and $N\equiv a+b\xi$ with $a$ and $b$ real constants. The solution has
\begin{equation}
{\mathcal P}_\xi=\pm N\partial_\xi N^{-1}. \label{sl2nullgeodesic2}
\end{equation}
where we choose to work with the positive sign in the following.
The field $\phi$, as it premultiplies an element of the Cartan sub-algebra $H$ embedded in $A_{D-3}^{+++}$, encodes the vielbein components. More precisely the vielbein ${e_{\hat{i}}}^j={(e^{-h})_{\hat{i}}}^k{(e^{-h})_k}^j$ where ${h_i}^j$ is the coefficient of the generator ${K^i}_j$ in the coset representative $g$ and $i<k\leq j$. We will discuss the construction of the vielbein in detail below but we also refer the reader to section 2 of \cite{Englert:2003zs} for a detailed argument. The dependece of $h$ on $\xi$ will depend crucially upon the embedding of the Cartan sub-algebra $H$ of $\mathfrak{sl}(2,\mathbb{R})$ into the Cartan sub-algebra of $A_{D-3}^{+++}$.

By establishing a dictionary one identifies ${\mathcal P}_\xi$ with a field strength whose index structure will depend upon the embedding of the positive generator $E$ in $A_{D-3}^{+++}$. Let us fix the prescription for identifying one-dimensional solutions in this way by considering the examples of the pp-wave and the KK$(D-5)$ brane.

\paragraph{The pp-wave.}To identify an $SL(2,\mathbb{R})$ root system we need find only a single real root in the root system of $A_{D-3}^{+++}$. Consider a positive root $\bm\alpha$ of the $SL(D,\mathbb{R})$ sub-group singled out under the decomposition of $A_{D-3}^{+++}$ carried out in section \ref{decompositionsection}. In the $\vec{e}_i$ basis we have
\begin{equation}
\bm\alpha=\vec{e}_i-\vec{e}_j \qquad \mbox{for } 1\leq i < j \leq D
\end{equation}
and it is associated to the generator ${K^i}_j$ and has associated Cartan sub-algebra element $H={K^i}_i-{K^j}_j$. Let us commence with the case where the index $i$ is timelike but all other indices are spacelike. This defines the involution on the $\mathfrak{sl}(D,\mathbb{R})$ algebra to be
\begin{equation}
\Omega({K^a}_{a+1})=\begin{cases}-{K^{a+1}}_a \quad &\mbox{for } 1\leq a<i\quad \mbox{and} \\ {K^{a+1}}_a \quad &\mbox{for } i\leq a<D.\end{cases}
\end{equation}
The sub-algebra fixed by $\Omega$ is $\mathfrak{so}(1,1)$. We can now use the solution for the null geodesic given in equations (\ref{sl2nullgeodesic1},\ref{sl2nullgeodesic2}), and orginally found in \cite{Houart:2009ya}, to read off the line element. 

It would however be helpful to illuminate why one may $``$read off$"$ the vielbein components. Consider the introduction of a translation generator\footnote{We have adopted a hatted index to indicate a curved space-time coordinate while an unhatted index to indicate a flat tangent space coordinate as in \cite{Houart:2009ya}.} ${P}_{\hat i}$ whose commutator with $\mathfrak{sl}(D,\mathbb{R})$ is 
\begin{equation}
[{P}_{\hat{i}}, {K^j}_k]=\delta_{\hat{i}}^j{P}_k.
\end{equation}
Conjugation of the translation generator by a representative element of the coset of $SL(D,\mathbb{R})$ having diagonal ${h_i}^i$ and off-diagonal ${A_j}^k$ fields non-zero gives\footnote{We remain in the Borel gauge for the group element so that $k>j$.} 
\begin{align}
g P_{\hat m} g^{-1}&=\mbox{exp}\bigg({h_i}^i{K^i}_i\bigg)\mbox{exp}\bigg({A_j}^k{K^j}_k\bigg)P_{\hat m}\mbox{exp}\bigg(-{A_j}^k{K^j}_k\bigg)\mbox{exp}\bigg(-{h_i}^i{K^i}_i\bigg)\\
&=\mbox{exp}\bigg({h_i}^i{K^i}_i\bigg)[P_{\hat m}-{A_j}^k\delta_{\hat{m}}^jP_k +\frac{1}{2!}{A_j}^k\delta_{\hat m}^j{A_l}^n\delta_{k}^lP_n-\ldots]\mbox{exp}\bigg(-{h_i}^i{K^i}_i\bigg)\\
&=\mbox{exp}\bigg({h_i}^i{K^i}_i\bigg)[{(e^{-A})_{\hat m}}^kP_k]\mbox{exp}\bigg(-{h_i}^i{K^i}_i\bigg)\\
&={(e^{-A})_{\hat m}}^k(P_k-\delta_k^i{h_i}^iP_i+\frac{1}{2!}\delta_k^i{h_i}^i\delta_i^j{h_j}^jP_j-\ldots)\\
&={(e^{-A})_{\hat m}}^k{(e^{-h})_{k}}^kP_k
\end{align}
Now we see that the combined exponentials act on $P_k$ as a vielbein:
\begin{equation}
{e_{\hat i}}^j\equiv {(e^{-A})_{\hat i}}^k{(e^{-h})_{k}}^j
\end{equation}
where $h$ is diagonal and $k>i$. If we had repeated the conjugation of $P_{\hat m}$ without any off-diagonal contributions to the vielbein (i.e. ${A_j}^k=0$) then we would have found 
\begin{equation}
{e_{\hat i}}^j\equiv {(e^{-h})_{\hat i}}^j\label{diagonalvielbein}.
\end{equation}

Returning to the example we may now read off the non-trivial components of the vielbein for the solution
\begin{equation}
{e_{\hat{i}}}^i=N^{-\frac{1}{2}}, \quad {e_{\hat{j}}}^j=N^{\frac{1}{2}}  \quad \mbox{and} \quad {e_{\hat{i}}}^j={(e^{-A})_{\hat i}}^j=-{A_{\hat{i}}}^j\end{equation}
where $i$ and $j$ now take fixed values given by the choice of root and $i<j$. The field ${A_i}^j$ is determined from the null geodesic motion on the coset. ${\mathcal P}_\xi=N\partial_{\xi}N^{-1}$ is identified with the components of a field strength for ${A_i}^j$ as follows
\begin{equation}
{F_{\xi i}}^j\equiv \partial_{[\xi} {A_{i]}}^j=N\partial_\xi N^{-1}.\label{ppfieldstrength}
\end{equation}
With this definition we have differentiated between the sets of (antisymmetrised) coordinates, after all one may wonder why we have assumed that the exterior derivative hits the first set of indices on $A$ and not the second. It is also worth emphasising that with this definition we are treating ${A_i}^j$ as a scalar object under the covariant derivative as the dictionary identifies components of $F$ with the components of a one-form. We only have a vector field strength, which we identified with a component of the Maurer-Cartan form. We do not construct a full [2,1] field strength tensor and for the moment the extra indices on $A$ play no role. We may put all sets of indices on an equal footing by forming the [2,2] field strength ${F_{\xi i}}^{\xi j}=D^{\xi}(N\partial_\xi N^{-1})\equiv D^{[\xi}(\partial_{[\xi} {A_{i]}}^{j]})$, where there is no summation over the repeated $\xi$ indices. However in order to find the off-diagonal vielbein components we will immediately remove the second derivative and the covariant derivative plays no role in this solution. It will however be important in later mixed symmetry solutions that we will develop. We may return to equation (\ref{ppfieldstrength}).
As ${A_i}^j={A_i}^j(\xi)$ then $\partial_{[\xi} A_{i]}^j=\partial_{\xi} {A_i}^j$. It is useful to embed the field strength  in space-time using the vielbein so that
\begin{equation}
{F_{\hat{\xi} \hat{i}}}^{\hat{j}}={e_{\hat i}}^{k}{e_l}^{\hat j}{F_{\hat{\xi}k}}^{l}=\partial_{\hat \xi} N^{-1}.
\end{equation}
Hence ${A_{\hat i}}^{\hat j}=N^{-1}+c$, with some constant $c$. Consequently ${A_{\hat i}}^{j}={e_{\hat k}}^j{A_{\hat i}}^{\hat{k}}=N^{-\frac{1}{2}}(1+cN)$ and so the off-diagonal component of the vielbein is
\begin{equation}
{e_{\hat{i}}}^j=-N^{-\frac{1}{2}}(1+cN).
\end{equation}
Imposing that the solution is asymptotically flat fixes $c=-1$ and writing $N=1+K$ we have:
\begin{equation}
{e_{\hat{i}}}^i=\frac{1}{\sqrt{1+K}}, \quad {e_{\hat{j}}}^j=\sqrt{1+K}  \quad \mbox{and} \quad {e_{\hat{i}}}^j=\frac{K}{\sqrt{1+K}}.
\end{equation}
This gives non-trivial metric components
\begin{align}
g_{\hat{i}\hat{i}}&=-\frac{1}{1+K}+\frac{K^2}{1+K}=-(1-K)\\
g_{\hat{i}\hat{j}}&=K\\
g_{\hat{j}\hat{j}}&=(1+K)
\end{align}
and the line element
\begin{align}
ds^2&=-(1-K)(dt^i)^2+2Kdt^idx^j+(1+K)(dx^j)^2+dy^kdy^l\eta^{(D-2)}_{kl}\\
&=2Kdu^2-2dudv+dy^kdy^l\eta^{(D-2)}_{kl}
\end{align}
where $u=\frac{1}{\sqrt{2}}(t+x)$ and $u=\frac{1}{\sqrt{2}}(t-x)$ are light cone coordinates. The linear function $N(\xi)$ once embedded in space-time becomes a linear function of one of the transverse $y^k$ coordinates. A crucial step in deriving the one-dimensional solution was the assumption that $N$ was a harmonic function of the single variable. Once the fields of the solution have been embedded in space time it is clear that all $D-2$ transverse coordinates are equivalent. Hence it is possible to promote the linear function $N$ to be a harmonic function in the transvers $D-2$-dimensional sub-space. Choosing $N$ to have the form
\begin{equation}
N=1+\frac{M}{r^{D-4}},
\end{equation}
where $r^2=\sum_{k=1}^{D-2}(y^k)^2$, gives a single centre pp-wave solution of eleven dimensional supergravity found in \cite{Hull:1984vh} and found in the case of general relativity in $D=4$ in \cite{Brinkmann:1925fr}.

\paragraph{The KK monopole.} Let us consider a second example of a solution derived from a single root. We will derive the KK$(D-5)$ monopole from a real root at level one in the decomposition of $A_{D-3}^{+++}$. This example will be the prototype for associating solutions to roots in the remainder of this paper. Let us consider the root whose generator has the Young table shown in equation (\ref{dualgravity}). The $SL(2,\mathbb{R})$ coset representative is again of the form shown in equation (\ref{sl2cosetrepresentative}) but now we take the level one root to be the single positve root of an $SL(2,\mathbb{R})$ embedded in $A_{D-3}^{+++}$. The generators are
\begin{equation}
H=-({K^1}_1+{K^2}_2+{K^3}_3)+{K^D}_D\quad \mbox{and} \quad E=R^{456\ldots D,D}.
\end{equation}
We will work with the involution with $\vec\epsilon = (-)$, where the sub-group is $SO(1,1)$, by choosing $x^4$ to be the time dimension. We note that we could have picked any one of $x^4,x^5,\ldots x^{(D-1)}$ to be the temporal coordinate while ensuring that $\Omega(E)=F$, however if we had picked $x^D$ to be temporal as the index appears twice in the generator $E$ we would have found $\Omega{(E)}=-F$ and the local sub-group would have been $SO(2)$. 

The field $A_{456\ldots D,D}$ is related by Hodge duality to a vielbein field ${A_{\hat i}}^D$ where $i\in\{1,2,3\}$. Once again we will use the null geodesic motion to find an expression for the off-diagonal components of the vielbein in terms of $N$, the linear function encoding the solution. As we suggested earlier we may simply take covariant derivatives, indicated here by ${\cal D}$, on all sets of indices
\begin{equation}
F_{\xi 456\ldots D,\xi D}={\cal D}_{\xi}(N{\cal D}_\xi N^{-1})\equiv {\cal D}_{\xi}{\cal D}_{\xi}A_{456\ldots D,D}. \label{sl2field}
\end{equation}
The non-trivial diagonal elements of the vielbein are
\begin{equation}
{e_{\hat i}}^i=N^{\frac{1}{2}} \quad \mbox{for } i\in\{1,2,3\} \quad \mbox{and} \quad {e_{\hat D}}^D=N^{-\frac{1}{2}}.
\end{equation}
Upon using the diagonal vielbein to embed the field strength in space-time we have
\begin{equation}
F_{\hat{\xi} \hat{4}\hat{5}\hat{6}\ldots \hat{D},\hat{\xi} \hat{D}}={\cal D}_{\hat\xi}({\cal D}_{\hat\xi} N^{-1})\equiv {\cal D}_{\hat{\xi}}{\cal D}_{\hat{\xi}}A_{\hat{4}\hat{5}\hat{6}\ldots \hat{D},\hat{D}}
\end{equation}
where we have made use of the identity ${\cal D}_{\hat{\xi}} ({e_{\hat i}}^i)=0$.
Before carrying out the Hodge dualisation it will be useful to pick an embedding of $\hat{\xi}$ identifying the parameter with one of the three tranverse directions labelled by $\{\hat{1},\hat{2},\hat{3}\}$ - for this example we will pick $\hat\xi=\hat 1$. Next we Hodge dualise the first set of indices to get
\begin{align}
\star_1 F_{\hat{1} \hat{4}\hat{5}\hat{6}\ldots \hat{D},\hat{\xi} \hat{D}}\equiv F_{\hat{2} \hat{3},\hat{1} \hat{D}}&=-{\cal D}_{\hat{1}}(N\partial_{\hat{1}} N^{-1})\equiv {\cal D}_{\hat{1}}\partial_{[\hat{2}}A_{\hat{3}],\hat{D}}\\
&\Rightarrow  \partial_{\hat{1}} N=\partial_{[\hat{2}}{A_{\hat{3}]}}^{\hat{D}}.
\end{align}
We have implicitly used ${\cal D}_{\hat{\xi}}(g_{\hat{\mu}\hat{\nu}})=0$ to derive the expressions above.
Had we embedded the solution with $\hat\xi=\hat 2$ or $\hat\xi=\hat 3$ we would have found
\begin{align}
-\partial_{\hat{2}} N&=\partial_{[\hat{1}}{A_{\hat{3}]}}^{\hat{D}}\qquad \mbox{ or}\\
\partial_{\hat{3}} N&=\partial_{[\hat{1}}{A_{\hat{2}]}}^{\hat{D}},
\end{align}
respectively. The three individual one-dimensional scalars ${A_{\hat{i}}}^{\hat{D}}$ may be collected together to form a three-dimensional vector $\vec{A}^D$ and simultaneously $N$ may be made a harmonic function both of which are now dependent on three coordinates. We emphasise that this enhancement of the fields is an application of the symmetry of the background metric in the three transverse directions labelled by $\{\hat{1},\hat{2},\hat{3}\}$. We are left with a single equation in the three dimensional subspace\begin{equation}
\bm\nabla N=\bm\nabla \wedge {\vec{A}}^{\hat{D}} \label{monopoleequation}
\end{equation}
which, in the Euclidean signature, is the equation defining a Taub-NUT \cite{Taub:1950ez,Newman:1963yy} or the generalised Gibbons-Hawking instanton metric \cite{Gibbons:1979zt,Gibbons:1979xm} on the four-dimensional Euclidean subspace spanned by $x^{\hat 1},x^{\hat 2},x^{\hat 3}$ and $x^{\hat D}$. Explicitly we suppose $N$ takes the form $N=1+\frac{2K}{r}$ and then upon changing to spherical coordinates we find $A_\phi=2K\cos{\theta}$ upto a function of $\phi$ and the non-trivial vielbein components are
\begin{equation}
{e_{\hat{r}}}^r=N^{\frac{1}{2}}, \quad {e_{\hat{\theta}}}^\theta=N^{\frac{1}{2}}, \quad {e_{\hat{\phi}}}^\phi=N^{\frac{1}{2}}, \quad {e_{\hat{\phi}}}^D=2KN^{-\frac{1}{2}}\cos{\theta} \quad \mbox{and} \quad {e_{\hat{D}}}^D=N^{-\frac{1}{2}}.
\end{equation}
The metric is the Euclidean Taub-NUT metric embedded in a $D$-dimensional Minkowski space-time discovered in \cite{Sorkin:1983ns,Gross:1983hb}:
\begin{equation}
ds^2=N(dr^2+r^2d\theta^2+r^2\sin^2\theta d\phi^2)+N^{-1}(dx^D+2K\cos\theta d\phi)^2+d\Sigma_{(1,D-5)}.
\end{equation}
To avoid a conical singularity $dx^D$ is periodically identified. When $D=5$ this is the KK-monopole metric, for $D>5$ we refer to this as the KK$(D-5)$ brane which upon dimensional reduction along $dx^D$ gives a $(D-5)$ brane metric. The Taub-NUT gravitational instanton in four Euclidean dimensions can be derived directly from the algebra $A_{1}^{+++}$ and the solution encodes the null geodesic on the coset $\frac{SL(2,\mathbb{R})}{SO(2)}$ for which ${\mathcal P}_\xi =iN\partial_\xi N^{-1}$ and the dictionary is modified to read $F_{\xi\mu|\xi\nu}=-i{\cal D}_\xi {\mathcal P}_\xi$.

The Young tableaux appearing at level $L$ and associated to real roots shown in equation (\ref{multidualgravity}) all have a two-dimensional transverse space. It will benefit us to return to equation (\ref{monopoleequation}) and investigate the solution when it is smeared down to two dimensions. Let us suppose we smear the solution along $x^3$ so that $N$ and ${A_{\hat i}}^{\hat D}$ depend only on $x^1$ and $x^2$. The components of equation (\ref{monopoleequation}) become:
\begin{equation}
\partial_{\hat 2}{A_{\hat 3}}^{\hat D}=\partial_{\hat 1}N, \quad \partial_{\hat 1}{A_{\hat 3}}^{\hat D}=-\partial_{\hat 2}N \quad \mbox{and} \quad \partial_{\hat 1}{A_{\hat 2}}^{\hat D}=\partial_{\hat{2}}{A_{\hat 1}}^{\hat D}.
\end{equation}
The last equation trivialises the field strength component ${F_{\hat{1}\hat{2}}}^{\hat D}$, while the first pair are the Cauchy-Riemann equations for an analytic function 
\begin{equation}
f=N+i{A_{\hat 3}}^{\hat D}.
\end{equation}
Smearing the harmonic function $N$ to two dimensions we have 
\begin{equation}
N=1+K\ln{(r^2)}=1+K\ln{(z\bar z)}=1+K\ln{(z)}+K\ln{(\bar z)}
\end{equation}
where $z=x^1+ix^2$ and 
\begin{equation}
f=1+K\ln{(z)}+K\ln{(\bar z)}+i{A_{\hat 3}}^{\hat D}
\end{equation}
will be holomorphic if ${A_{\hat 3}}^{\hat D}=-iK\ln{(\frac{z}{\bar z})}=2K\theta$, where $\theta$ is the argument of $z$. We note that $|f|^2=N^2+4K^2\theta^2$. Given the distinguished transverse space the appearance of holomorphic functions to describe co-dimension two solutions is not surprising - they will be a feature of the higher level solutions as well. 

\subsection{$SL(2,\mathbb{R})/SO(1,1)$: higher level solutions} \label{singlesolutions2}
In this section we will truncate $A_{D-3}^{+++}$ to $\mathfrak{sl}(2,\mathbb{R})$ sub-algebras using roots of higher level. The null geodesic motion on these cosets will encode the solutions generated by the Geroch group - in particular we will reproduce the infinite tower of gravitational solutions found in \cite{Englert:2007qb} and in so doing we shall understand the appearance of holomorphic functions that describe the solution.

The $A^{+++}_{D-3}$ algebra contains infinitely many positive real roots associated with generators having the symmetries of the Young table in equation (\ref{multidualgravity}). There are a set of roots of this type occurring at each level $L$ which can be expressed in the $\vec{e}_i$ basis using only $\vec{e}_3,\vec{e}_4,\ldots \vec{e}_{D}$ as
\begin{equation}
\bm\alpha = \vec e_{i}-\vec e_{j} + L\left(\sum^{D}_{k=3}\vec e_{k}\right) \label{levelLroot}
\end{equation}
\noindent where $i,j \in \left\{3,\ldots, D\right\}$ and $i\neq j$. Let the specific $\bm\alpha$ with $i=D$ and $j=3$ be the positive simple root of an $SL(2,\mathbb{R})$ embedded within $A^{+++}_{D-3}$. The generators of $SL(2,\mathbb{R})$ in terms of the generators of $A^{+++}_{D-3}$ are
\begin{align}
H&=-L({K^1}_1+{K^2}_2)-{K^3}_3+{K^D}_D, \label{Cartanmutlidual}\\
E&=R^{345\ldots D|\;\ldots \;|345\ldots D|456\ldots D|D} \label{positivemutlidual} \qquad \mbox{and}\\
F&=R_{345\ldots D|\;\ldots \;|345\ldots D|456\ldots D|D}\label{negativemutlidual}. 
\end{align}
\noindent In the previous section the level one root of the above type was shown to be associated with the KK$(D-5)$ brane. In that case the solution was constructed from the level one root
\begin{equation}
\bm\alpha = \vec e_{4}+\vec e_{5} + \ldots +\vec{e}_{D-1}+2\vec{e}_{D} 
\end{equation}
and one of the set of coordinates $\{x^4,x^5,\ldots , x^{D-1}\}$ was chosen to be timelike. The involution $\Omega$ is chosen so that it acts on the generator $R^{45\ldots (D-1)D|D}$ as $\Omega(R^{45\ldots (D-1)D|D})=R_{45\ldots (D-1)D|D}$ and the involution invariant sub-algebra is $SO(1,1)$. For the higher level roots
\begin{equation}
\bm\alpha = \vec (L-1)e_{3}+L(\vec e_{4}+\vec e_{5} + \ldots +\vec{e}_{D-1})+(L+1)\vec{e}_{D} 
\end{equation}
choosing one of the coordinates $\{x^4,x^5,\ldots , x^{D-1}\}$ to be timelike implies that the involution acts on the associated element of the algebra as
\begin{equation}
\Omega(E)=(-1)^{L+1}F \qquad \mbox{where } \quad t\in\{x^4,x^5,\ldots , x^{D-1}\}
\end{equation}
where we have used the notation of equations (\ref{positivemutlidual}) and (\ref{negativemutlidual}) to indicate the positive and negative generators. The sub-algebra left invariant under the involution will consequently be $SO(1,1)$ for odd $L$ and $SO(2)$ for even $L$. Of course by choosing $x^3$ or $x^{D}$ to be the sole timelike coordinate the situation is reversed as the involution then acts as
\begin{equation}
\Omega(E)=(-1)^{L}F \qquad \mbox{where } \quad t\in\{x^3,x^D\}.
\end{equation}
We will focus our attention on null geodesics on $\frac{SL(2,\mathbb{R})}{SO(1,1)}$ and will differentiate in the following between odd and even level roots where needed. 

To construct the solutions in the previous section we made use of a coset model dictionary which identified the field multiplying the level one generator in the Maurer-Cartan form with the $[(D-2)|1]$-form field strength $F_{\xi 34\ldots D|D}$ which could then be dualised. We therefore propose an extension of the dictionary which includes higher level objects appearing at arbitrary level $L$ and identifies them with a democratic field strength. These would be associated with a $[(D-1)|\;\ldots\;|(D-1)|(D-2)|2]$-form field strength which we can convert into a $[2|2]$-form field strength through by dualisation of the first $L$ sets of indices. The procedure for this is simply the application of covariant derivatives which are anti-symmetrised with each set of indices giving the dictionary 
\begin{equation}
F_{\xi 3\ldots D|\ldots |\xi 3\ldots D|\xi 4\ldots D|\xi D} = {\mathcal P}_{\xi 3\ldots D|\ldots |\xi 3\ldots D|\xi 4\ldots D|\xi D} \equiv {\cal D}^{L}_{\xi}{\mathcal P}_{\xi}.
\end{equation}  
\noindent In order to Hodge dualise this field strength we identify the coset coordinate $\xi$ with a dimension in the transverse space $(x^1,x^2)$ and embed the field strength in space-time using the diagonalised vielbein (see equation (\ref{diagonalvielbein})) which is derived from the Cartan element in equation (\ref{Cartanmutlidual}). The non-trivial vielbein components are 
\begin{equation}
{e_{\hat{1}}}^1=N^{\frac{L}{2}}, {e_{\hat{2}}}^2=N^\frac{L}{2}, {e_{\hat{3}}}^3=N^{\frac{1}{2}} \quad \mbox{and} \quad {e_{\hat{D}}}^D=N^{-\frac{1}{2}}. \label{levelLvielbein}
\end{equation}
The field strength is then dualised over its first $L$ sets of indices, and the remaining $D$ index is raised. The dualisation is sensitive to the choice of temporal coordinate in the background space-time. All the solutions associated to the positive root $\vec{\alpha}$ give product space-time manifolds of the form ${\cal M}_4\otimes {\cal N}_{D-4}$, where ${\cal M}_4$ is a four-dimensional manifold and ${\cal N}_{D-4}$ is a $(D-4)$-dimensional manifold which is not warped in the solution. The coordinates of ${\cal N}_{D-4}$ are $\{x^4,x^5,\ldots, x^{D-1}\}$ and the dualised field strength depends upon whether ${\cal N}_{D-4}$ has Euclidean or Minkowski signature. When ${\cal N}$ is Minkowski the dualised field strength is given by
\begin{eqnarray}
&{F_{\hat{2}|\ldots|\hat{2}|\hat{2}\hat{3}|\hat{1}}}^{\hat{D}} = (-1)^{(D-1)(L-1)}{\cal D}^{L}_{\hat{1}}(N {\mathcal P}_{\hat{1}}) & \quad \mbox{for }\; \hat{\xi}=\hat{1} \quad \mbox{and} \\
&{F_{\hat{1}|\ldots|\hat{1}|\hat{1}\hat{3}|\hat{2}}}^{\hat{D}} = (-1)^{D(L-1)}{\cal D}^{L}_{\hat{2}}(N{\mathcal P}_{\hat{2}}) & \quad \mbox{for }\;\hat{\xi}=\hat{2}.
\end{eqnarray}
While if ${\cal N}$ is Euclidean (so that either $x^3$ or $x^D$ is the sole temporal coordinate) then the dual field strength is
\begin{eqnarray}
&{F_{\hat{2}|\ldots|\hat{2}|\hat{2}\hat{3}|\hat{1}}}^{\hat{D}} = (-1)^{D(L-1)}{\cal D}^{L}_{\hat{1}}(N {\mathcal P}_{\hat{1}}) & \quad \mbox{for }\; \hat{\xi}=\hat{1} \quad \mbox{and} \\
&{F_{\hat{1}|\ldots|\hat{1}|\hat{1}\hat{3}|\hat{2}}}^{\hat{D}} = 
(-1)^{(D-1)(L-1)}{\cal D}^{L}_{\hat{2}}(N{\mathcal P}_{\hat{2}}) & \quad \mbox{for }\;\hat{\xi}=\hat{2}.
\end{eqnarray}
The signs of the pair of equations in each case have switched if ${\cal N}$ is chosen to have Euclidean rather than Minkowski signature. From the null geodesic motion on $\frac{SL(2,\mathbb{R})}{SO(1,1)}$ we have ${\mathcal P}_{\xi}=N\partial_\xi N^{-1}$ and the set of equations above may be summarised by
\begin{align}
{F_{\hat{j}|\ldots|\hat{j}|\hat{j}\hat{3}|\hat{i}}}^{\hat{D}} &= \kappa (\epsilon_{\hat{i}\hat{j}})^L {\cal D}_{\hat{i}}^{L+1}N \qquad \mbox{where }\begin{cases} \kappa = -(-1)^{D(L-1)} & \mbox{if } {\cal N} \mbox{ is Euclidean}\\
\kappa =(-1)^{(D-1)(L-1)} & \mbox{if } {\cal N} \mbox{ is Minkowski}
\end{cases}\label{dictionary}
\end{align}
where $\hat{i},\hat{j}\in \{\hat{1},\hat{2}\}$ and $\epsilon_{\hat{i}\hat{j}}$ is the Levi-Civita symbol in the two-dimensional sub-space with coordinates $(x^1,x^2)$, normalised such that $\epsilon_{\hat{1}\hat{2}}=1$. The dual field strengths are derivatives of the off-diagonal components of the vielbein ${A_{\hat{3}|}}^{\hat{D}}$
\begin{equation}
{F_{\hat{j}|\ldots|\hat{j}|\hat{j}\hat{3}|\hat{i}}}^{\hat{D}} = {\cal D}_{\hat{i}}{\cal D}_{\hat{j}}^{L-1}{\cal D}_{[\hat{j}}{A_{\hat{3}]}}^{\hat{D}}={\cal D}_{\hat{i}}{\cal D}_{\hat{j}}^L{A_{\hat{3}}}^{\hat{D}} \label{fieldstrength}
\end{equation} 
\noindent where we have assumed that ${A_{\hat{3}}}^{\hat{D}}$ is dependent on only the transverse coordinates $x^1$ and $x^2$. The combination of equations (\ref{dictionary}) and (\ref{fieldstrength})   
give a monopole-like partial differential equation of order $L+1$ to solve for ${A_{\hat{3}}}^{\hat{D}}$ which may be trivially solved for a one-dimensional harmonic function $N$. 

We may unsmear the one-dimensional equation to two dimensions\footnote{In section \ref{singlesolutions1} we were able to unsmear the one-dimensional KK-monopole solution to three dimensions due to the isometries of the metric in $\{x^1,x^2,x^3\}$.} by taking advantage of the symmetry between the $x^1$ and $x^2$ coordinates in the metric (see equation (\ref{levelLvielbein})) for the objects associated to arbitrary level $L$ generators of $A_{D-3}^{+++}$. The constraint, coming from the null geodesic motion on the coset, that $N$ is a harmonic function is maintained so that $N$ is a harmonic function in the two transverse dimensions $(x^1,x^2)$ and takes the form $N=a+b\ln(r)$ where $r^2\equiv (x^1)^2+(x^2)^2$. The consistent two-dimensional version of equations (\ref{dictionary}) and (\ref{fieldstrength}) gives
\begin{equation}
{\cal D}_{\hat{i}}{\cal D}_{\hat{j}_1}{\cal D}_{\hat{j}_2}\ldots {\cal D}_{\hat{j}_L}{A_{\hat{3}}}^{\hat{D}} =\kappa (\epsilon_{\hat{i}_1\hat{j}_1})(\epsilon_{\hat{i}_2\hat{j}_2})\ldots (\epsilon_{\hat{i}_L\hat{j}_L}){\cal D}_{\hat{i}} {\cal D}_{\hat{i}_1}{\cal D}_{\hat{i}_2}\ldots {\cal D}_{\hat{i}_L}N \label{covariantD}
\end{equation} 
where $\hat{i}_n,\hat{j}_n\in \{\hat{1},\hat{2}\}$ for $0<n\leq L$ and $n\in \mathbb{Z}$. For guidance in determining solutions to this equation we isolate the terms with only partial derivatives, having integrated both sides with repsect to $x^i$ and setting the constant to zero, we have
\begin{equation}
\partial_{\hat{j}_1}\partial_{\hat{j}_2}\ldots \partial_{\hat{j}_L}{A_{\hat{3}}}^{\hat{D}} =\kappa (\epsilon_{\hat{i}_1\hat{j}_1})(\epsilon_{\hat{i}_2\hat{j}_2})\ldots (\epsilon_{\hat{i}_L\hat{j}_L})\partial_{\hat{i}_1}\partial_{\hat{i}_2}\ldots \partial_{\hat{i}_L}N\label{partialde}
\end{equation} 
\noindent which have a convenient set of solutions that can be summarised for the odd and even levels as
\begin{eqnarray}
\mbox{Odd } L &:& {A_{\hat{3}}}^{\hat{D}} = \kappa (-1)^{\frac{L+1}{2}} B \label{OddSoln} \\
\mbox{Even } L &:& {A_{\hat{3}}}^{\hat{D}} = \kappa(-1)^{\frac{L}{2}} N \label{EvenSoln}
\end{eqnarray}
\noindent where $B$ is the harmonic conjugate of $N$ such that $\partial_{i}B = \epsilon_{ij}\partial_{j}N$. Equation (\ref{partialde}) gives $2^L$ equations to solve each of which is idential to one of the $L+1$ equations of the form
\begin{equation}
\partial^n_{\hat{1}}\partial^{L-n}_{\hat{2}}{A_{\hat{3}}}^{\hat{D}} =\kappa (-1)^{L-n}\partial^n_{\hat{2}}\partial^{L-n}_{\hat{1}}N \label{generalpde}
\end{equation} 
where $0\leq n \leq L$ for $n\in \mathbb{Z}$. Consider first the case when $L$ is even: upon substitution of ${A_{\hat{3}}}^{\hat{D}} = \kappa (-1)^{\frac{L}{2}} N$, by virtue of the commutativity of the partial derivative, there remain only $\frac{L}{2}$ independent equations to solve (those for which $0\leq n < \frac{L}{2}$). As $N$ is harmonic in $(x^1,x^2)$ we have $\partial_{\hat{1}}^2N=-\partial_{\hat{2}}^2N$ and by applying this identity $m=\frac{L}{2}-n$ times we see the $\frac{L}{2}$ equations are all solved identically:
\begin{equation}
\partial^n_{\hat{1}}\partial^{L-n}_{\hat{2}}{A_{\hat{3}}}^{\hat{D}} =\kappa (-1)^{\frac{L}{2}}\partial^n_{\hat{1}}\partial^{L-n}_{\hat{2}}N=\kappa (-1)^{\frac{L}{2}+m} \partial^{n+2m}_{\hat{1}}\partial^{L-n-2m}_{\hat{2}}N=\kappa (-1)^{L-n}\partial^n_{\hat{2}}\partial^{L-n}_{\hat{1}}N. \label{evenproof}
\end{equation} 
For odd values of $L$ we substitute equation (\ref{OddSoln}) into (\ref{partialde}) to obtain
\begin{equation}
\kappa(\epsilon_{\hat{j}_L\hat{i}_L}) (-1)^{\frac{L+1}{2}}\partial_{\hat{j}_1}\partial_{\hat{j}_2}\ldots \partial_{\hat{j}_{L-1}}\partial_{\hat{i}_{L}}N =\kappa (\epsilon_{\hat{i}_1\hat{j}_1})(\epsilon_{\hat{i}_2\hat{j}_2})\ldots (\epsilon_{\hat{i}_L\hat{j}_L})\partial_{\hat{i}_1}\partial_{\hat{i}_2}\ldots \partial_{\hat{i}_L}N
\end{equation} 
which simplifies to (dropping constant terms)
\begin{equation}
 (-1)^{\frac{L-1}{2}}\partial_{\hat{j}_1}\partial_{\hat{j}_2}\ldots \partial_{\hat{j}_{L-1}}N = (\epsilon_{\hat{i}_1\hat{j}_1})(\epsilon_{\hat{i}_2\hat{j}_2})\ldots (\epsilon_{\hat{i}_{L-1}\hat{j}_{L-1}})\partial_{\hat{i}_1}\partial_{\hat{i}_2}\ldots \partial_{\hat{i}_{L-1}}N
\end{equation} 
where, as $L-1$ is even, these equations are identical to the set obtained for even $L$ and shown to be identities in equation (\ref{evenproof}).

The proof that equations (\ref{OddSoln}) and (\ref{EvenSoln}) are solutions of equation (\ref{partialde}) relied solely upon the fact that partial derivatives commute. The covariant derivatives of the full equation (\ref{covariantD}) do not in general commute. However if the harmonic function $N$ is further constrained to be a holomorphic or anti-holomorphic function in the complex variables $z=x^1+ix^2$ or $\bar{z}=x^1-ix^2$ then the component of the curvature tensor $R_{\hat{1}\hat{2}\hat{1}\hat{2}}$ in the transverse space vanishes for arbitrary level (see Appendix \ref{appriemann}) and the covariant derivatives in these coordinates do commute. This observation guarantees that the equations for the dual field strength may be rearranged so that they take the form 
\begin{equation}
{\cal D}^n_{\hat{1}}{\cal D}^{L-n}_{\hat{2}}{A_{\hat{3}}}^{\hat{D}} =\kappa (-1)^{L-n}{\cal D}^n_{\hat{2}}{\cal D}^{L-n}_{\hat{1}}N \label{generalcovde}
\end{equation} 
where $0\leq n \leq L$ for $n\in \mathbb{Z}$. We note that ${\cal D}_{\hat{1}}{\cal D}_{\hat{1}}N=-{\cal D}_{\hat{2}}{\cal D}_{\hat{2}}N$ and so the arguments for the partial derivatives acting on the harmonic function carry across to the covariant derivatives acting on the (anti-)holomorphic function and the vielbein component ${A_{\hat{3}}}^{\hat{D}}$ is given by equations (\ref{OddSoln}) and (\ref{EvenSoln}) for the odd and even level fields.

The form of the full metric depends on the whether the dual gravtion field appears at an even or odd level in the decomposition of $A_{D-3}^{+++}$.

\subsubsection*{$SL(2,\mathbb{R})/SO(1,1)$: arbitrary even levels}
The even level root given in equation (\ref{levelLroot}) is associated with a coset model $SL(2,\mathbb{R})/SO(1,1)$ if the temporal coordinate is $x^i$ or $x^j$ and consequently the flat $D-4$-dimensional space $\cal N$ is Euclidean. Hence from equation (\ref{dictionary}) we have $\kappa=(-1)^{(1-D)}$ and from equation (\ref{EvenSoln}) ${A_{\hat{3}}}^{\hat{D}}=(-1)^{(1-D+\frac{L}{2})}N$. For the example with $i=D$ and $j=3$ we have
\begin{equation}
ds^{2} = N^{L}\left((dx^{\hat{1}})^2+(dx^{\hat{2}})^2\right)+N(dx^{\hat{3}})^2-N^{-1}(dx^{\hat{D}}
-{A_{\hat{3}}}^{\hat{D}}dx^{\hat{3}})^{2}+d\Sigma^{2}_{(D-4)} \quad \end{equation}
when $x^D$ is the temporal coordinate, and 
\begin{equation}
ds^{2} = N^{L}\left((dx^{\hat{1}})^2+(dx^{\hat{2}})^2\right)-N(dx^{\hat{3}})^2+N^{-1}(dx^{\hat{D}}
-{A_{\hat{3}}}^{\hat{D}}dx^{\hat{3}})^{2}+d\Sigma^{2}_{(D-4)}
\end{equation}
when $x^3$ is the temporal coordinate. The only non-zero components of the Einstein tensor for this metric are proportional to $\left(\left(\partial_{1}N\right)^{2}+\left(\partial_{2}N\right)^{2}\right)$ when $N$ is harmonic. For all levels, except the level $0$ solution which trivially satisfies the Einstein equations, this becomes a vacuum solution when $N$ is (anti-)holomorphic. 

\subsubsection*{$SL(2,\mathbb{R})/SO(1,1)$: arbitrary odd levels}
The odd level root given in equation (\ref{levelLroot}) is associated with a coset model $SL(2,\mathbb{R})/SO(1,1)$ if the temporal coordinate is neither $x^i$ or $x^j$ but one of the set $\{x^4,x^5,\ldots, x^{D-1}\}$ and the flat $D-4$-dimensional space-time $\cal N$ is Minkowski. Hence from equation (\ref{dictionary}) we have $\kappa=1$ and from equation (\ref{OddSoln}) ${A_{\hat{3}}}^{\hat{D}}=(-1)^{\frac{L+1}{2}}B$ where $\partial_{\hat{1}} B = \partial_{\hat{2}} N$ and $\partial_{\hat{2}} B = -\partial_{\hat{1}} N$. For the example with $i=D$ and $j=3$ we have
\begin{equation}
ds^{2} = N^{L}\left((dx^{\hat{1}})^2+(dx^{\hat{2}})^2\right)+N(dx^{\hat{3}})^2+N^{-1}\left(dx^{\hat{D}}-{A_{\hat{3}}^{\hat{D}}} dx^{\hat{3}}\right)^{2}+d\Sigma^{2}_{(1,D-5)}.
\end{equation}
\noindent The only non-zero components of the Einstein tensor for this metric are again proportional to $\left(\left(\partial_{1}N\right)^{2}+\left(\partial_{2}N\right)^{2}\right)$. Taking $N$ to be (anti-)holomorphic ensures this is a solution to the vacuum Einstein equations. 

\subsection{$SL(3,\mathbb{R})/SO(1,2)$: composite gravitational solutions} \label{boundstateofdualgravitons}
In this section we will construct what we will refer to as bound states of KK-monopoles consisting of pairs of the solutions described in section 4.2. To do this we will truncate $A_{D-3}^{+++}$ to $\mathfrak{sl}(3,\mathbb{R})$ sub-algebras whose simple positive roots consist of two real roots which individually are the simple positive roots of two of the $\mathfrak{sl}(2,\mathbb{R})$ sub-algebras discussed in section \ref{singlesolutions1} and \ref{singlesolutions2}. The resulting bound states will interpolate continuously between solutions generated by the Geroch group.

Bound states of two KK-monopoles, just as for the dyonic membrane - the bound state of the membrane and fivebrane in supergravity \cite{Izquierdo:1995ms} discussed in the context of $E_{11}$ in \cite{Cook:2009ri,Houart:2009ya}, correspond to null geodesics on the coset of $\frac{SL(3,\mathbb{R})}{SO(1,2)}$. The algebra $\mathfrak{sl}(3,\mathbb{R})$ has non-trivial commutators
\begin{align}
\nonumber [H_1,E_1]&=2E_1, \quad [H_1,E_2]=-E_2, \quad [H_1,E_{12}]=E_{12}\\
[H_2,E_1]&=-E_1,  \quad [H_2,E_2]=2E_2, \quad [H_1,F_{12}]=-F_{12}\\
\nonumber [E_1,E_2]&=E_{12} \\
\nonumber [F_1,F_2]&=-F_{12}
\end{align}
and the sub-algebra $\mathfrak{so}(1,2)$ is invariant under the involution $\Omega$ which acts as $\Omega(E_{1})=F_{1}$, $\Omega(E_{2})=-F_{2}$, $\Omega(E_{12})=F_{12}$, $\Omega(H_1)=-H_1$ and $\Omega(H_2)=-H_2$.  There are three canonical $\mathfrak{sl}(2,\mathbb{R})$ sub-algebras within $\mathfrak{sl}(3,\mathbb{R})$ and truncation to any of these three sub-algebras leaves one of two cosets either $\frac{SL(2,\mathbb{R})}{SO(1,1)}$ or $\frac{SL(2,\mathbb{R})}{SO(2)}$. For the involution given above the subsets of generators $\{E_1, H_1, F_1\}$ and $\{E_{12}, H_1+H_2, F_{12}\}$ defines the coset $\frac{SL(2,\mathbb{R})}{SO(1,1)}$ and correspond to KK-monopole solutions and other solutions as described in section \ref{singlesolutions1} and \ref{singlesolutions2}, while the generators $\{E_2, H_2, F_2\}$ and the involution define an $\frac{SL(2,\mathbb{R})}{SO(2)}$ coset whose geodesics we have not discussed in detail in the present paper. In this way the full solution related to the null geodesic on the full coset  $\frac{SL(3,\mathbb{R})}{SO(1,2)}$ is understood to correspond to a bound state of a pair of KK-monopole and similar objects. There are two harmonic functions $N_1$ and $N_2$ which are used to define the solution and related to each other by 
\begin{equation}
N_{2} = \mbox{sin}^{2}(\beta)+\mbox{cos}^{2}(\beta)N_{1}\label{harmoniclimits}
\end{equation} 
where $\beta\in \mathbb{R}$, each of which may be thought of as the harmonic function defining the $SL(2,\mathbb{R})$ coset solutions of the previous sections. The parameter $\beta$ encodes the action of the generator of the compact symmetry in $\mathfrak{so}(1,2)$ which transforms the charges of the harmonic functions\footnote{See section 3.6 of \cite{Houart:2009ya} for a discussion of the action of the compact transformations of the sub-group $SO(1,2)$.}. 

\paragraph{Bound states consisting of two KK-monopoles} We will construct the bound state solutions which possess a common two-dimensional transverse space whose simple positive roots both appear at level one in the decomposition of $A_{D-3}^{+++}$. Since $\mathfrak{sl}(3,\mathbb{R})$ cannot be constructed from level one roots in $A^{+++}_{1}$ this is only applicable for algebras with $D \geq 5$ and the solutions we find will be five-dimensional solutions embedded in a $D$-dimensional background, that is space-time manifolds found are product manifolds of the form ${\cal M}_5 \otimes {\cal N}_{D-5}$, where the dimension of the manifold is indicated by the subscript label and ${\cal N}_{D-5}$ is either a Euclidean or a Minkowski space. Consider the pair of real level one roots of $A_{D-3}^{+++}$ given by
\begin{align}
\bm\alpha_{1} &= \vec e_{D}-\vec e_{3}+\sum^{D}_{i=3}\vec e_{i}  \\
\bm\alpha_{2} &= \vec e_{D-1}-\vec e_{D}+\sum^{D}_{i=3}\vec e_{i}
\end{align}
\noindent and note that these satisfy
\begin{equation}
\left\langle \bm\alpha_{1},\bm\alpha_{2}\right\rangle = -1
\end{equation}
and so are the simple positive roots of an $\mathfrak{sl}(3,\mathbb{R})$ sub-algebra whose third positive root is
\begin{equation}
\bm\alpha_{1}+\bm\alpha_{2} = \vec e_{D-1}-\vec e_{3}+2\sum^{D}_{i=3}\vec e_{i}.
\end{equation} 
\noindent They have the Cartan elements
\begin{eqnarray}
H_{1} &=& -\left({K^1}_{1}+{K^2}_{2}+{K^3}_{3}\right) + {K^D}_{D}  \\
H_{2} &=& -\left({K^1}_{1}+{K^2}_{2}+{K^D}_{D}\right) + {K^{(D-1)}}_{(D-1)}
\end{eqnarray}
\noindent and the three positive generators for this example are
\begin{equation}
E_{1}=R^{4\ldots D|D}, \, E_{2}=R^{3\dots (D-1)|(D-1)}\quad \mbox{ and } \quad E_{12}= [E_{1},E_{2}]=-R^{3\ldots D|4\ldots D|(D-1)}.
\end{equation}
\noindent  Imposing the involution to act as $\Omega(E_1)=F_1$ and $\Omega(E_2)=-F_2$ is equivalent to setting $x^{D-1}$ to be the temporal coordinate. Bound state solutions are identified with null-geodesics on cosets of $\frac{SL(3,\mathbb{R})}{SO(1,2)}$ where the representative coset group element is expressed as
$$g=\exp({\phi_1 H_1 +\phi_2 H_2})\exp(C_1 E_1 + C_2E_2 +C_{12}E_{12})$$
where $\phi_1$, $\phi_2$, $C_1$, $C_2$ and $C_{12}$ are functions of the null geodesic parameter $\xi$. These brane coset models were first solved for bound state solutions in \cite{Houart:2009ya} where the ansatz  $\phi_1=\frac{1}{2}\ln{N_1}$ and $\phi_2=\frac{1}{2}\ln{N_2}$ is used. The diagonal portion of the metric for our bound state as
\begin{equation}
ds^{2}_{diagonal}=N_{1}N_{2}\left((dx^{\hat{1}})^2+(dx^{\hat{2}})^2\right)+N_{1}(dx^{\hat{3}})^2-N^{-1}_{2}(dx^{\widehat{D-1}})^2+\frac{N_{2}}{N_{1}}(dx^{\hat{D}})^2 + d\Omega^{2}_{(D-5)}.
\end{equation}
where $N_{1}=a+b \xi$, $N_{2}=c+d \xi$ and $\xi$ is the parameter labelling translations along the null geodesic. As for the $\mathfrak{sl}(2,\mathbb{R})$ case described in detail earlier the Maurer-Cartan form is split as $\partial_\xi g g^{-1} = {\mathcal P}_\xi + Q_\xi$ where ${\mathcal Q}_\xi$ indicates generators of the $\mathfrak{so}(2,1)$ algebra of the isometry group of the coset. The remainder of the coset is spanned by the Cartan elements $H_1$, $H_2$ and $H_{12}$ and $S_1\equiv\frac{1}{2}(E_1-F_1)$, $S_2\equiv \frac{1}{2}(E_2+F_2)$ and $S_{12}\equiv \frac{1}{2}(E_{12}-F_{12})$, where $F_1$, $F_2$ and $F_{12}$ are the generators associated to the negative roots $-\bm\alpha_1$, $-\bm\alpha_2$ and $-\bm\alpha_{12}$. The coefficients of the $S_1$, $S_2$ and $S_{12}$ generators in the Maurer-Cartan form are found \cite{Houart:2009ya}  to be
\begin{eqnarray}
{\mathcal P}_{\xi,1} = -\sqrt{\frac{\alpha}{b}}\frac{\partial_{\xi}N_{1}}{N_{1}\sqrt{N_{2}}}, \; {\mathcal P}_{\xi,2} = -\sqrt{\frac{\alpha}{d}}\frac{\partial_{\xi}N_{2}}{N_{2}\sqrt{N_{1}}} \quad \mbox{and} \quad {\mathcal P}_{\xi,12} = -\sqrt{\frac{d}{b}}\frac{\partial_{\xi}N_{1}}{\sqrt{N_{1}N_{2}}},
\end{eqnarray}
\noindent where $\alpha=N_{2}\partial_{\xi}N_{1}-N_{1}\partial_{\xi}N_{2}=bc-ad$. The form of these ${\mathcal P}_{\xi,i}$ are chosen so that terms in the square root correspond to functions of the parameter $\beta$ which is mapped between $[0,\pi/2]$ by the compact local symmetry. Specifically the harmonic functions are 
$N_1=1+q\xi$ and $N_2=1+q\xi \cos^2{\beta}$ so that $\alpha=q\sin^2{\beta}$. Each ${\mathcal P}_\xi$ corresponds to a dual-gravity field whose construction is given by the same dictionary used in the previous sections:
\begin{eqnarray}
F_{\xi4\ldots D|D} &=& {\mathcal P}_{\xi,1}, \nonumber \\
F_{\xi3\ldots D-1|D-1} &=& {\mathcal P}_{\xi,2} \qquad \mbox{and} \label{sl3fields} \\
F_{\xi4\ldots D|\xi3\ldots D|D-1} &=& D_{\xi}{\mathcal P}_{\xi,12}.\nonumber
\end{eqnarray}
\noindent We first identify $\xi$ with one of the transverse coordinates $x^1$ or $x^2$. After dualisations and transformations, which are identical to those performed in previous sections, the dual field strengths are, when $\xi$ is identified with $x^1$,
\begin{eqnarray}
{F_{\hat{2}\hat{3}}}^{\hat{D}} &=&  \sin(\beta) \frac{\partial_{\hat{1}}N_{1}}{N_{2}} = D_{\hat{2}}{(A_{1})_{\hat{3}}}^{\hat{D}},  \\
{F_{\hat{2}\hat{D}}}^{\widehat{D-1}} &=& (-1)^{D}\tan(\beta) \frac{\partial_{\hat{1}}N_{2}}{N_{1}} = D_{\hat{2}}{(A_{2})_{\hat{D}}}^{\widehat{(D-1)}} \qquad \mbox{and} \\
{F_{\hat{2}|\hat{2}\hat{3}}}^{\widehat{D-1}} &=& (-1)^{D}\cos(\beta) D_{\hat{1}}D_{\hat{1}}N_{1}= D_{\hat{2}}D_{\hat{2}}{(A_{12})_{\hat{3}}}^{\widehat{(D-1)}}.
\end{eqnarray}
\noindent As with the previous examples, the one-dimensional solutions may be unsmeared using the symmetry of the transverse directions to two-dimensional fields which are functions of $x^1$ and $x^2$. The full equations are then
\begin{eqnarray}
{F_{\hat{i}\hat{3}}}^{\hat{D}} &=& \sin(\beta)\frac{\epsilon_{\hat{i}\hat{j}}\partial_{\hat{j}}N_{1}}{N_{2}} = D_{\hat{i}}{(A_{1})_{\hat{3}}}^{\hat{D}}\\
{F_{\hat{i}\hat{D}}}^{\widehat{D-1}}&=&(-1)^D\tan(\beta)\frac{\epsilon_{\hat{i}\hat{j}}\partial_{\hat{i}}N_{2}}{N_{1}} = D_{\hat{i}}{(A_{2})_{\hat{D}}}^{\widehat{D-1}}\\
{F_{\hat{i}\hat{3}}}^{\widehat{D-1}} &=& (-1)^D\cos(\beta)\epsilon_{\hat{i}_1\hat{j}_1}\epsilon_{\hat{i}_2\hat{j}_2}D_{\hat{j}_1}D_{\hat{j}_2}N_{1} = D_{\hat{i}_1}D_{\hat{i}_2}{(A_{12})_{\hat{3}}}^{\widehat{D-1}}.
\end{eqnarray}
An integrability condition on these equations is provided, in the two dimensional space, by
\begin{eqnarray}
\mbox{d}A_1&=&\sin(\beta)\frac{*\mbox{d}N_{1}}{N_{2}} \\
\Rightarrow \quad 0\;=\;\mbox{dd}A_1&=&\sin(\beta)\,\mbox{d}\bigg(\frac{*\mbox{d}N_{1}}{N_{2}} \bigg)
\end{eqnarray}
or, in components,
\begin{equation}
0=\epsilon^{ki}\partial_{k}\partial_{i}{(A_1)_{\hat{3}}}^{\hat{D}}=\sin(\beta)\epsilon^{ki}\epsilon_{ij}\partial_{k}\left(\frac{\partial_{j}N_{1}}{N_{2}}\right)
\end{equation}
\noindent so that, in order for this system to be integrable we need:
\begin{equation}
\partial_{i}\left(\frac{\partial^{i}N_{1}}{N_{2}}\right)=0
\end{equation}
\noindent with summation over $i$ implied. If we take $N_{1}$ to be harmonic and $N_{2}$ to limit to $N_{1}$ as described by (\ref{harmoniclimits}) the integrability condition reduces to:
\begin{equation}
\left(\partial_{1}N_{1}\right)^{2}+\left(\partial_{2}N_{1}\right)^{2}=0.
\end{equation}
\noindent This is solved by taking $N_{1}$, and subsequently $N_{2}$, to be (anti-)holomorphic\footnote{This integrability condition can also be satisfied for arbitrary harmonic functions $N_{1}$ and $N_{2}$ when they are harmonic conjugates. However, the only pair of harmonic functions which would limit to each other as described by equation (\ref{harmoniclimits}) are constant functions.}, which was the same condition required for the component KK-monopole solutions to be vacuum Einstein solutions. The holomorphic functions $N_{1}=1+f(z)$ and $N_{2}=1+\mbox{cos}^{2}(\beta)f(z)$ have dual gravity fields given by
\begin{align}
{(A_1)_{\hat{3}}}^{\hat{D}} &=B_1 \sin{\beta} \label{dualfield1}\\
{(A_2)_{\hat{D}}}^{\widehat{D-1}} &=(-1)^D B_2 \tan{\beta} \label{dualfield2}\\
{(A_{12})_{\hat{3}}}^{\widehat{D-1}} &=(-1)^{(D+1)} N_1\cos{\beta}
\end{align}
where
\begin{align}
\partial_iB_1 &=\epsilon_{ij}\frac{\partial_j N_1}{N_2} \\
\partial_iB_2 &=\epsilon_{ij}\frac{\partial_j N_2}{N_1}
\end{align}
When $\beta=\frac{\pi}{2}$ this reduces to a single KK-monopole solution, while when $\beta=0$ we find a level two object as described section \ref{singlesolutions1}. One can solve equations (\ref{dualfield1}) and (\ref{dualfield2}) to find
\begin{eqnarray}
{(A_1)_{\hat{3}}}^{\hat{D}} &=& i\,\mbox{tan}(\beta)\mbox{sec}(\beta)\ln (N_{2}) \quad \mbox{and}\\
{(A_{2})_{\hat{D}}}^{\widehat{D-1}} &=& i(-1)^D\,\mbox{sin}(\beta)\mbox{cos}(\beta)\ln (N_1).
\end{eqnarray}
\noindent In the case of anti-holomorphic functions where $f=f(\bar{z})$ the $A_{1}$ and $A_{2}$ solutions are simply the opposite sign of those above. Including the off-diagonal dual gravity fields we find the metric
\begin{align}
ds^{2}=&N_{1}N_{2}\left((dx^{\hat{1}})^2+(dx^{\hat{2}})^2\right)+N_{1}(dx^{\hat{3}})^2 + d\Omega^{2}_{|D-5|}\label{2kk6metric}\\
\nonumber &-N^{-1}_{2}\left(dx^{\widehat{D-1}}-{(A_{2})_{\hat{D}}}^{\widehat{D-1}}dx^{\hat{D}}-{(A_{12})_{\hat{3}}}^{\widehat{D-1}}dx^{\hat{3}}\right)^{2}+\frac{N_{2}}{N_{1}}\left(dx^{\hat{D}}-{(A_1)_{\hat{3}}}^{\hat{D}}dx^{\hat{3}}\right)^{2}. 
\end{align}
\noindent We recall the construction of the harmonic functions in equation (\ref{harmoniclimits}) of the model and find that when $\beta=\pi/2$ the harmonic function $N_{2}=1$ and we recover the individual root solution corresponding to $\bm\alpha_{1}=( D,3)_{1}$. When $\beta=0$ only the $A_{12}$ gauge field is present and $N_{1}=N_{2}\equiv N$ leaves us with precisely the $\mathfrak{sl}(3,\mathbb{R})$ model solution with the level 2 root $\vec \alpha_{12} = \bm\alpha_{1}+\bm\alpha_{2}$. The metric in equation (\ref{2kk6metric}) with (anti-)holomorphic $N_{1}$ and $N_{2}$ has a vanishing Ricci scalar for all $\beta$ values (see Appendix B) and has a vanishing curvature tensor at the endpoints which correspond with the level 1 and level 2 solutions. However the bound state is not a solution to the vacuum Einstein equations for $\beta$ taking values in the open set $(0,\frac{\pi}{2})$. In sections 5 and 6, by considering the lift of the sigma-model to $D$ dimensions, we will investigate this obstruction to the full interpolating bound state being a solution to the Einstein-Hilbert action.

\subsection{$SL(3,\mathbb{R})/SO(1,2)$: composite gravitational solutions with arbitrary levels}

The above construction can be generalised to include arbitrary level roots and we present here the $\frak{sl}(3,\mathbb{R})$ model which has simple real roots found at arbitrary level $L_{1}$ and $L_{2}=1$ which will possess an $\frak{so}(1,2)$ invariant sub-algebra and whose associated solutions have a common two dimensional transverse space. Taking 
\begin{eqnarray}
\bm\alpha_{1}&=&\vec e_{D}-\vec e_{3}+L_{1}\sum^{D}_{i=3}\vec{e_{i}} \mbox{ and}\\ 
\bm\alpha_{2}&=&\vec e_{D-1}-\vec e_{D}+\sum^{D}_{i=3}\vec{e_{i}},
\end{eqnarray}
\noindent whose inner product is $-1$, the algebra associated with these roots
\begin{eqnarray}
H_{\bm\alpha_1} = -({K^{1}}_{1}+{K^{2}}_{2})+{K^{3}}_{3}-{K^{D}}_{D},&& H_{\bm\alpha_2} = -({K^{1}}_{1}+{K^{2}}_{2}+{K^{D}}_{D})-{K^{D-1}}_{D-1}\nonumber \\
E_{1} = R^{D|4\ldots D|(3\ldots D)_{1}|\ldots |(3\ldots D)_{L_{1}-1}},&& E_{2}=R^{D-1|3\ldots D-1}\\
 && E_{12}=R^{D-1|4\ldots D|(3\ldots D)_{1}|\ldots |(3\ldots D)_{L_{1}}}\nonumber
\end{eqnarray}
\noindent leads us to identify the ${\mathcal P}_{i}$ fields from the coset model equations of motion with fields
\begin{eqnarray}
F_{D|\xi 4\ldots D|(\xi 3\ldots D)_1|\ldots|(\xi 3\ldots D)_{L_{1}-1}} &=& D^{L_{1}-1}_{\xi}{\mathcal P}_{\xi,1} \\
F_{D-1|\xi 3\ldots D-1} &=& {\mathcal P}_{\xi,2} \\
F_{D-1|\xi 4\ldots D|(\xi 3\ldots D)_1|\ldots|(\xi 3\ldots D)_{L_{1}}} &=& D^{L_{1}}_{\xi}{\mathcal P}_{\xi,12}.
\end{eqnarray}
\noindent We may now employ the same techniques from previous sections to obtain the differential equations which describe our dual gravity fields and unsmear these to find
\begin{eqnarray}
{F_{\hat{i}\hat{3}|(\hat{i})_{1}|\ldots|(\hat{i})_{L_{1}-1}}}^{\hat{D}} &=& \mbox{sin}(\beta)D^{L_{1}-1}_{\hat{i}}\left(\frac{\partial_{\hat{i}}N_{1}}{N_{2}}\right) = (\epsilon_{\hat{i}\hat{j}})^{L_{1}}D^{L_{1}}_{\hat{j}}(A_{1})_{\hat{3}}^{\hat{D}}\\
{F_{\hat{i}\hat{D}}}^{\hat{D-1}}&=&\mbox{tan}(\beta)\frac{\partial_{\hat{i}}N_{2}}{N_{1}} = \epsilon_{\hat{i}\hat{j}}D_{\hat{j}}(A_{2})_{\hat{D}}^{\hat{D-1}}\\
{F_{\hat{i}\hat{3}|(\hat{i})_{1}|\ldots|(\hat{i})_{L_{1}}}}^{\hat{D-1}} &=& \mbox{cos}(\beta)D^{L_{1}+1}_{\hat{i}}N_{1} = (\epsilon_{\hat{i}\hat{j}})^{L_{1}+1}D^{L_{1}+1}_{\hat{j}}(A_{12})_{\hat{3}}^{\hat{D-1}}.
\end{eqnarray}
\noindent As we had found in the $\frak{sl}(3,\mathbb{R})$ model above, these equations only have complex solutions where $N_{1}$ and $N_{2}$ are (anti-)holomorphic functions. The methods used to find arbitrary level solutions in previous sections are still valid since $D_{\hat{1}}D_{\hat{1}}N=-D_{\hat{2}}D_{\hat{2}}N$ for harmonic functions $N$ and $R_{\hat{1}\hat{2}\hat{1}\hat{2}}=0$. We therefore find that
\begin{eqnarray}
A_{1} &=& (i)^{L_{1}}\mbox{tan}(\beta)\mbox{sec}(\beta)\mbox{ Log}N_{2} \\
A_{2} &=& i\mbox{sin}(\beta)\mbox{cos}(\beta)\mbox{ Log}N_{1} \\
A_{12} &=& (i)^{L_{1}+1}\mbox{cos}(\beta)N_{1}
\end{eqnarray}
\noindent when $N_{1}$ and $N_{2}$ are holomorphic. When they are anti-holomorphic every $i$ is replaced by $-i$. We note that these solutions are valid for odd and even $L_{1}$ when the involution is correctly chosen so that $\Omega (E_{1})=F_{1}$ and $\Omega (E_{2})=-F_{2}$. 

We must now specify the level $L_{1}$ and find the appropriate involution in order to build the full solutions. This requires us to consider the odd and even $L_{1}$ separately. For odd $L_{1}$ the involution required has $t=x_{D-1}$ and for even $L_{1}$ $t=x_{D}$ so that the full set of solutions of this form are given by
\begin{eqnarray}
ds^{2}&=&N^{L_{1}}_{1}N_{2}\left(dx^{2}_{1}+dx^{2}_{2}\right)+N_{1}dx^{2}_{3}+(-1)^{L_{1}}N^{-1}_{2}\left(dx_{D-1}+A_{2}dx_{D}+A_{12}dx_{3}\right)^{2}\nonumber \\
&&+(-1)^{L_{1}+1}\frac{N_{2}}{N_{1}}\left(dx_{D}+A_{1}dx_{3}\right)^{2} + d\Omega^{2}_{(D-5)}.
\end{eqnarray}

\section{The supergravity dictionary and multiforms.} \label{multiforms}

The dimensional reduction to three-dimensions of a $D$-dimensional theory allows the remaining part of the theory to be expressed in terms of scalars that parameterise a coset. The scalars of a general theory arise from both the gravity and the matter sectors of the theory, however upon dimensional reduction some information and structure of the $D$-dimensional theory is lost. In particular there is no information in the three-dimensional theory concerning the index-structure of the $D$-dimensional field strength which sources any scalar field. In the present paper we face the problem of lifting the one-dimensional coset invariant Lagrangian to $D$-dimensions where the fields are mixed-symmetry tensors of $GL(D,\mathbb{R})$. For the coset $\frac{SL(3,\mathbb{R})}{SO(1,2)}$ the invariant Lagrangian (\ref{CosetModelLagrangian}) is
\begin{equation}
{\cal L}=-2(\partial_\xi \phi_1)^2 -2(\partial_\xi \phi_2)^2+2(\partial_\xi \phi_1)(\partial_\xi \phi_2) +\tfrac{1}{2}({\mathcal P}_{\xi,1})^2-\tfrac{1}{2}({\mathcal P}_{\xi,2})^2+\tfrac{1}{2}({\mathcal P}_{\xi,12})^2
\end{equation}
where we have set the lapse function $\eta$ to minus one and $\xi$ dentotes the single spatial coordinate. The $\phi_i$ encode the diagonal components of the vielbein from which the diagonal entries of the $D$-dimensional metric may be reconstructed. The $\phi_i$ terms lift to the Ricci scalar constructed from the diagonal entries of the $D$-dimensional metric as shown in appendix \ref{scalar}. The remaining ${\mathcal P}_{\xi,i}$ terms correspond to kinetic terms for the three mixed symmetry fields associated with the Borel sub-algebra of $\mathfrak{sl}(3,\mathbb{R})$ being considered. As commented upon in the earlier sections there is an ambiguity in identifying ${\mathcal P}_{\xi,i}$ with $D$-dimensional field strengths: the supergravity dictionary for mixed symmetry fields remains to be written. 
Consider the example embedded at levels zero, one and two of the decomposition of the $A_{D-3}^{+++}$ algebra where the mixed-symmetry gauge fields are $A_{\mu_1\ldots \mu_{D-3}|\nu}$ and $A_{\mu_1\ldots \mu_{D-2}|\nu_1\ldots \nu_{D-3}|\rho}$ (indices labelled with the same letter are implicitly antisymmetrised). Let the notation $\Omega^{[a_1|a_2|\ldots |a_n]}$ denote the space of mixed-symmetry forms having the symmetry of the Young tableau with $n$ columns of heights $a_1$, $a_2$, $\ldots$ $a_n$ where $a_1 \geq a_2 \geq \ldots a_n$ i.e. $A_{\mu_1\ldots \mu_{D-3}|\nu}\in \Omega^{[D-3|1]}$ and $A_{\mu_1\ldots \mu_{D-2}|\nu_1\ldots \nu_{D-3}|\rho}\in \Omega^{[D-2|D-3|1]}$. The exterior derivative acts on the space of mixed-symmetry tensors as 
\begin{equation}
\d :\Omega^{[a_1|a_2|\ldots |a_n]}\rightarrow \Omega^{[a_1+1|a_2|\ldots |a_n]}\oplus \Omega^{[a_1|a_2+1|\ldots |a_n]} \oplus \ldots \Omega^{[a_1|a_2|\ldots |a_n+1]} \oplus \Omega^{[a_1|a_2|\ldots |a_n|1]}
\end{equation}
by introducing a partial derivative which is projected with the symmetries of each of the multi-form spaces \cite{1982olver,1987olver,DuboisViolette:1999rd,DuboisViolette:2001jk,Bekaert:2002dt}. We include the possibility indicated by the last multi-form space in the sequence above that the derivative is not-antisymmetrised with respect to any of the indices of the multi-form that it acts on. For example the exterior derivative acts on level one multiform components to define a sequence of field strength components
\begin{equation}
\d:(A_{\mu_2\ldots\mu_{D-2}|\nu_2})\rightarrow \partial_{\mu_1}A_{\mu_2\mu_3|\nu_2} +\partial_{\nu_1}A_{\mu_2\mu_3|\nu_2}+\partial_{\rho}A_{\mu_2\mu_3|\nu_2}.
\end{equation}
Unlike form fields, where the exterior derivative takes $p$-form gauge fields to $p+1$-form fields strengths, a multi-form gauge field is mapped to multiple multi-fprm field strengths. It is therefore ambiguous which higher-dimensional field strength components should be preferred in the supergravity dictionary and equated with ${\mathcal P}_{\xi,1}$ and ${\mathcal P}_{\xi,2}$. We will argue that there is a minimal consistent way to identify ${\mathcal P}_{\xi,i}$ which is indicated by the embedding of the $\mathfrak{sl}(3,\mathbb{R})$ into the algebra of $A_{D-3}^{+++}$ and our guiding principle will be to ensure that the equations for ${\mathcal P}_\xi$ which reflect the $\mathfrak{sl}(3,\mathbb{R})$ structure are maintained by the dictionary. 

For a representative coset element
\begin{equation}
g=\exp(\phi_1 H_1 +\phi_2 H_2)\exp(C_1E_1+C_2E_2+C_{12}E_{12})
\end{equation}
we compute
\begin{align}
\nonumber {\mathcal P}_{\xi,1}&= \exp{(2\phi_1 -\phi_2)}\partial_\xi C_1,\label{dictionaryterms}\\
{\mathcal P}_{\xi,2}&= \exp{(2\phi_2 -\phi_1)}\partial_\xi C_2 \quad \mbox{and}\\
\nonumber {\mathcal P}_{\xi,12}&= \exp{(\phi_1 +\phi_2)}(\partial_\xi C_{12}-\frac{1}{2}\partial_\xi C_1 C_2+\frac{1}{2}\partial_\xi C_2 C_1)\\
\nonumber &= \exp{(\phi_1 +\phi_2)}\partial_\xi C_{12}-\frac{1}{2}{\mathcal P}_{\xi,1} C_2+\frac{1}{2}{\mathcal P}_{\xi,2}C_1.
\end{align}
Our proposal will be most simply motivated by first considering a simple alternative dictionary. Suppose that, contrary to our proposition, the multiforms were treated as form fields by declaring that a set of their antisymmetric indices are in the priveleged position of being space-time form indices while the remaining indices are treated as internal indices. For example suppose that we treat $A_{\mu_1\ldots \mu_{D-3}|\nu}$ as a $D-3$ form (carrying an internal vector index) and $A_{\mu_1\ldots \mu_{D-2}|\nu_1\ldots \nu_{D-3}|\rho}$ as a $D-2$ form with corresponding field strengths given by
\begin{align}
F_{[D-2|1]}&\equiv (D-2)\partial_{\mu_1}A_{\mu_2\ldots \mu_{D-2}|\nu} \qquad \mbox{and}\\
\nonumber G_{[D-1|D-3|1]}&\equiv G_{\mu_1\ldots \mu_{D-1}|\nu_1\ldots \nu_{D-3}|\rho} \\
\nonumber&=(D-1)(\partial_{\mu_1}A_{\mu_2\ldots \mu_{D-1}|\nu_1\ldots \nu_{D-3}|\rho}
-\tfrac{(D-2)}{2}\partial_{\mu_1}A_{\nu_1\ldots \nu_{D-3}|\mu_{D-1}}A_{\mu_2\mu_3\ldots \mu_{D-2}|\rho}\\
& \qquad +\tfrac{(D-2)}{2}\partial_{\mu_1}A_{\mu_2\mu_3\ldots \mu_{D-2}|\rho}A_{\nu_1\ldots \nu_{D-3}|\mu_{D-1}}).
\end{align}
The definition of $G_{[D-1,D-3,1]}$ is found using 
\begin{equation}
[R^{a_1a_2\ldots a_{D-3}|b},R^{c_1c_2\ldots c_{D-3}|d}]=R^{a_1\ldots a_{D-3}d|c_1\ldots c_{D-3}|b}-R^{c_1\ldots c_{D-3}b|a_1\ldots a_{D-3}|d}+\ldots
\end{equation}
where the ellipsis indicates level two generators in the full $A_{D-3}^{+++}$ algebra beyond the truncation to the $\mathfrak{sl}(3,\mathbb{R})$ algebra encoding the bound state we are considering here. By comparison with the expression for ${\mathcal P}_{\xi,12}$ in equation (\ref{dictionaryterms}) we see that while ${\mathcal P}_{\xi,2}$ is identified with a component of $F_{[D-2|1]}$, ${\mathcal P}_{\xi,1}$ would be identified with $2\partial_{\mu_1} A_{\nu_1\ldots \nu_{D-3}|\mu_{2}}\equiv F_{[D-3|2]}$. The dictionary is not well defined at level one as the two level one fields are treated differently. Consider instead the proposition that the exterior derivative acts on level two multiform fields in the following minimal manner
\begin{equation}
\d:\Omega^{[D-2|D-3|1]}\rightarrow \Omega^{[D-1|D-3|1]} \oplus \Omega^{[D-2|D-2|1]} \oplus \Omega^{[D-2|D-3|2]}
\end{equation}
we refer to this as a minimal action when we neglect the mapping into the multiforms with the symmetries of four column wide Young tableaux. At level one we consider the full mapping:
\begin{equation}
\d:\Omega^{[D-3|1]}\rightarrow \Omega^{[D-2|1]} \oplus \Omega^{[D-3|2]} \oplus \Omega^{[D-3|1|1]}
\end{equation}
so that in both cases the derivative is distributed across three sets of indices. 
Consider a five-dimensional example\footnote{As the bound states are product spaces in which only a five-dimensional sub-manifold has a non-trivial metric, we will not lose any generality by focussing on a five-dimensional example.} constructed using the Borel sub-algebra
\begin{align}
\nonumber H_1&=-({K^1}_1+{K^2}_2)-{K^3}_3+{K^5}_5, \\
H_2&=-({K^1}_1+{K^2}_2)-{K^5}_5+{K^4}_4, \\
\nonumber E_1&=R^{45|5}, \quad E_2=R^{34|4} \quad \mbox{and} \quad E_{12}=R^{345|45|4}
\end{align}
where the involution $\Omega$ is chosen to be consistent with taking $x^4$ as the single temporal coordinate. The field strengths are exterior derivatives of multiform tensors $A_{\mu_1\mu_2|\nu}$ and $A_{\mu_1\mu_2\mu_3|\nu_1\nu_2|\rho}$ and the dictionary identifies ${\mathcal P}_\xi$ with components across different multiform spaces which reduce to a sum of vectors as
\begin{align}
{\mathcal P}_{\xi,1}&=F_{\xi 4 5|5}+F_{ 4 5|\xi 5}+F_{ 4 5|5|\xi}\equiv \d_\xi A_{45|5},\\
{\mathcal P}_{\xi,2}&=F_{\xi 3 4|4}+F_{ 3 4|\xi 4}+F_{ 3 4|4|\xi}\equiv \d_\xi A_{34|4} \quad \mbox{and} \\
{\mathcal P}_{\xi,12}&=G_{\xi 3 4 5|4 5|4}+G_{3 4 5|\xi 4 5|4}+G_{3 4 5|4 5|\xi 4}\equiv \d_\xi A_{345|45|4}.
\end{align}
Now, as vectors,
\begin{align}
F_{\xi 4 5|5} = F_{ 4 5|\xi 5} = F_{ 4 5|5|\xi}=\partial_\xi (A_{45|5})
\end{align}
whereas upon the lift to five dimensions these components, while all equal, arise from three different multiform field strengths. While at level two we have 
\begin{align}
G_{\xi 3 4 5|4 5|4}&\equiv\partial_{\xi} A_{3 4 5|4 5 |4}-\tfrac{1}{2} F_{45|\xi 5}A_{34|4}+\tfrac{1}{2} F_{\xi 34|4}A_{45|5}\\
=G_{3 4 5|\xi 4 5|4}&\equiv \partial_{\xi} A_{3 4 5|4 5 |4}-\tfrac{1}{2} F_{\xi 45| 5}A_{34|4}+\tfrac{1}{2} F_{34|4|\xi}A_{45|5}\\
=G_{3 4 5|4 5|\xi 4}&\equiv \partial_{\xi} A_{3 4 5|4 5 |4}-\tfrac{1}{2} F_{45| 5|\xi}A_{34|4}+\tfrac{1}{2} F_{34|4|\xi}A_{45|5}
\end{align}
consequently
\begin{align}
\nonumber {\mathcal P}_{\xi,12}&= \d_{\xi} A_{3 4 5]|4 5 |4}\\
& \quad -\tfrac{1}{2} (F_{\xi 45|5}
+F_{45|\xi 5}+ F_{ 4 5|5|\xi})A_{34|4}+\tfrac{1}{2} (F_{\xi 34|4}+F_{34|\xi 4}+ F_{ 3 4|4|\xi})A_{45|5}\\
\nonumber &= \d_\xi C_{12} -\tfrac{1}{2}{\mathcal P}_{\xi,1} C_2+\tfrac{1}{2}{\mathcal P}_{\xi,2} C_1
\end{align}
where the partial derivative is understood to distribute as the component of an exterior derivative across the multiform fields. While the dictionary definition contains a redundancy in the three-fold generation of field strengths from a single gauge field it has the advantage that it reproduces the non-trivial structure equation of the $\mathfrak{sl}(3,\mathbb{R})$ sub-algebra. 

The redundancy in the dictionary permits us to prefer a set of field strengths, that is as the components of each field strength are equal, for example $F_{\xi 45|5}=F_{45|\xi 5}=F_{45|5|\xi}=-\frac{1}{3}\sin{\beta}\partial_\xi N_1^{-1}$, we may eliminate field strengths algebraically in the action. In practise we may return to treating a set of indices in a priveleged manner, at least for identifying a $D$-dimensional action whose equations of motion are satisfied by the null geodesic on $\frac{SL(3,\mathbb{R})}{SO(1,2)}$. For example the case where the derivative is antisymmetrised with the leading column of each Young tableau has the action 
\begin{equation}
S_1=\int R \star \mathbb{I} -\tfrac{1}{2} F_{[D-2|1]}\wedge \star F_{[D-2|1]} -\tfrac{1}{2} G_{[D-1|D-3|1]}\wedge \star G_{[D-1|D-3|1]} \label{action}
\end{equation}
where $F_{\xi 4\dots D|D]}={\mathcal P}_{\xi,1}$, $F_{\xi 3\dots (D-1)|(D-1)]}={\mathcal P}_{\xi,2}$, $G_{[\xi 3\ldots D|4\ldots D|(D-1)]}={\mathcal P}_{\xi|12}$, $\star$ denotes the Hodge dual on the form indices, while the remaining internal indices on the kinetic terms are contracted with the metric. The equations of motion for the metric, $A_{[D-3|1]}$ and $A_{[D-2|D-3|1]}$ from (\ref{action}) are satisfied for all points on the interpolating bound state described by a null-geodesic on $\frac{SL(3,\mathbb{R})}{SO(1,2)}$, that is by the non-zero field strength components
\begin{align}
\nonumber F_{\xi 4\ldots (D-1)D|D}&=-\sin{\beta}\, \partial_\xi N_1^{-1},\\
F_{\xi 3\ldots (D-2)(D-1)|(D-1)}&=-\tan{\beta}\, \partial_\xi N_2^{-1} \qquad \mbox{and} \label{1dsolution}\\
\nonumber G_{\xi 3\ldots D|4\ldots D| (D-1)}&= -\cos{\beta}\, \frac{\partial_\xi N_1}{N_1 N_2}
\end{align}
where $N_1=1+Q\xi$, $N_2=1+Q\xi\cos^2{\beta}$ and $\xi$ labels a single transverse direction either $\xi=1$ or $\xi=2$, see appendix \ref{EinsteinEquations} to see the equations of motions satisfied in the five-dimensional case. Furthermore the field strength components may be unsmeared to two-dimensions and still satisfy the equations of motion of the above action. The unsmearing takes advantage of the spherical symmetry in the two transerse directions, in this case the solution is described as above in equation (\ref{1dsolution}) but allowing $\xi\in \{1,2\}$ and redefining $N_1=1+Q \ln{(r)}$ and $N_2=1+Q\ln{(r)}\cos^2{\beta}$ where $r\equiv \sqrt{(x^1)^2+(x^2)^2}$. 

Similarly one could have considered the actions
\begin{equation}
S_2=\int R \star_2 \mathbb{I} -\tfrac{1}{2} F_{[D-3|2]}\wedge \star_2 F_{[D-3|2]} -\tfrac{1}{2} G_{[D-2|D-2|1]}\wedge \star_2 G_{[D-2|D-2|1]}. 
\end{equation}
where $\star_2$ indicates the Hodge dual on the second set of indices of the field strength, or
\begin{equation}
S_3=\int R \star_3 \mathbb{I} -\tfrac{1}{2} F_{[D-3|1|1]}\wedge \star_3 F_{[D-3|1|1]} -\tfrac{1}{2} G_{[D-2|D-3|2]}\wedge \star_3 G_{[D-2|D-3|2]}. 
\end{equation}
and $\star_3$ indicates the Hodge dual on the third set of indices of the field strength. Each action has equations of motion solved by fields encoded in the null geodesic on $\frac{SL(3,\mathbb{R})}{SO(1,2)}$. 

\section{An obstruction to an $A_{D-3}^{+++}$ symmetry of Einstein-Hilbert action.} \label{obstruction}
The algebra $A_{8}^{+++}\in E_{11}$ was first identified with an extended symmetry of gravity in \cite{Lambert:2001he}. An action for the $D$-dimensional dual graviton was given in \cite{West:2001as} and investigated further in \cite{West:2002jj}. The reason $A_{D-3}^{+++}$ is relevant to gravity is apparent: the fields associated with the fundamental generators of the algebra are of the correct index type to be associated with the vielbein at level zero and with the dual graviton at level one, all other generators in the algebra are constructed by taking commutators of this pair. It is also evident that a traceless, massless field associated with $R^{\mu_1\ldots \mu_{D-3}|\nu}$ carries $\frac{D}{2}(D-3)$ degrees of freedom as does the $D$-dimensional graviton. But one may wonder whether a theory containing both fields is a theory of a single graviton or a pair of gravitons. In the first case one would expect to identify components of $g_{\mu\nu}$ and $A_{\mu_1\ldots \mu_{D-3}|\nu}$ by duality relations, and solutions of exotic gravity and matter actions would be mapped to solutions of the Einstein-Hilbert action. Consequently we face the puzzle of how to dualise the exotic actions of the previous section, which admit the full interpolating bound state of dual gravitons as a solution, to the Einstein-Hilbert action. We are aware from the discussion in the first half of the paper that the dualisation of the bound state solution is not a solution of the Einstein-Hilbert action apart from at the end points of the interpolation. We therefore expect to find that the action (\ref{action}) is only equivalent to the Einstein-Hilbert action when $\beta=0,\frac{\pi}{2}$ that is at the end points of the interpolation. In this section we will show that this is the case and show that the full interpolating solution is preserved when we treat the mixed-symmetry fields as multiforms.

The prototype bound state solution encoded as a null geodesic on $\frac{SL(3,\mathbb{R})}{SO(1,2)}$ is the dyonic membrane of supergravity \cite{Izquierdo:1995ms}. The equation of motion for $A_{\mu_1\mu_2\mu_3}$ has contributions from the Chern-Simons term and it is precisely the interpolating parts of the bound state where the Chern-Simons term has an active role. In the present context, where we have shown that the end-points of the interpolating gravitational bound state are solutions to Einstein-Hilbert gravity but the interpolating points are not solutions, we anticipate that the full interpolating bound state will be a solution of the Einstein-Hilbert action with an additional Chern-Simons-like term.

Commencing with the $D$-dimensional action:
\begin{equation}
S_1=\int R \star \mathbb{I} -\tfrac{1}{2} F_{[D-2|1]}\wedge \star F_{[D-2|1]} -\tfrac{1}{2} G_{[D-1|D-3|1]}\wedge \star G_{[D-1|D-3|1]}
\end{equation}
where $R$ is the Ricci curvature formed from the fields associated with the Cartan sub-algebra, i.e. from the diagonal part of the metric, $\star$ indicates the Hodge dual and $\wedge$ the exterior product on the form indices (all other indices are contracted using the metric). To make the connection with the Einstein-Hilbert action one must dualise the higher rank mixed symmetry field strengths at the level of the action to find an action for the vielbein. The dualisation would be carried out in two steps with the first step eliminating $A_{[D-2,D-3,1]}$ and introducing a new non-zero component of $A_{[D-3,1]}$, the second step would replace $A_{[D-3,1]}$ with vielbein components corresponding to off-diagonal components of the metric. To dualise $G_{[D-1|D-3|1]}$ on its first set of indices alone we introduce a Lagrange multiplier $\chi$ via the term
\begin{equation}
-\chi^{[D-3|1]}(d G_{[D-1|D-3|1]} -Y\cdot(F_{[2|D-3]} F_{[D-2|1]})) \label{Lagrangemultiplier}
\end{equation}
to the action (\ref{action}), where $F_{[2|D-3]}$ has components $2\partial_{\mu_1}A_{\nu_1\ldots \nu_{D-3}|\mu_2}$, $d$ indicates an exterior derivative which acts only on the first set of form indices in $G$, i.e. 
$$dG=\partial_{\mu_1} G_{\mu_2\ldots \mu_D|\nu_1\ldots \nu_{D-3}|\rho} dx^{\mu_1}\wedge dx^{\mu_2}\wedge \ldots dx^{\mu_D}\otimes dx^{\nu_1}\wedge \ldots dx^{\nu_{D-3}}\otimes dx^{\rho}$$
and $Y$ denotes the young projector that projects into $\Omega^{[D|D-3|1]}$.
Varying the action with respect to $\chi$ gives the term in brackets above which is identically zero (when treating the fields as forms). Varying the action with respect to $G$ gives the algebraic identity:
\begin{equation}
{F_{[1|}}^{D-3|1]}\equiv d\chi^{[D-3|1]}= \star {G_{[D-1|}}^{D-3|1]}.
\end{equation}
This is a component of a new one-form field strength and not a component of $F_{[D-2|1]}$ and is related to $A_{[D-3|1]}$ by
\begin{equation}
F_{\mu_1|\nu_1\ldots \nu_{D-3}|\rho}\equiv \partial_{\mu_1} A_{\nu_1\ldots \nu_{D-3}|\rho}.
\end{equation}
This observation is already enough to motivate treating the fields as multiforms. However we understand from the previous section that components of $F_{\mu_1|\nu_1\ldots \nu_{D-3}|\rho}$ and $F_{\mu_1\nu_1\ldots \nu_{D-3}|\rho}$ are equal and one may take advantage of this to identify a new non-zero component of $F_{[D-2|1]}$ in the action. Substituting our dualisation and identity into the action we find 
\begin{equation}
S_1=\int R \star \mathbb{I} -\tfrac{1}{2} F_{[D-2|1]}\wedge \star F_{[D-2|1]} +A^{[D-3|1]}Y\cdot(F_{[2|D-3]} F_{[D-2|1]})\label{action2}.
\end{equation}
After dualising $G$, the bound state solution has non-zero field strength components given by (see equation (\ref{1dsolution}))
\begin{align}
\nonumber F_{\xi 4\ldots (D-1)D|D}&=-\sin{\beta}\, \partial_\xi N_1^{-1},\\
F_{\xi 3\ldots (D-2)(D-1)|(D-1)}&=-\tan{\beta}\, \partial_\xi N_2^{-1} \qquad \mbox{and} \label{1dsolution2}\\
\nonumber F_{\xi'4\ldots D| (D-1)}&= \cos{\beta}\, \frac{\partial_\xi N_1}{N_1 N_2}
\end{align}
where $\xi'\neq \xi$ and $\xi,\xi'\in\{1,2\}$. These components now fail to solve the metric's equation of motion for the full interpolation, but do solve it at the end points. We have erred in our dualisation. The source of our mistake is the elimination of $A_{[D-3|D-2|1]}$ in the Bianchi identity for $G$. While we have attempted to treat it as a $(D-3)$-form so that $d^2 A_{[D-3|D-2|1]}=0$ the structure of the algebra necessitates that it is a multi-form field such that ${\mathfrak d}^n A_{[D-3|D-2|1]}=0$ only for $n\geq 4$. Modifying the Lagrange multiplier term in (\ref{Lagrangemultiplier}) to include the second derivatives on $A_{[D-2|D-3|1]}$ and integrating by parts gives 
\begin{equation}
+{F_{[1|}}^{D-3|1]}\bigg(G_{[D-1|D-3|1]} -Y\cdot(dA_{[D-2|D-3|1]}- \tfrac{1}{2}F_{[D-3|2]} A_{[D-3|1]}+\tfrac{1}{2}F_{[D-2|1]}A_{[D-3|1]})\bigg). \label{Lagrangemultiplier2}
\end{equation}
Repeating the dualisation fails to eliminate $A_{[D-2|D-3|1]}$ from the action and we are left with
\begin{equation}
S'_1=\int R \star \mathbb{I} -\tfrac{1}{2} F_{[D-2|1]}\wedge \star F_{[D-2|1]} -{F_{[1|}}^{D-3|1]}dA_{[D-2|D-3|1]}-A^{[D-3|1]}Y\cdot (F_{[2|D-3]} F_{[D-2|1]})\label{action3}
\end{equation}
where for the bound state we have
\begin{equation}
A_{[D-2|D-3|1]}=\frac{1}{2}\cos{\beta}(\frac{1}{N_2}+\frac{1}{N_1\cos^2{\beta}}).
\end{equation}
This action does admit the full interpolating bound state as a solution to its equations of motion, see appendix \ref{EinsteinEquations} for the five-dimensional equations of motion. We note that due to the contraction of indices the Chern-Simons term contributes to the metric equation of motion. 

Following the proposal that the mixed-symmetry fields must be treated as multiforms to preserve the solutions under dualisation, there is no possibility to remove $A_{[D-2|D-3|1]}$ without considering a higher derivative action. The required Bianchi identity is fourth order in derivatives:
\begin{equation}
{\mathfrak d}^3(G_{[D-2|D-3|1]})=\frac{1}{2}{\mathfrak d}(Y\cdot(-F_{[D-2|2]} F_{[D-3|2]}+F_{[D-2|2]} F_{[D-2|1]})).
\end{equation}
The Bianchi identity is trivially zero for functions of one variable but contributes for functions of two variables. Generically as ${\mathfrak d}^2 G \in \Omega^{[D-1|D-2|2]}$ then ${\mathfrak d}^3 G \in \Omega^{[D|D-2|2]}\oplus \Omega^{[D-1|D-1|2]} \oplus \Omega^{[D-1|D-2|3]}$. A Lagrange multiplier would exist in the (thrice) dual space $\Omega^{[0|2|D-2]}\oplus \Omega^{[1|1|D-2]} \oplus \Omega^{[1|2|D-3]}$ which is the same space in which ${\mathfrak d}^2 \Omega^{[D-3|1|0]}$ exists. The appropriate Lagrange multiplier term is 
\begin{equation}
{\mathfrak d}^2\chi_{[D-3|1|0]}\bigg({\mathfrak d}^3 (G_{[D-1|D-2|2]})-\frac{1}{2}Y\cdot(F_{[D-2|2]} F_{[D-3|2]}-F_{[D-2|2]} F_{[D-2|1]})\bigg)
\end{equation}
where $Y$ acts so that the appropriate Young tableau symmetries are projected onto the product of multiforms. Now the Lagrange multiplier term is a six-derivative term and the non-trivial fields for the bound state will only solve the equations of motion of an action in which all terms are also six-derivative terms, that is the level one field strength would be $F_{[D-2|2|1]}$ with components $\partial_{\rho}\partial_{\nu_1}\partial_{\mu_1}A_{\mu_2\ldots \mu_{D-2}|\nu_2}$ and the level zero field associated to the diagonal part of the vielbein would have an equivalent six-derivative term. The precise form of the term can be reconstructed from a six-derivative sigma-model\footnote{In the same way that the Ricci scalar is found from the two-derivative sigma-model in appendix \ref{scalar}.} where the Lagrangian is 
\begin{equation}
{\cal L}'=-({\mathfrak d}^2 {\cal P}_\xi|{\mathfrak  d}^2 {\cal P}_\xi)
\end{equation}
we find 
\begin{align}
R_6\equiv & -2\partial_\mu\partial^\mu\partial_\nu\partial^\nu\partial_\kappa\partial^\lambda {h_\lambda}^\kappa+2\partial_\mu\partial^\mu\partial_\nu\partial^\nu\partial_\kappa\partial^\kappa {h_\lambda}^\lambda+4\partial_\mu\partial_\nu\partial^\kappa {h_\kappa}^\lambda \partial^\mu\partial^\nu\partial_\lambda {h_\sigma}^\sigma\\
\nonumber &-4\partial_\mu\partial_\nu\partial_\kappa {h_\lambda}^\sigma \partial^\mu\partial^\nu\partial^\lambda {h_\sigma}^\kappa+2\partial_\mu\partial_\nu\partial_\kappa {h_\lambda}^\kappa \partial^\mu\partial^\nu\partial^\sigma {h_\sigma}^\lambda-\partial_\mu\partial_\nu\partial_\kappa {h_\lambda}^\sigma \partial^\mu\partial^\nu\partial^\kappa {h_\sigma}^\lambda\\
\nonumber & -\partial_\mu\partial_\nu\partial_\kappa {h_\lambda}^\lambda \partial^\mu\partial^\nu\partial^\kappa {h_\sigma}^\lambda.
\end{align}
It is not clear to the authors that this term has a simple geometrical expression. For example the equivalent four-derivative set of terms are not related to the Gauss-Bonnet gravity terms.

The $D$-dimensional six-derivative action after dualisation is
\begin{align}
S=\int\int\int & R_6 \star_1 \star_2 \star_3 \mathbb{I}-\frac{1}{2}F_{[D-2|2|1]}\wedge^3 \star_1 \star_2 \star_3  F_{[D-2|2|1]}\\
\nonumber &+\frac{1}{2}F_{[D-2|2|1]}\wedge^3 (Y\cdot (F_{[D-2|2]}F_{[D-3|2]}F_{[D-2|2]}F_{[D-2|1]}))
\end{align}
where $Y$ is the appropriate Young tableau projector, $\wedge^3$ denotes the triple wedge product applied to each of the three sets of indices and $\star_i$ denotes the Hodge dual on the $i$'th set of antisymmetric indices. 
One might hope to continue the dualisation in the same manner by constructing the field strength with derivatives on each set of indices of $A_{[D-3|1]}$, i.e. $F_{[D-2|2]}$ and construct a four derivative term, nested within a further two derivatives, in the action with a Lagrange multiplier and the Bianchi identity for $F_{[D-2|2]}$ generically 
\begin{equation}
\int {\mathfrak d}^2 \int \int {\mathfrak d}\chi_{[1|D-3]}{\mathfrak d}(F_{[D-2|2]})
\end{equation}
The field is self-dual when dualisation is carried out over all sets of indices. 
Restricting to a dualisation over a single index requires the term
\begin{equation}
\int \int {\mathfrak d}^4  \int {\chi_{[1|}}^{1]}({\mathfrak d}(F_{[D-2|1]})-F_{[D-3|2]})
\end{equation}
so that $d{\chi_{[1|}}^{1]}=\star_1 {F_{[D-2|}}^{1]}$ but the field $A_{[D-3|1]}$ remains present in the action. 

\section{Discussion}
In this paper we have investigated the gravitational solutions associated to real roots of $A_{D-3}^{+++}$ algebras. For the case when $D=11$ the solutions we have presented here are also present in the non-linear realisation of  $\mathfrak{e}_{11}$ which is conjectured to be the extension of supergravity relevant to M-theory \cite{West:2001as}. Such affine classes of solutions have been studied before \cite{Kleinschmidt:2005bq,Englert:2007qb} and it has been shown that the Geroch group of solutions, defined using the Weyl reflections of $A_1^+$, are not only relevant to the gravitational sector of M-theory but also to the M2-M5 branes as well \cite{Englert:2007qb}. This work established solutions for every co-dimension two object which is predicted to exist in M-theory from $E_{11}$. In the present work we have understood each solution within the gravity tower sub-sector of  \cite{Englert:2007qb} in terms of a null geodesic motion on coset $\frac{SL(2,\mathbb{R})}{SO(1,1)}$. At low levels of  $A_{D-3}^{+++}$ the solutions include the pp-wave and the KK$(D-5)$-brane. For a different choice of real form of  $A_{D-3}^{+++}$ the KK$(D-5)$-brane solution is the Euclidean Taub-NUT solution, which is a solution of four dimensional gravity theory trivially embedded in a $D$-dimensional Minkowski spacetime. Similarly the co-dimension two solutions investigated in this paper all possess a large transverse isometry and we may regard them as five-dimensional gravity solutions trivially embedded in a $D$-dimensional background. 

The observation of \cite{Cook:2009ri} that bound-state solutions, such as the dyonic membrane, could be described by an $E_{11}$ group element led to the investigation of Lagrangians on cosets of groups of rank two and greater \cite{Houart:2009ya}. It was shown that the bound state solutions could also be understood as encoding a null geodesic motion on a coset. In both  \cite{Cook:2009ri,Houart:2009ya} the solutions were characterised by a continuous interpolating parameter that moved the from one $\frac{1}{2}$-BPS solution to another. By considering the $\mathfrak{sl}(3,\mathbb{R})$ sub-algebras embedded within the algebra of $A_{D-3}^{+++}$ model we have constructed interpolating solutions which move between any two gravitational solutions which appear at adjacent levels in the level decomposition of $A_{D-3}^{+++}$ into representations of $A_{D-1}$. Such a construction  allows the possibility to, with a combination of several models in succession, interpolate between the solutions occurring at any positive levels by means of a smooth interpolating parameter. Consider the combination of roots
\begin{equation}
\bm\alpha_{i} = \begin{cases} \vec e_{D}-\vec e_{3} +\sum^{D}_{i=3}\vec e_{i} & \mbox{if } i=1 \\
\vec e_{D-1}-\vec e_{D} +\sum^{D}_{i=3}\vec e_{i} & \mbox{if } i>1 \mbox{ and } i \mbox{ even} \\
\vec e_{D}-\vec e_{D-1} +\sum^{D}_{i=3}\vec e_{i} & \mbox{if } i>1 \mbox{ and } i \mbox{ odd} \end{cases}
\end{equation}

\noindent with $t=x_{D-1}$. Any root $\bm\alpha_{12\ldots n}=\bm\alpha_{1}+\bm\alpha_{2}+\ldots+\bm\alpha_{n}$ will have inner product of $-1$ with $\bm\alpha_{n+1}$. Choosing an involution such that the generator associated with the first root is involution invariant while the second is not, so that the involution invariant sub-algebra is $\mathfrak{so}(1,2)$. If we begin at any level $m$ we may create a coset model which will limit to the level $m+1$ solution from which we can construct a new model. This shows us that by advancing the parameters in each $\mathfrak{sl}(3,\mathbb{R})$ to produce the next level object we may telescopically reach an arbitrary positive level from a series of $\mathfrak{sl}(3,\mathbb{R})$ models. However the bound states of $A_{D-3}^{+++}$ constructed as null geodesics on $\frac{SL(3,\mathbb{R})}{SO(1,2)}$ do not lift to interpolating solutions of the Einstein-Hilbert action. Instead, as shown in \cite{Englert:2007qb}, only the end points of the bound states are solutions of the Einstein-Hilbert action. As shown explicitly in appendix \ref{scalar} the bound state is a solution to a gravity and matter theory, which may be simply constructed using the index structure of the mixed-symmetry fields and the one-dimensional sigma-model Lagrangian. This is puzzling as one would expect to be able to dualise the action on fields of $A_{D-3}^{+++}$ to and action written in terms of the vielbein, while preserving the full interpolating solution. The loss of the full interpolating solution is a consequence of the vanishing of terms such as $d^2 A_{[D-2|\ldots |D-2|D-3|1]}$, while if the field is retained in the action the full interpolating solution persists. This observation, discussed in section \ref{obstruction}, motivates the consideration of an extension of the exterior derivative to a derivative which distibutes over all the indices of multiform fields \cite{1982olver,1987olver,DuboisViolette:1999rd,DuboisViolette:2001jk,Bekaert:2002dt}. We argued in section \ref{multiforms} that such a treatment of multiform fields maintained the structure of the equations for the null geodesic on the coset and gave a simple extension of the supergravity dictionary to multiform field-strengths. The brane coset model remains to be extended to include a generalised vielbein as appears in the non-linear realisation of $l_1\ltimes E_{11}$ \cite{West:2011mm} and such an extension may suggest an alternative interpretation for the mixed-symmetry field strengths in the supergravity dictionary

It was argued in section \ref{obstruction} that the exotic gravity and matter action could not be dualised to an action of just the Einstein-Hilbert term alone, instead Chern-Simons-like terms will remain. It is interesting to wonder, in the context of recent observations in massive gravity \cite{deRham:2010kj,deRham:2010ik}, whether the Chern-Simons term retained from the $\mathfrak{sl}(3,\mathbb{R})$ sigma-model action might be consistently identified with a product of vielbein components as seen in the actions of \cite{Hinterbichler:2012cn} - at first glance this would seem unlikely due to the presence of derivatives in the sigma-model terms. If such a link were made the dual graviton would be reinterpreted as a second graviton.

The initial motivation for studying $A_{D-3}^{+++}$ algebras was that $E_{11}$ is the dimensional reduction of $A_9^{+++}$ and contains $A_8^{+++}$. Consequently the M-theory bound states of branes encoded as null geodesics on cosets of sub-groups of $E_{11}$ should lift to bound states described within the gravitational algebra $A_9^{+++}$. As argued in this paper there are no interpolating solutions of the Einstein-Hilbert term alone, suggesting that $A_{D-3}^{+++}$ is a continuous symmetry of an extension of the Einstein-Hilbert action. We have given examples in section \ref{obstruction} of both matter terms and Chern-Simons terms which can be added to the Einstein-Hilbert term so that the extended action admits full interpolating bound state solutions. The class of solutions that we have considered has not included the dimensional lift of the dyonic membrane. The embedding of dyonic membrane into the twelve-dimensional theory merits a closer examination as it provides a link between bound states involving mixed-symmetry fields in twelve dimensions and form fields in eleven dimensions and one expects to recover a full interpolating solution in twelve dimensions. The level one and level two generators of ${\mathfrak e}_{11}$ are lifted to the $A_9^{+++}$ generators
\begin{align}
R_{\mu_9\mu_{10}\mu_{11}}&=\frac{1}{8!}\epsilon_{\mu_1\ldots \mu_{11}}R^{\mu_1\ldots \mu_8}\longrightarrow R^{\mu_1\ldots \mu_9|\mu_9}\\
R_{\mu_6\mu_7\mu_8\mu_9\mu_{10}\mu_{11}}&=\frac{1}{5!11!}\epsilon_{\mu_1\ldots \mu_{11}}\epsilon_{\nu_1\ldots \nu_{11}}R^{\nu_1\ldots \nu_{11}|\mu_1\ldots \mu_5}\longrightarrow R^{\nu_1\ldots \nu_{12}|\mu_1\ldots \mu_6|\rho|\sigma}.
\end{align}
The M2-M5 tower of solutions \cite{Englert:2007qb} relevant to M-theory may also be re-interpreted in the context of $A_{D-3}^{+++}$ algebras. Via the dimensional reduction of $A_{12}^{+++}$ it is possible to understand the affine multiplet of states including the M2 and M5 brane uncovered in \cite{Englert:2007qb} as part of the Geroch group associated to the twelve-dimensional gravitational theory. The dimensional reduction of the solutions related by the Geroch group in twelve dimensions includes all the states recognised in both the gravity tower and the M2-M5 tower of \cite{Englert:2007qb}. 

It is to be hoped that the analysis of brane solutions as null geodesics on cosets of finite sub-groups of Kac-Moody algebras will be extended to null geodesics on cosets of affine sub-algebras. The difference between an affine algebra and the series of $\mathfrak{sl}(3,\mathbb{R})$ sub-algebras we have investigated in the present work is seemingly small. After all the affine coset model would be expected to describe solutions with continuous parameters that move directly between any pair of generators appearing at any level in the algebraic decomposition, while the sequence of $\mathfrak{sl}(3,\mathbb{R})$ sub-algebras we have investigated only interpolate directly between generators appearing at adjacent levels. However suitable combinations of interpolating solutions encoded in $\mathfrak{sl}(3,\mathbb{R})$ 
sub-algebras may be found that interpolate between any two levels in the algebra. The algebra $A^{+}_{2}$ formed by the roots $\bm\alpha_{1}$, $\bm\alpha_{2}$ and $\bm\alpha_{0}=\vec e_{3}-\vec e_{10}$ contains all of the roots $\bm\alpha_{i}$ listed above and would include them in one model forming a substantial subsector of M-theory. 

\section*{Acknowledgements}
It is a pleasure to thank Ling Bao, Nicolas Boulanger, Lisa Carbone, Axel Kleinschmidt and Peter West for discussions during the course of this work. PPC would like to thank the Isaac Newton Institute and the organisers of ``The Mathematics and Applications of Branes in String and M-Theory" workshop where part of this work was carried out. This work has been supported by the STFC rolling grant (ST/G00395/1) of the theoretical physics group at King's College London.

\newpage

\appendix
\section{Low-level roots of $A_8^{+++}$.}
\begin{longtable}{|r|r@{\ }r@{\ }r@{\ }r@{\ }r@{\ }r@{\ }r@{\ }r@{\ }r@{\ }r|r@{\ }r@{\ }r@{\ }r@{\ }r@{\ }r@{\ }r@{\ }r@{\ }r@{\ }r@{\ }r|r|r|r|r|} 
\caption{$A_{10}^{}$ representations in $A_{8}^{+++}$} \label{rootsofA8+++}\\ 
\hline 
\multicolumn{1}{|c|}{$m_{11}$} & 
\multicolumn{10}{|c|}{${\mathcal P}_r$} & 
\multicolumn{11}{|c|}{$root vector, \beta$} & 
\multicolumn{1}{|c|}{$\beta^2$} & 
\multicolumn{1}{|c|}{$dimension$} & 
\multicolumn{1}{|c|}{$mult$} & 
\multicolumn{1}{|c|}{$mu$}\\ 
\hline 
\hline 
0 & 0 & 0 & 0 & 0 & 0 & 0 & 0 & 0 & 0 & 0 & 0 & 0 & 0 & 0 & 0 & 0 & 0 & 0 & 0 & 0 & 0 & 0 & 1 & 11 & 1\\ 
0 & 1 & 0 & 0 & 0 & 0 & 0 & 0 & 0 & 0 & 1 & -1 & -1 & -1 & -1 & -1 & -1 & -1 & -1 & -1 & -1 & 0 & 2 & 120 & 1 & 1\\ 
\hline 
1 & 0 & 1 & 0 & 0 & 0 & 0 & 0 & 0 & 0 & 0 & 0 & 0 & 1 & 1 & 1 & 1 & 1 & 1 & 1 & 1 & 1 & 0 & 55 & 8 & 0\\ 
1 & 0 & 0 & 1 & 0 & 0 & 0 & 0 & 0 & 0 & 1 & 0 & 0 & 0 & 0 & 0 & 0 & 0 & 0 & 0 & 0 & 1 & 2 & 1760 & 1 & 1\\ 
\hline 
2 & 0 & 0 & 0 & 1 & 0 & 0 & 0 & 0 & 0 & 0 & 1 & 2 & 3 & 2 & 2 & 2 & 2 & 2 & 2 & 2 & 2 & -4 & 330 & 185 & 0\\ 
2 & 1 & 0 & 1 & 0 & 0 & 0 & 0 & 0 & 0 & 0 & 0 & 1 & 2 & 2 & 2 & 2 & 2 & 2 & 2 & 2 & 2 & -2 & 1485 & 44 & 1\\ 
2 & 0 & 2 & 0 & 0 & 0 & 0 & 0 & 0 & 0 & 0 & 0 & 0 & 2 & 2 & 2 & 2 & 2 & 2 & 2 & 2 & 2 & 0 & 1210 & 8 & 0\\ 
2 & 0 & 0 & 0 & 0 & 1 & 0 & 0 & 0 & 0 & 1 & 1 & 2 & 3 & 2 & 1 & 1 & 1 & 1 & 1 & 1 & 2 & -2 & 4752 & 40 & 1\\ 
2 & 1 & 0 & 0 & 1 & 0 & 0 & 0 & 0 & 0 & 1 & 0 & 1 & 2 & 1 & 1 & 1 & 1 & 1 & 1 & 1 & 2 & 0 & 33033 & 8 & 1\\ 
2 & 0 & 0 & 0 & 0 & 0 & 1 & 0 & 0 & 1 & 0 & 1 & 2 & 3 & 2 & 1 & 0 & 0 & 0 & 0 & 1 & 2 & 0 & 20328 & 6 & 0\\ 
2 & 0 & 0 & 0 & 0 & 0 & 1 & 0 & 0 & 0 & 2 & 1 & 2 & 3 & 2 & 1 & 0 & 0 & 0 & 0 & 0 & 2 & 2 & 25740 & 1 & 1\\ 
2 & 0 & 1 & 1 & 0 & 0 & 0 & 0 & 0 & 0 & 1 & 0 & 0 & 1 & 1 & 1 & 1 & 1 & 1 & 1 & 1 & 2 & 2 & 57200 & 1 & 1\\ 
2 & 1 & 0 & 0 & 0 & 1 & 0 & 0 & 0 & 1 & 0 & 0 & 1 & 2 & 1 & 0 & 0 & 0 & 0 & 0 & 1 & 2 & 2 & 214500 & 1 & 1\\ 
\hline 
3 & 0 & 0 & 0 & 0 & 0 & 1 & 0 & 0 & 0 & 0 & 2 & 4 & 6 & 5 & 4 & 3 & 3 & 3 & 3 & 3 & 3 & -12 & 462 & 19852 & 2\\ 
3 & 0 & 0 & 0 & 0 & 0 & 0 & 1 & 0 & 0 & 1 & 2 & 4 & 6 & 5 & 4 & 3 & 2 & 2 & 2 & 2 & 3 & -10 & 3168 & 6376 & 4\\ 
3 & 1 & 0 & 0 & 0 & 1 & 0 & 0 & 0 & 0 & 0 & 1 & 3 & 5 & 4 & 3 & 3 & 3 & 3 & 3 & 3 & 3 & -10 & 4620 & 7000 & 6\\ 
3 & 0 & 0 & 0 & 0 & 0 & 0 & 0 & 1 & 1 & 0 & 2 & 4 & 6 & 5 & 4 & 3 & 2 & 1 & 1 & 2 & 3 & -8 & 5445 & 1816 & 2\\ 
3 & 0 & 0 & 0 & 0 & 0 & 0 & 0 & 1 & 0 & 2 & 2 & 4 & 6 & 5 & 4 & 3 & 2 & 1 & 1 & 1 & 3 & -6 & 7722 & 584 & 2\\ 
3 & 0 & 1 & 0 & 1 & 0 & 0 & 0 & 0 & 0 & 0 & 1 & 2 & 4 & 3 & 3 & 3 & 3 & 3 & 3 & 3 & 3 & -8 & 13068 & 2332 & 5\\ 
3 & 0 & 0 & 0 & 0 & 0 & 0 & 0 & 0 & 2 & 1 & 2 & 4 & 6 & 5 & 4 & 3 & 2 & 1 & 0 & 1 & 3 & -4 & 7865 & 120 & 1\\ 
3 & 2 & 0 & 0 & 1 & 0 & 0 & 0 & 0 & 0 & 0 & 0 & 2 & 4 & 3 & 3 & 3 & 3 & 3 & 3 & 3 & 3 & -6 & 17160 & 691 & 2\\ 
3 & 0 & 0 & 2 & 0 & 0 & 0 & 0 & 0 & 0 & 0 & 1 & 2 & 3 & 3 & 3 & 3 & 3 & 3 & 3 & 3 & 3 & -6 & 9075 & 712 & 1\\ 
3 & 0 & 0 & 0 & 0 & 0 & 0 & 0 & 0 & 1 & 3 & 2 & 4 & 6 & 5 & 4 & 3 & 2 & 1 & 0 & 0 & 3 & 0 & 8008 & 8 & 0\\ 
3 & 1 & 0 & 0 & 0 & 0 & 1 & 0 & 0 & 0 & 1 & 1 & 3 & 5 & 4 & 3 & 2 & 2 & 2 & 2 & 2 & 3 & -8 & 47190 & 2116 & 7\\ 
3 & 1 & 1 & 1 & 0 & 0 & 0 & 0 & 0 & 0 & 0 & 0 & 1 & 3 & 3 & 3 & 3 & 3 & 3 & 3 & 3 & 3 & -4 & 37752 & 192 & 3\\ 
3 & 0 & 3 & 0 & 0 & 0 & 0 & 0 & 0 & 0 & 0 & 0 & 0 & 3 & 3 & 3 & 3 & 3 & 3 & 3 & 3 & 3 & 0 & 15730 & 8 & 0\\ 
3 & 1 & 0 & 0 & 0 & 0 & 0 & 1 & 0 & 1 & 0 & 1 & 3 & 5 & 4 & 3 & 2 & 1 & 1 & 1 & 2 & 3 & -6 & 135135 & 558 & 4\\ 
3 & 1 & 0 & 0 & 0 & 0 & 0 & 1 & 0 & 0 & 2 & 1 & 3 & 5 & 4 & 3 & 2 & 1 & 1 & 1 & 1 & 3 & -4 & 177870 & 162 & 3\\ 3 & 1 & 0 & 0 & 0 & 0 & 0 & 0 & 2 & 0 & 0 & 1 & 3 & 5 & 4 & 3 & 2 & 1 & 0 & 1 & 2 & 3 & -4 & 94380 & 113 & 1\\ 
3 & 0 & 1 & 0 & 0 & 1 & 0 & 0 & 0 & 0 & 1 & 1 & 2 & 4 & 3 & 2 & 2 & 2 & 2 & 2 & 2 & 3 & -6 & 205920 & 657 & 8\\ 
3 & 2 & 0 & 0 & 0 & 1 & 0 & 0 & 0 & 0 & 1 & 0 & 2 & 4 & 3 & 2 & 2 & 2 & 2 & 2 & 2 & 3 & -4 & 261360 & 182 & 3\\ 
3 & 1 & 0 & 0 & 0 & 0 & 0 & 0 & 1 & 1 & 1 & 1 & 3 & 5 & 4 & 3 & 2 & 1 & 0 & 0 & 1 & 3 & -2 & 394240 & 29 & 1\\ 
3 & 0 & 0 & 1 & 1 & 0 & 0 & 0 & 0 & 0 & 1 & 1 & 2 & 3 & 2 & 2 & 2 & 2 & 2 & 2 & 2 & 3 & -4 & 297297 & 185 & 4\\ 
3 & 1 & 0 & 0 & 0 & 0 & 0 & 0 & 1 & 0 & 3 & 1 & 3 & 5 & 4 & 3 & 2 & 1 & 0 & 0 & 0 & 3 & 2 & 314600 & 1 & 1\\ 
3 & 1 & 1 & 0 & 1 & 0 & 0 & 0 & 0 & 0 & 1 & 0 & 1 & 3 & 2 & 2 & 2 & 2 & 2 & 2 & 2 & 3 & -2 & 970200 & 44 & 4\\ 
3 & 0 & 1 & 0 & 0 & 0 & 1 & 0 & 0 & 1 & 0 & 1 & 2 & 4 & 3 & 2 & 1 & 1 & 1 & 1 & 2 & 3 & -4 & 926640 & 158 & 4\\ 
3 & 2 & 0 & 0 & 0 & 0 & 1 & 0 & 0 & 1 & 0 & 0 & 2 & 4 & 3 & 2 & 1 & 1 & 1 & 1 & 2 & 3 & -2 & 1156155 & 41 & 2\\ 
3 & 1 & 0 & 2 & 0 & 0 & 0 & 0 & 0 & 0 & 1 & 0 & 1 & 2 & 2 & 2 & 2 & 2 & 2 & 2 & 2 & 3 & 0 & 731808 & 8 & 1\\ 
3 & 0 & 1 & 0 & 0 & 0 & 1 & 0 & 0 & 0 & 2 & 1 & 2 & 4 & 3 & 2 & 1 & 1 & 1 & 1 & 1 & 3 & -2 & 1179750 & 40 & 3\\ 
3 & 2 & 0 & 0 & 0 & 0 & 1 & 0 & 0 & 0 & 2 & 0 & 2 & 4 & 3 & 2 & 1 & 1 & 1 & 1 & 1 & 3 & 0 & 1470150 & 8 & 1\\ 
3 & 0 & 2 & 1 & 0 & 0 & 0 & 0 & 0 & 0 & 1 & 0 & 0 & 2 & 2 & 2 & 2 & 2 & 2 & 2 & 2 & 3 & 2 & 880880 & 1 & 1\\ 
3 & 0 & 1 & 0 & 0 & 0 & 0 & 1 & 1 & 0 & 0 & 1 & 2 & 4 & 3 & 2 & 1 & 0 & 0 & 1 & 2 & 3 & -2 & 1359072 & 28 & 1\\ 
3 & 2 & 0 & 0 & 0 & 0 & 0 & 1 & 1 & 0 & 0 & 0 & 2 & 4 & 3 & 2 & 1 & 0 & 0 & 1 & 2 & 3 & 0 & 1681680 & 7 & 0\\ 
3 & 0 & 0 & 1 & 0 & 1 & 0 & 0 & 0 & 1 & 0 & 1 & 2 & 3 & 2 & 1 & 1 & 1 & 1 & 1 & 2 & 3 & -2 & 2265120 & 40 & 3\\ 
3 & 0 & 1 & 0 & 0 & 0 & 0 & 1 & 0 & 1 & 1 & 1 & 2 & 4 & 3 & 2 & 1 & 0 & 0 & 0 & 1 & 3 & 0 & 4459455 & 6 & 1\\ 
3 & 0 & 0 & 1 & 0 & 1 & 0 & 0 & 0 & 0 & 2 & 1 & 2 & 3 & 2 & 1 & 1 & 1 & 1 & 1 & 1 & 3 & 0 & 2837835 & 8 & 2\\ 
3 & 0 & 0 & 0 & 2 & 0 & 0 & 0 & 0 & 1 & 0 & 1 & 2 & 3 & 1 & 1 & 1 & 1 & 1 & 1 & 2 & 3 & 0 & 1486485 & 8 & 1\\ 
3 & 2 & 0 & 0 & 0 & 0 & 0 & 1 & 0 & 1 & 1 & 0 & 2 & 4 & 3 & 2 & 1 & 0 & 0 & 0 & 1 & 3 & 2 & 5505500 & 1 & 1\\ 
3 & 1 & 1 & 0 & 0 & 1 & 0 & 0 & 0 & 1 & 0 & 0 & 1 & 3 & 2 & 1 & 1 & 1 & 1 & 1 & 2 & 3 & 0 & 6795360 & 8 & 2\\ 
3 & 0 & 0 & 0 & 2 & 0 & 0 & 0 & 0 & 0 & 2 & 1 & 2 & 3 & 1 & 1 & 1 & 1 & 1 & 1 & 1 & 3 & 2 & 1849848 & 0 & 0\\ 
3 & 1 & 1 & 0 & 0 & 1 & 0 & 0 & 0 & 0 & 2 & 0 & 1 & 3 & 2 & 1 & 1 & 1 & 1 & 1 & 1 & 3 & 2 & 8494200 & 1 & 1\\ 
3 & 0 & 0 & 1 & 0 & 0 & 1 & 0 & 1 & 0 & 0 & 1 & 2 & 3 & 2 & 1 & 0 & 0 & 0 & 1 & 2 & 3 & 0 & 5813808 & 6 & 0\\ 
3 & 1 & 0 & 1 & 1 & 0 & 0 & 0 & 0 & 1 & 0 & 0 & 1 & 2 & 1 & 1 & 1 & 1 & 1 & 1 & 2 & 3 & 2 & 10900890 & 1 & 1\\ 
3 & 1 & 1 & 0 & 0 & 0 & 1 & 0 & 1 & 0 & 0 & 0 & 1 & 3 & 2 & 1 & 0 & 0 & 0 & 1 & 2 & 3 & 2 & 16816800 & 1 & 1\\ 
3 & 0 & 0 & 1 & 0 & 0 & 1 & 0 & 0 & 1 & 1 & 1 & 2 & 3 & 2 & 1 & 0 & 0 & 0 & 0 & 1 & 3 & 2 & 17571840 & 1 & 1\\ 
3 & 0 & 0 & 0 & 1 & 1 & 0 & 0 & 1 & 0 & 0 & 1 & 2 & 3 & 1 & 0 & 0 & 0 & 0 & 1 & 2 & 3 & 2 & 8305440 & 1 & 1\\ 
\hline 
\end{longtable}

\section{$\frak{sl}(3,\mathbb{R})/\frak{so}(1,2)$ gravitational 
solutions Ricci scalar} \label{appriemann}
In this paper we produce several solutions from the coset model which 
describe a 5 dimensional metric of the form
\begin{equation}
ds^{2} = f(x_{1}+ i x_{2})(dx^{2}_{1}+dx^{2}_{2})+g_{\mu\nu}(x_{1}+ 
ix_{2})dx^{\mu}dx^{\nu}
\end{equation}
\noindent where Greek indices run over coordinates $x_{3}$ to $x_{5}$ and we let Latin indices represent the coordinates $x_{1}$ and $x_{2}$. It can be easily shown that particular components of the Riemann tensor vanish namely
\begin{align}
R_{a\mu\nu\rho}&=0 \qquad \mbox{and}\\
R_{\mu\nu\rho\sigma}&=0.
\end{align}
The first equality is true for metrics where $f$ and the 
components of $g_{\mu\nu}$ are functions of $x_{1}$ and $x_{2}$, while 
the second equality requires $f$ and $g_{\mu\nu}$ to be (anti-)holomorphic functions. 
It is also easy to verify that, due to the fact that the components of 
the metric are holomorphic,
\begin{equation}
R_{\mu 1 \nu 1} = -R_{\mu 2 \nu 2}\label{R1122}.
\end{equation}
\noindent Therefore the only non-zero components of the Ricci tensor are 
$R_{a b}$. Another direct result of equation (\ref{R1122}) is that $R_{11}=-R_{22}$ 
so that all of these solutions have zero Ricci scalar.
For the solution presented in the first $\mathfrak{sl}(3,\mathbb{R})$ model (\ref{2kk6metric}) it is easiest to transform to $z=x_{1}+ix_{2}$ and $\bar{z}=x_{1}-ix_{2}$ where the only non-zero Ricci tensor component is 
\begin{eqnarray}
R_{zz} &=& -N_{2}^{3}A_{1}'-N_{2}^{2}(N_{1}')^{2}+N_{1}^{2}\left(N_{1}A_{2}'-(N_{2}')^{2}\right)\nonumber \\
&&+N_{1}N_{2}\left((A_{12}')^{2}-2A_{1}A_{12}'A_{2}'+A_{1}^{2}(A_{2}')^{2}+N_{1}'N_{2}'\right)
\end{eqnarray}
where the prime indicates a derivative with respect to $z$. For the dual gravity fields we found in section \ref{boundstateofdualgravitons} each term cancels except for those which contain factors of $A_{1}$. While these terms vanish for the limits where $\beta=0$ or $\tfrac{\pi}{2}$ they are non-zero for the interpolating metrics. 

\section{The Ricci scalar in the sigma model}\label{scalar}
The metric is related to the fields ${h_a}^b$ which appear at level zero in the decomposition of the $A_{D-3}^{+++}$ algebra by:
\begin{equation}
g_{\mu\nu}={(e^{-h})_\mu}^a{(e^{-h})_\nu}^b\eta_{ab}
\end{equation}
the Chistoffel symbols are
\begin{equation}
\Gamma_{\mu\nu}^\rho=-\partial_{\mu}{h_{\nu}}^\rho-\partial_{\nu}{h_{\mu}}^\rho + \partial^{\rho}{h_{\mu\nu}}
\end{equation}
and the Ricci scalar is
\begin{align}
R=&-2\partial_\kappa\partial^\lambda  {h_\lambda}^\kappa+2\partial_\kappa\partial^\kappa  {h_\lambda}^\lambda+ 2\partial_\sigma  {h_\lambda}^\sigma\partial^\kappa  {h_\kappa}^\lambda + 4\partial^\kappa  {h_\kappa}^\lambda\partial_\lambda  {h_\sigma}^\sigma -4\partial_\sigma  {h_\lambda}^\kappa\partial^\lambda  {h_\kappa}^\sigma\\
\nonumber &  -\partial^\kappa  {h_\lambda}^\sigma\partial^\kappa  {h_\sigma}^\lambda-\partial_\lambda  {h_\kappa}^\kappa\partial^\lambda  {h_\sigma}^\sigma.
\end{align}
The bound state of dual gravitons that we have studied in this paper have representative coset group elements
\begin{equation}
g=\exp(\phi_1 H_1 +\phi_2 H_2)\exp(C_1 E_1 + C_2 E_2 +C_{12}E_{12}).
\end{equation}
The bound state constructed from real roots at levels one and two in the algebra decomposition has
\begin{align}
H_1&= -({K^1}_1+{K^2}_2+{K^3}_3)+{K^D}_D \\
H_2&= -({K^1}_1+{K^2}_2+{K^D}_D)+{K^{D-1}}_{D-1} 
\end{align}
and hence the non-zero components of ${h_\mu}^\nu$ are
\begin{align}
{h_1}^1={h_2}^2=-(\phi_1+\phi_2), \quad {h_3}^3=-\phi_1, \quad {h_{D-1}}^{D-1}=\phi_2 \quad \mbox{and} \quad {h_D}^D=\phi_1-\phi_2
\end{align}
and consequently
\begin{equation}
R=-2\partial_\xi \phi_1 \partial^\xi \phi_1+2\partial_\xi \phi_1 \partial^\xi \phi_2-2\partial_\xi \phi_2 \partial^\xi \phi_2-2\partial_\xi\partial^\xi \phi_1 -2\partial_\xi\partial^\xi \phi_2 
\end{equation}
where $\phi_1$ and $\phi_2$ are functions $\xi$ which is identified with $x^1$ or $x^2$ (and $\xi$ is not summed over). The Ricci scalar is identical to the set of terms in $\phi_1$ and $\phi_2$ that appear in the brane sigma-model Lagrangian for the $\frac{SL(3,\mathbb{R})}{SO(1,2)}$ coset when the lapse function $\eta$ is set to minus one and the total derivative terms in the above are discarded.
\section{The Einstein Equations for the $D=5$ bound state.}\label{EinsteinEquations}
The equations of motion for the action in equation (\ref{action}) when $D=5$ are
\begin{align}
0=&R_{\mu\nu}-\frac{1}{2}g_{\mu\nu}R+g_{\mu\nu}\frac{1}{24}F_{\mu_1\mu_2\mu_3|\nu}F^{\mu_1\mu_2\mu_3|\nu}+g_{\mu\nu}\frac{1}{192}G_{\mu_1\mu_2\mu_3\mu_4|\nu_1\nu_2|\rho}G^{\mu_1\mu_2\mu_3\mu_4|\nu_1\nu_2|\rho}\label{level0equation}\\
\nonumber &-\frac{1}{4}F_{\mu\mu_2\mu_3|\nu_1}{F_\nu}^{\mu_2\mu_3|\nu_1}-\frac{1}{12}F_{\mu_1\mu_2\mu_3|\mu}{F^{\mu_1\mu_2\mu_3|}}_{\nu}\\
\nonumber &-\frac{1}{24}G_{\mu\mu_2\mu_3\mu_4|\nu_1\nu_2|\rho}{G^\nu}^{\mu_2\mu_3\mu_4|\nu_1\nu_2|\rho}-\frac{1}{48}G_{\mu_1\mu_2\mu_3\mu_4|\mu\nu_2|\rho}{{G^{\mu_1\mu_2\mu_3\mu_4|}}_{\nu}}^{\nu_2|\rho}\\
\nonumber &-\frac{1}{96}G_{\mu_1\mu_2\mu_3\mu_4|\nu_1\nu_2|\rho}{G^{\mu_1\mu_2\mu_3\mu_4|\nu_1\nu_2|}}_\nu \\
0=&\partial_{\mu_3}(\sqrt{-g} F^{\mu_1\mu_2\mu_3|\nu})-\frac{1}{2}\partial_{\nu_1}(\sqrt{-g}G^{\nu_1\nu_2\nu_3\nu|\mu_1\mu_2|\rho})A_{\nu_2\nu_3|\rho}\label{level1equation}\\
\nonumber &+\frac{1}{2}\partial_{\mu_3}(\sqrt{-g}G^{\mu_1\mu_2\mu_3\mu_4|\nu_1\nu_2|\nu})A_{\nu_1\nu_2|\mu_4}
+\frac{\sqrt{-g}}{4}G^{\mu_1\mu_2\mu_3\mu_4|\nu_1\nu_2|\nu}F_{\nu_1\nu_2|\mu_3\mu_4}\\
\nonumber &-\frac{\sqrt{-g}}{6}G^{\nu_1\nu_2\nu_3\nu|\mu_1\mu_2|\rho}F_{\nu_1\nu_2\nu_3|\rho}\\
0=&\partial_{\mu_1} (\sqrt{-g}G^{\mu_1\mu_2\mu_3\mu_4|\nu_1\nu_2|\rho})\label{level2equation}.
\end{align}
The null-geodesic motion on the coset $\frac{SL(3,\mathbb{R})}{SO(1,2)}$ parameterised by $\xi=x^1$ is given by the volume element
\begin{equation}
ds^2=N_1N_2\bigg((dx^1)^2+(dx^2)^2+\tfrac{1}{N_2}(dx^3)^2-\tfrac{1}{N_1N_2^2}(dx^4)^2+\tfrac{1}{N_1^2}(dx^5)^2\bigg) \label{diagonalmetric}
\end{equation}
where $N_1=1+Qx^1$ and $N_2=1+Qx^1 \cos^2{\beta}$ and the non-zero field strength components
\begin{align}
\nonumber F_{\hat{1} \hat{4} \hat{5}|\hat{5}}&=-\sin{\beta}\, \partial_{\hat{1}} N_1^{-1},\\
F_{\hat{1} \hat{3}\hat{4}|\hat{4}}&=-\tan{\beta}\, \partial_{\hat{1}} N_2^{-1} \qquad \mbox{and} \\
\nonumber G_{{\hat{1}} \hat{3}\hat{4}\hat{5}|\hat{4}\hat{5}| \hat{4}}&= -\cos{\beta}\, \frac{\partial_{\hat{1}} N_1}{N_1 N_2}
\end{align}
implying the non-zero gauge-field components are
\begin{align}
\nonumber A_{\hat{4} \hat{5}|\hat{5}}&=-\sin{\beta}\,  N_1^{-1},\\
A_{\hat{3}\hat{4}|\hat{4}}&=-\tan{\beta}\, N_2^{-1} \qquad \mbox{and} \\
\nonumber A_{\hat{3}\hat{4}\hat{5}|\hat{4}\hat{5}| \hat{4}}&=\frac{1}{2}\cos{\beta}(\frac{1}{N_2}+\frac{1}{N_1\cos^2{\beta}}).
\end{align}
Equation (\ref{level2equation}) is satisfied as $N_1$ is a harmonic function:
\begin{equation}
0=\partial_{\hat{1}} (\sqrt{-g}G^{\hat{1}\hat{3}\hat{4}\hat{5}|\hat{4}\hat{5}|\hat{4}})=\cos{\beta} \partial_{\hat{1}}\partial_{\hat{1}}N_1=0.
\end{equation}
Splitting equation (\ref{level1equation}) into the two non-trivial equations gives as the coefficient of the variations $\delta A_{\hat{4}\hat{5}|\hat{5}}$ 
\begin{align}
\nonumber \partial_{\hat{1}}(\sqrt{-g} F^{\hat{1}\hat{4}\hat{5}|\hat{5}})-\sqrt{-g}G^{\hat{1}\hat{3}\hat{4}\hat{5}|\hat{4}\hat{5}|\hat{4}}F_{\hat{1}\hat{3}\hat{4}|\hat{4}}&=-\sin{\beta} \partial_{\hat{1}} (\frac{\partial_{\hat{1}}N_1}{N_2})-\cos{\beta}\partial_{\hat{1}}N_1 (-\tan{\beta}\, \partial_{\hat{1}} N_2^{-1})\\
 &=0
\end{align}
and for $\delta A_{\hat{3}\hat{4}|\hat{4}}$ we have 
\begin{align}
\partial_{\hat{1}}(\sqrt{-g} F^{\hat{1}\hat{3}\hat{4}|\hat{4}})+\sqrt{-g}G^{\hat{1}\hat{3}\hat{4}\hat{5}|\hat{4}\hat{5}|\hat{4}}F_{\hat{4}\hat{5}|\hat{1}\hat{5}}&=\tan{\beta} \partial_{\hat{1}} (\frac{\partial_{\hat{1}}N_2}{N_1})+\cos{\beta}\partial_{\hat{1}}N_1 (-\sin{\beta}\, \partial_{\hat{1}} N_1^{-1})\\
\nonumber &=0
\end{align}
where in the final equation we note that $\partial_{\hat{1}}N_2=\cos^2{\beta}\partial_{\hat{1}}N_1$. For the Einstein equations (\ref{level0equation}) it is notationally useful to write $F_1^2\equiv 6F_{\hat{1}\hat{4}\hat{5}|\hat{5}}F^{\hat{1}\hat{4}\hat{5}|\hat{5}}$, $F_2^2\equiv 6F_{\hat{1}\hat{3}\hat{4}|\hat{4}}F^{\hat{1}\hat{3}\hat{4}|\hat{4}}$, $G^2\equiv 48G_{\hat{1}\hat{3}\hat{4}\hat{5}|\hat{4}\hat{5}|\hat{4}}G^{\hat{1}\hat{3}\hat{4}\hat{5}|\hat{4}\hat{5}|\hat{4}}$ and $\hat{G}_{\mu\nu}=R_{\mu\nu}-\frac{1}{2}g_{\mu\nu}R$, so that the five non-trivial Einstein equations are
\begin{align}
\hat{G}_{\hat{1}\hat{1}}&=-g_{\hat{1}\hat{1}}\frac{1}{4}(F_1^2+F_2^2+G^2)+g_{\hat{1}\hat{1}}\frac{1}{2}(F_1^2+F_2^2+G^2)\\
\nonumber &=\frac{-1}{4N_1^2N_2^2}(N_2^2(\partial_{\hat{1}}N_1)^2-N_1N_2\partial_{\hat{1}}N_1\partial_{\hat{1}}N_2+N_1^2(\partial_{\hat{1}}N_2)^2)\\
\hat{G}_{\hat{2}\hat{2}}&=-g_{\hat{2}\hat{2}}\frac{1}{4}(F_1^2+F_2^2+G^2)\\
\nonumber &=\frac{1}{4N_1^2N_2^2}(N_2^2(\partial_{\hat{1}}N_1)^2-N_1N_2\partial_{\hat{1}}N_1\partial_{\hat{1}}N_2+N_1^2(\partial_{\hat{1}}N_2)^2)\\
\hat{G}_{\hat{3}\hat{3}}&=-g_{\hat{3}\hat{3}}\frac{1}{4}(F_1^2+F_2^2+G^2)+g_{\hat{3}\hat{3}}\frac{1}{2}(F_2^2+G^2)\\
\nonumber &=\frac{1}{4N_1^2N_2^3}(N_2^2(\partial_{\hat{1}}N_1)^2-N_1^2(\partial_{\hat{1}}N_2)^2-N_1N_2\partial_{\hat{1}}N_1\partial_{\hat{1}}N_2)\\
\hat{G}_{\hat{4}\hat{4}}&=-g_{\hat{4}\hat{4}}\frac{1}{4}(F_1^2+F_2^2+G^2)+g_{\hat{4}\hat{4}}\frac{1}{2}(F_1^2+2F_2^2+3G^2)\\
\nonumber &=\frac{1}{4N_1^3N_2^4}(N_2^2(\partial_{\hat{1}}N_1)^2+3N_1^2(\partial_{\hat{1}}N_2)^2+N_1N_2\partial_{\hat{1}}N_1\partial_{\hat{1}}N_2)\\
\hat{G}_{\hat{5}\hat{5}}&=-g_{\hat{5}\hat{5}}\frac{1}{4}(F_1^2+F_2^2+G^2)+g_{\hat{5}\hat{5}}\frac{1}{2}(2F_1^2+2G^2)\\
\nonumber &=\frac{1}{4N_1^4N_2^2}(-3N_2^2(\partial_{\hat{1}}N_1)^2+N_1^2(\partial_{\hat{1}}N_2)^2-N_1N_2\partial_{\hat{1}}N_1\partial_{\hat{1}}N_2)
\end{align}
which gives the components of the Einstein tensor corresponding to the metric of equation (\ref{diagonalmetric}).

The dualised action, where $A_{[D-2|D-3|1]}$ has not been eliminated is
\begin{equation}
S'_1=\int R \star \mathbb{I} -\tfrac{1}{2} F_{[D-2|1]}\wedge \star F_{[D-2|1]} -{F_{[1|}}^{D-3|1]}dA_{[D-2|D-3|1]}-A^{[D-3|1]}F_{[2|D-3]}\wedge F_{[D-2|1]}\label{action4}
\end{equation}
where the metric is unchanged from equation (\ref{diagonalmetric}).
The equations of motion for the action in equation (\ref{action4}) when $D=5$ are
\begin{align}
0=&R_{\mu\nu}-\frac{1}{2}g_{\mu\nu}R+g_{\mu\nu}\frac{1}{24}F_{\mu_1\mu_2\mu_3|\nu}F^{\mu_1\mu_2\mu_3|\nu}\label{duallevel0equation}\\
\nonumber &-\frac{1}{4}F_{\mu\mu_2\mu_3|\nu_1}{F_\nu}^{\mu_2\mu_3|\nu_1}-\frac{1}{12}F_{\mu_1\mu_2\mu_3|\mu}{F^{\mu_1\mu_2\mu_3|}}_{\nu}\\
\nonumber &+\tfrac{1}{2\sqrt{-g}}{F_{\mu_1|\mu}}^{\nu_2|\rho}\bigg(\partial_{\mu_2}A_{\mu_3\mu_4 \mu_5|\nu\nu_2|\rho}-A_{\mu_2|\nu\nu_2}F_{\mu_3\mu_4\mu_5|\rho} \bigg) \\
\nonumber &+\tfrac{1}{\sqrt{-g}}{{F_{\mu_1|}}^{\nu_1\nu_2|}}_\mu\bigg(\partial_{\mu_2}A_{\mu_3\mu_4 \mu_5|\nu_1\nu_2|\nu}  -A_{\mu_2|\nu_1\nu_2}F_{\mu_3\mu_4\mu_5|\nu} \bigg) \\
0=&\partial_{\mu_3}(\sqrt{-g} F^{\mu_1\mu_2\mu_3|\nu})-\frac{1}{2}\partial_{\nu_1}({F_{\nu_2|}}^{\mu_1\mu_2|\rho}A_{\nu_3\nu_4|\rho} )-\frac{1}{2}{F_{\nu_2|}}^{\mu_1\mu_2|\rho}F_{\nu_1\nu_3\nu_4|\rho} \label{duallevel1equation} \\
0=&\partial_{\mu_1} ({{F_{\mu_2|}}^{\nu_1\nu_2|\rho}})
\label{duallevel2equation}.
\end{align}
The dualised bound state has non-trivial field strength and gauge field components
, the non-trivial field strength components are
\begin{align}
\nonumber F_{\hat{1} \hat{4} \hat{5}|\hat{5}}&=-\sin{\beta}\, \partial_{\hat{1}} N_1^{-1},\\
F_{\hat{1} \hat{3}\hat{4}|\hat{4}}&=-\tan{\beta}\, \partial_{\hat{1}} N_2^{-1} \qquad \mbox{and} \\
\nonumber  F_{\hat{2}\hat{4}\hat{5}|\hat{4}}\equiv F_{{\hat{2}}|\hat{4}\hat{5}|\hat{4}}&= \cos{\beta}\, \frac{\partial_{\hat{1}} N_1}{N_1 N_2}
\end{align}
implying the non-zero gauge-field components are
\begin{align}
\nonumber A_{\hat{4} \hat{5}|\hat{5}}&=-\sin{\beta}\,  N_1^{-1},\\
A_{\hat{3}\hat{4}|\hat{4}}&=-\tan{\beta}\, N_2^{-1} \qquad \mbox{and} \\
\nonumber A_{\hat{3}\hat{4}\hat{5}|\hat{4}\hat{5}| \hat{4}}&=\frac{1}{2}\cos{\beta}(\frac{1}{N_2}+\frac{1}{N_1\cos^2{\beta}}).
\end{align}
Equation (\ref{duallevel2equation}) is trivially satisfied for all field strength components, the least trivial equation being
\begin{equation}
\partial_{\hat{1}} ({{F_{\hat{2}|}}^{\hat{4}\hat{5}|\hat{4}}})=\partial_{\hat{1}} (\cos{\beta} \partial_{\hat{1}}N_1)=0
\end{equation}
which holds as $N_1$ is a harmonic function in $x^1$.
Equation (\ref{duallevel1equation}) splits into two non-trivial equations, the coefficient of the variations $\delta A_{\hat{4}\hat{5}|\hat{5}}$  gives
\begin{align}
\partial_{\hat{1}}(\sqrt{-g} F^{\hat{1}\hat{4}\hat{5}|\hat{5}})-\frac{1}{2}\partial_{\hat{1}}({F_{\hat{2}|}}^{\hat{4}\hat{5}|\hat{4}}A_{\hat{3}\hat{4}|\hat{4}})-\frac{1}{2}{F_{\hat{2}|}}^{\hat{4}\hat{5}|\hat{4}}F_{\hat{1}\hat{3}\hat{4}|\hat{4}}&=0
\end{align}
and the coefficient of $\delta A_{\hat{3}\hat{4}|\hat{4}}$ 
\begin{align}
\partial_{\hat{1}}(\sqrt{-g} F^{\hat{1}\hat{3}\hat{4}|\hat{4}})+\frac{1}{2}{F_{\hat{2}|}}^{\hat{4}\hat{5}|\hat{4}}F_{\hat{1}\hat{4}\hat{5}|\hat{5}}
+\frac{1}{2}\partial_{\hat{1}}({F_{\hat{2}|}}^{\hat{4}\hat{5}|\hat{4}}A_{\hat{4}\hat{5}|\hat{5}})&=0
\end{align}
For the Einstein equations (\ref{duallevel0equation}) it is notationally useful to write $F_{12}^2\equiv 6F_{\hat{2}\hat{4}\hat{5}|\hat{4}}F^{\hat{2}\hat{4}\hat{5}|\hat{4}}$ so that the five non-trivial Einstein equations are\footnote{For comparison with the previous set of Einstein equations it is useful to note that $F_{12}^2=-G^2$.}
\begin{align}
\hat{G}_{\hat{1}\hat{1}}=&-g_{\hat{1}\hat{1}}\frac{1}{4}(F_1^2+F_2^2+F_{12}^2)+g_{\hat{1}\hat{1}}\frac{1}{2}(F_1^2+F_2^2)\\
\nonumber =&\frac{-1}{4N_1^2N_2^2}(N_2^2(\partial_{\hat{1}}N_1)^2-N_1N_2\partial_{\hat{1}}N_1\partial_{\hat{1}}N_2+N_1^2(\partial_{\hat{1}}N_2)^2)\\
\hat{G}_{\hat{2}\hat{2}}=&-g_{\hat{2}\hat{2}}\frac{1}{4}(F_1^2+F_2^2+F_{12}^2)+g_{\hat{2}\hat{2}}\frac{1}{2}(F_{12}^2)\\
\nonumber =&\frac{1}{4N_1^2N_2^2}(N_2^2(\partial_{\hat{1}}N_1)^2-N_1N_2\partial_{\hat{1}}N_1\partial_{\hat{1}}N_2+N_1^2(\partial_{\hat{1}}N_2)^2)\\
\hat{G}_{\hat{3}\hat{3}}=&-g_{\hat{3}\hat{3}}\frac{1}{4}(F_1^2+F_2^2+F_{12}^2)+g_{\hat{3}\hat{3}}\frac{1}{2}(F_2^2)\\
\nonumber =&\frac{1}{4N_1^2N_2^3}(N_2^2(\partial_{\hat{1}}N_1)^2-N_1^2(\partial_{\hat{1}}N_2)^2-N_1N_2\partial_{\hat{1}}N_1\partial_{\hat{1}}N_2)\\
\hat{G}_{\hat{4}\hat{4}}=&-g_{\hat{4}\hat{4}}\frac{1}{4}(F_1^2+F_2^2+F_{12}^2)+g_{\hat{4}\hat{4}}\frac{1}{2}(F_1^2+2F_2^2+2F_{12}^2)\\
\nonumber &-g_{\hat{4}\hat{4}}\frac{2}{\sqrt{-g}}{F_{\hat{2}|}}^{\hat{4}\hat{5}|\hat{4}}(\partial_{\hat{1}}A_{\hat{3}\hat{4}\hat{5}|\hat{4}\hat{5}|\hat{4}}+\frac{1}{2}F_{\hat{1}\hat{3}\hat{4}|\hat{4}}A_{\hat{4}\hat{5}|\hat{5}}-\frac{1}{2}F_{\hat{1}\hat{4}\hat{5}|\hat{5}}A_{\hat{3}\hat{4}|\hat{4}})\\
\nonumber =&-g_{\hat{4}\hat{4}}\frac{1}{4}(F_1^2+F_2^2+F_{12}^2)+g_{\hat{4}\hat{4}}\frac{1}{2}(F_1^2+2F_2^2+2F_{12}^2)-2g_{\hat{4}\hat{4}}(F_{12}^2)\\
\nonumber =&\frac{1}{4N_1^3N_2^4}(N_2^2(\partial_{\hat{1}}N_1)^2+3N_1^2(\partial_{\hat{1}}N_2)^2+N_1N_2\partial_{\hat{1}}N_1\partial_{\hat{1}}N_2)\\
\hat{G}_{\hat{5}\hat{5}}=&-g_{\hat{5}\hat{5}}\frac{1}{4}(F_1^2+F_2^2+F_{12}^2)+g_{\hat{5}\hat{5}}\frac{1}{2}(2F_1^2+F_{12}^2)\\
\nonumber &-g_{\hat{5}\hat{5}}\frac{1}{\sqrt{-g}}{F_{\hat{2}|}}^{\hat{4}\hat{5}|\hat{4}}(\partial_{\hat{1}}A_{\hat{3}\hat{4}\hat{5}|\hat{4}\hat{5}|\hat{4}}+\frac{1}{2}F_{\hat{1}\hat{3}\hat{4}|\hat{4}}A_{\hat{4}\hat{5}|\hat{5}}-\frac{1}{2}F_{\hat{1}\hat{4}\hat{5}|\hat{5}}A_{\hat{3}\hat{4}|\hat{4}})\\
\nonumber =&-g_{\hat{5}\hat{5}}\frac{1}{4}(F_1^2+F_2^2+F_{12}^2)+g_{\hat{5}\hat{5}}\frac{1}{2}(2F_1^2+F_{12}^2)-g_{\hat{5}\hat{5}}(F_{12}^2)\\
\nonumber =&\frac{1}{4N_1^4N_2^2}(-3N_2^2(\partial_{\hat{1}}N_1)^2+N_1^2(\partial_{\hat{1}}N_2)^2-N_1N_2\partial_{\hat{1}}N_1\partial_{\hat{1}}N_2)
\end{align}
which gives the non-zero components of the Einstein tensor corresponding to the metric of equation (\ref{diagonalmetric}).

\bibliographystyle{utphys}
\bibliography{A+++cosetmodelsolutions}
\end{document}